\documentclass[fleqn,usenatbib]{mnras}

\usepackage[margin=5pt]{subfig}
\usepackage{comment}
\usepackage{longtable}
\usepackage{caption}
\usepackage{threeparttable}
\usepackage{booktabs}
\usepackage[figuresright]{rotating}
\usepackage{textcase}
\newcommand{\be}{\begin{equation}}
\newcommand{\ee}{\end{equation}}
\newcommand{\bea}{\begin{eqnarray}}
\newcommand{\eea}{\end{eqnarray}}
\newcommand{\hunit}{$\rm{km \ s^{-1} \ Mpc^{-1}}$}
\newcommand{\lcdm}{$\Lambda$CDM}
\newcommand{\pcdm}{$\phi$CDM}
\usepackage{array}
\makeatletter
\newcommand{\thickhline}{%
    \noalign {\ifnum 0=`}\fi \hrule height 1pt
    \futurelet \reserved@a \@xhline
}
\newcolumntype{"}{@{\hskip\tabcolsep\vrule width 1pt\hskip\tabcolsep}}
\makeatother

\newcommand{\hii}{H\,\textsc{ii}}
\newcommand{\Om}{\Omega_{\rm m0}}
\newcommand{\Ok}{\Omega_{\rm k0}}
\newcommand{\om}{$\Omega_{\rm m0}$}
\newcommand{\ok}{$\Omega_{\rm k0}$}
\newcommand{\wx}{$w_{\rm X}$}

\newcommand{\obhs}{$\Omega_{b}h^2$}
\newcommand{\ochs}{$\Omega_{c}h^2$}

\usepackage[T1]{fontenc}
\usepackage{scalerel}
\usepackage{tikz}
\usetikzlibrary{svg.path}

\definecolor{orcidlogocol}{HTML}{A6CE39}
\tikzset{
  orcidlogo/.pic={
    \fill[orcidlogocol] svg{M256,128c0,70.7-57.3,128-128,128C57.3,256,0,198.7,0,128C0,57.3,57.3,0,128,0C198.7,0,256,57.3,256,128z};
    \fill[white] svg{M86.3,186.2H70.9V79.1h15.4v48.4V186.2z}
                 svg{M108.9,79.1h41.6c39.6,0,57,28.3,57,53.6c0,27.5-21.5,53.6-56.8,53.6h-41.8V79.1z M124.3,172.4h24.5c34.9,0,42.9-26.5,42.9-39.7c0-21.5-13.7-39.7-43.7-39.7h-23.7V172.4z}
                 svg{M88.7,56.8c0,5.5-4.5,10.1-10.1,10.1c-5.6,0-10.1-4.6-10.1-10.1c0-5.6,4.5-10.1,10.1-10.1C84.2,46.7,88.7,51.3,88.7,56.8z};
  }
}

\newcommand\orcidicon[1]{\href{https://orcid.org/#1}{\mbox{\scalerel*{
\begin{tikzpicture}[yscale=-1,transform shape]
\pic{orcidlogo};
\end{tikzpicture}
}{|}}}}

\usepackage{hyperref} 

\DeclareRobustCommand{\VAN}[3]{#2}
\let\VANthebibliography\thebibliography
\def\thebibliography{\DeclareRobustCommand{\VAN}[3]{##3}\VANthebibliography}

\usepackage{graphicx}	
\usepackage{amsmath}	
\usepackage{amssymb}	
\usepackage{fnpct}

\title[GRB cosmological parameter constraints]{Standardizing Dainotti-correlated gamma-ray bursts, and using them with standardized Amati-correlated gamma-ray bursts to constrain cosmological model parameters}

 \author[Cao, Khadka, \& Ratra]{
 Shulei Cao,$^{\orcidicon{0000-0003-2421-7071}1}$\thanks{E-mail: shulei@phys.ksu.edu}
 Narayan Khadka,$^{\orcidicon{0000-0001-5512-2716}1}$\thanks{E-mail: nkhadka@phys.ksu.edu}
 Bharat Ratra$^{\orcidicon{0000-0002-7307-0726}1}$\thanks{E-mail: ratra@phys.ksu.edu}
 \\
 $^{1}$Department of Physics, Kansas State University, 116 Cardwell Hall, Manhattan, KS 66502, USA
 }

\date{Accepted XXX. Received YYY; in original form ZZZ}

\pubyear{2021}

\begin{document}
\label{firstpage}
\pagerange{\pageref{firstpage}--\pageref{lastpage}}
\maketitle

\begin{abstract}
We show that each of the three Dainotti-correlated gamma-ray burst (GRB) data sets recently compiled by Wang et al.\ and Hu et al., that together probe the redshift range $0.35 \leq z \leq 5.91$, obey cosmological-model-independent Dainotti correlations and so are standardizable. We use these GRB data in conjunction with the best currently-available Amati-correlated GRB data, that probe $0.3399 \leq z \leq 8.2$, to constrain cosmological model parameters. The resulting cosmological constraints are weak, providing lower limits on the non-relativistic matter density parameter, mildly favoring non-zero spatial curvature, and largely consistent with currently accelerated cosmological expansion as well as with constraints determined from better-established data.
\end{abstract}




\section{Introduction} \label{sec:intro}

The observed currently accelerated cosmological expansion indicates that --- if general relativity provides an accurate description of gravitation on cosmological scales --- dark energy must contribute significantly to the current cosmological energy budget. The simpler spatially-flat $\Lambda$CDM model \citep{peeb84} is consistent with this and other observations. Fits of this model to most better-established cosmological data suggest that a time-independent cosmological constant $(\Lambda)$ provides $\sim 70\%$ of the current cosmological energy budget, non-relativistic cold dark matter (CDM) provides $\sim 25\%$, and non-relativistic baryonic matter provides most of the remaining $\sim 5\%$ \citep[see, e.g.][]{Farooq_Ranjeet_Crandall_Ratra_2017, scolnic_et_al_2018, planck2018b, eBOSS_2020}. While the spatially-flat $\Lambda$CDM model is consistent with most observations \citep[see, e.g.][]{DiValentinoetal2021a, PerivolaropoulosSkara2021}, observational data do not strongly rule out a little spatial curvature or dynamical dark energy. In this paper, in addition to the spatially-flat $\Lambda$CDM model, we also study spatially non-flat and dynamical dark energy models.  

Observational astronomy now provides many measurements that can be used to test cosmological models. Largely, these data are either at low or at high redshift. So cosmological models are mostly tested at low and high redshifts, remaining poorly tested in the intermediate redshift regime. The highest redshift of the better-established low-redshift data, $\sim 2.3$, is reached through baryon acoustic oscillation (BAO) observations; the high redshift region, $z \sim 1100$, is probed by better-established cosmic microwave background anisotropy data. Fits of these better-established data to cosmological models provide mostly mutually consistent results. However, for a better understanding of our Universe, it is necessary to also test cosmological models in the intermediate redshift range of $2.3 \lesssim z \lesssim 1100$. Some progress has been achieved: methods that test cosmological models in the intermediate redshift region include the use of \hii\ starburst galaxy measurements which reach to $z \sim 2.4$ \citep{Mania_2012, Chavez_2014, GonzalezMoran2019, GonzalezMoranetal2021, Caoetal_2020, Caoetal_2021c, Caoetal_2021a, Johnsonetal2021}, quasar angular size measurements which reach to $\sim 2.7$ \citep{Cao_et_al2017a, Ryanetal2019, Caoetal_2020, Caoetal_2021a, Zhengetal2021, Lian_etal_2021}, and quasar flux measurements which reach to $\sim 7.5$ \citep{RisalitiLusso2015, RisalitiLusso2019, KhadkaRatra2020a, KhadkaRatra2020b, KhadkaRatra2021a, KhadkaRatra2021b, Yangetal2020, Lussoetal2020, ZhaoXia2021, Lietal2021, Lian_etal_2021, Rezaeietal2021, Luongoetal2021}.\footnote{Note that in the latest \cite{Lussoetal2020} quasar flux compilation, their assumed UV--X-ray correlation is valid only to a much lower redshift, $z \sim 1.5-1.7$, and so these quasars can be used to derive only lower-$z$ cosmological constraints \citep{KhadkaRatra2021a, KhadkaRatra2021b}.}  

Gamma-ray burst (GRB) measurements are another high redshift probe and reach to $z \sim 8.2$ \citep{Amati2008, Amati2019, Salvaterraetal2009, Tanviretal2009, samushia_ratra_2010, Cardoneetal2010, Dainottietal2013a, Wangetal2015, Wang_2016, DainottiaDelVecchio2017, Dirirsa2019, KhadkaRatra2020c, Demianskietal_2021, Khadkaetal_2021b, Luongoetal2021, galaxies9040077}. While there are quite a few Amati correlation long GRBs that have been used to constrain cosmological parameters, currently only a smaller fraction of 118 such GRBs (hereafter A118) that cover the redshift range $0.3399 \leq z \leq 8.2$ \citep{KhadkaRatra2020c, Khadkaetal_2021a} are reliable enough to be used to constrain cosmological parameters. To date, this is the lower-$z$ data set used to constrain cosmological parameters that spans the widest range of redshifts. These A118 data provide cosmological constraints which are consistent with those obtained from the better-established cosmological probes but the GRB constraints are significantly less restrictive. To obtain tighter cosmological constraints using GRB data, we need to make use of more GRBs.

Recently \cite{Wangetal_2021} and \cite{Huetal_2021} have compiled smaller GRB data sets that together probe the redshift range $0.35 \leq z \leq 5.91$. These are GRBs whose plateau phase luminosity $L_0$ and spin-down characteristic time $t_b$ are correlated through the Dainotti ($L_0-t_b$) correlation \citep{Dainottietal2013b,Dainottietal2017}. This correlation between $L_0$ and $t_b$ allows one to use these GRBs for cosmological purposes. These GRBs can be classified in two categories depending on whether the plateau phase is dominated by magnetic dipole (MD) radiation or gravitational wave (GW) emission \citep{Wangetal_2021, Huetal_2021}. In this paper we use long and short GRBs whose plateau phase is dominated by MD radiation (hereafter MD-LGRBs and MD-SGRBs) and long GRBs whose plateau phase is dominated by GW emission (hereafter GW-LGRBs). All three sets of GRBs obey the Dainotti correlation but each set can have different correlation parameters. We use the three individual GRB data sets, as well as some combinations of them, to constrain cosmological model parameters and Dainotti correlation parameters simultaneously.\footnote{The advantage of fitting cosmological and GRB correlation parameters simultaneously is that the fitting process is free from the circularity problem. More specifically, this procedure allows us to determine whether the GRB correlation parameters depend on the assumed cosmological model and so determine whether the GRBs are standardizable.} We find that these GRBs are standardizable, as was assumed in \cite{Wangetal_2021} and \cite{Huetal_2021}. However, cosmological constraints obtained from these Dainotti correlation GRB data sets are very weak. 

When we combine the MD-LGRB or GW-LGRB data sets with the 115 non-overlapping Amati correlation GRBs from the A118 data set, they slightly tighten the constraints from the 115 Amati correlation GRBs, but not significantly so. Each of the individual Amati or Dainotti correlation GRB data sets, as well as combinations of these GRB data sets, mostly provide only lower limits on the current value of the non-relativistic matter energy density parameter $\Omega_{\rm m0}$ and the resulting cosmological parameter constraints are mostly consistent with those obtained from better-established cosmological data. 

In this paper, we use a combination of Hubble parameter ($H(z)$) and BAO data, $H(z)$ + BAO, results as a proxy for better-established data results, to compare with our GRB data results. Qualitatively, results from the individual GRB data sets, as well as those from combinations of GRB data sets, are consistent with those from the $H(z)$ + BAO data which favor $\Omega_{\rm m0} \sim 0.3$, but there are a few combinations of GRB data sets with constraints on \om\ being more than $2\sigma$ away from 0.3 in the \lcdm\ models.

This paper is structured as follows. In Sec.\ \ref{sec:model} we summarize the cosmological models we use. In Sec.\ \ref{sec:data} we describe the data sets we analyze. In Sec.\ \ref{sec:analysis} we summarize our analyses techniques. In Sec.\ \ref{sec:results} we present our results. We conclude in Sec.\ \ref{sec:conclusion}.

\section{Cosmological models}
\label{sec:model}

In this paper we derive cosmological parameter constraints in six different general-relativity cosmological dark energy models. Three of them assume spatially-flat geometry while the other three allow non-flat spatial geometry.\footnote{For recent discussions of constraints on non-flat models see \cite{Chen_et_al_2016}, \cite{rana_jain_mahajan_mukherjee_2017}, \cite{ooba_etal_2018a, ooba_etal_2018c}, \cite{yu_etal_2018}, \cite{park_ratra_2019a, park_ratra_2019c}, \cite{wei_2018}, \cite{DES_2019}, \cite{li_etal_2020}, \cite{Handley2019}, \cite{efstathiou_gratton_2020}, \cite{DiValentinoetal2021b}, \cite{Velasquez-Toribio_2020}, \cite{Vagnozzi_2020a, Vagnozzi_2020b}, \cite{KiDSCollaboration2021}, \cite{ArjonaNesseris2021}, \cite{Dhawanetal2021}, and references therein.} These models are used to predict the luminosity distance for a GRB at a given redshift. For this purpose the fundamental quantity is the expansion rate of the Universe, or the Hubble parameter, $H(z)$, a function of cosmological parameters and redshift. 

The Hubble parameter in all six models we use can be written as
\begin{equation}
\label{eq:hubpar}
    H(z) = H_0\sqrt{\Omega_{\rm m0}(1 + z)^3 + \Omega_{\rm k0}(1 + z)^2 + \Omega_{\rm DE}(z)},
\end{equation}
where $H_0$ is the Hubble constant and $\Omega_{\rm k0}$ is the current value of the spatial curvature energy density parameter. In analyses of the $H(z)$ + BAO data set we express $\Omega_{\rm m0}$ in terms of the current value of the baryonic matter energy density parameter $(\Omega_{b})$ and the current value of the cold dark matter energy density parameter $(\Omega_{c})$ through the equation $\Omega_{\rm m0}$ = $\Omega_{b}$ + $\Omega_{c}$. In four of the six models, the dark energy density parameter term is expressed as $\Omega_{\rm DE}(z)$ = $\Omega_{\rm DE0}(1+z)^{1+w_{\rm DE}}$, where $\Omega_{\rm DE0}$ is the current value of the dark energy density parameter and $w_{\rm DE}$ is the dark energy equation of state parameter.

In the $\Lambda$CDM models $w_{\rm DE} = -1$ and $\Omega_{\rm DE}$ = $\Omega_{\rm DE0}$ = $\Omega_{\Lambda}$ is the cosmological constant dark energy density parameter and is time-independent. The current values of the three energy density parameters are related by the energy budget equation, $\Omega_{\rm m0} + \Omega_{\rm k0} + \Omega_{\Lambda} = 1$. In the spatially-flat $\Lambda$CDM model we choose to constrain $\Omega_{\rm m0}$ and $H_0$ while in the spatially non-flat $\Lambda$CDM model we constrain $\Omega_{\rm m0}$, $\Omega_{\rm k0}$, and $H_0$. For analyses which involve $H(z)$ + BAO data, instead of $\Omega_{\rm m0}$ we constrain $\Omega_b h^2$ and $\Omega_c h^2$; here $h$ is the Hubble constant in units of 100 km s$^{-1}$ Mpc$^{-1}$.

In the XCDM parametrizations the equation of state for the dynamical dark energy X-fluid is $P_{\rm X} = w_{\rm X} \rho_{\rm X}$ where $P_{\rm X}$, $\rho_{\rm X}$, and $w_{\rm X}$ are the pressure, energy density, and equation of state parameter for the dynamical dark energy X-fluid, and $\Omega_{\rm DE0}$ = $\Omega_{\rm X0}$ is the current value of the X-fluid dynamical dark energy density parameter. In this case the current values of the three energy density parameters are related by $\Omega_{\rm m0} + \Omega_{\rm k0} + \Omega_{\rm X0} = 1$. The X-fluid energy density decreases with time when \wx\ satisfies the conditions $-1 < w_{\rm X} < 0$. In the spatially-flat XCDM parametrization we choose to constrain $\Omega_{\rm m0}$, \wx, and $H_0$ while in the non-flat XCDM parametrization we constrain $\Omega_{\rm m0}$, $\Omega_{\rm k0}$, \wx, and $H_0$. For analyses which involve $H(z)$ + BAO data, instead of $\Omega_{\rm m0}$ we constrain $\Omega_b h^2$ and $\Omega_c h^2$. When $w_{\rm X} = -1$ the XCDM parametrizations reduce to the $\Lambda$CDM models.

In the $\phi$CDM models the dynamical dark energy is a scalar field $\phi$ (\citealp{peebrat88, ratpeeb88, pavlov13}).\footnote{Discussions of observational constraints on the \pcdm\ model can be traced through \cite{chen_etal_2017}, \cite{Zhaietal2017}, \cite{SolaPeracaulaetal2018, SolaPercaulaetal2019}, \cite{ooba_etal_2018b, ooba_etal_2019}, \cite{park_ratra_2018, park_ratra_2019b, park_ratra_2020}, \cite{Sangwanetal2018}, \cite{singh_etal_2019}, \cite{UrenaLopezRoy2020}, \cite{SinhaBanerjee2021}, \cite{Caoetal_2021b}, \cite{Khadkaetal_2021a}, \cite{Xuetal2021}, \cite{deCruzetal2021} and references therein.} In this model $\Omega_{\rm DE}$ = $\Omega_{\phi}(z, \alpha)$, the scalar field dynamical dark energy density parameter, is determined by the scalar field potential energy density, for which we assume an inverse power law form,
\begin{equation}
\label{eq:phiCDMV}
    V(\phi) = \frac{1}{2}\kappa m_{p}^2 \phi^{-\alpha}.
\end{equation}
Here $m_{p}$ is the Planck mass, $\alpha$ is a positive parameter, and $\kappa$ is a constant whose value is determined by using the shooting method to ensure that the current energy budget equation $\Omega_{\rm m0} + \Omega_{\rm k0} + \Omega_{\phi}(z = 0, \alpha) = 1$ is satisfied.

In the $\phi$CDM models the dynamics of a spatially homogeneous scalar field $\phi$ is governed by two coupled non-linear ordinary differential equations. The first is the dark energy scalar field equation of motion
\be\label{em}
\ddot{\phi}+3\bigg(\frac{\dot{a}}{a}\bigg)\dot{\phi}-\frac{1}{2}\alpha\kappa m_p^2\phi^{-\alpha-1}=0,
\ee
and the second is the Friedmann equation
\be\label{fe}
\bigg(\frac{\dot{a}}{a}\bigg)^2=\frac{8\pi}{3m_p^2}(\rho_{\rm m}+\rho_{\phi})-\frac{k}{a^2},
\ee
where $a$ is the scale factor and an overdot denotes a time derivative. In eq.\ \eqref{fe}, ${-k}/{a^2}$ is the spatial curvature term with $\Omega_{\rm k0} = 0$, $>0$, $<0$ corresponding to $k = 0$, $-1$, $+1$, respectively, and $\rho_{\rm m}$ and $\rho_{\phi}$ are the non-relativistic matter and scalar field energy densities where
\be\label{rp}
\rho_{\phi}=\frac{m_p^2}{32\pi}\bigg(\dot{\phi}^2+\kappa m_p^2\phi^{-\alpha}\bigg).
\ee
By solving eqs.\ \eqref{em} and \eqref{fe} numerically we can compute $\rho_{\phi}$ and then compute $\Omega_{\phi}(z, \alpha)$ by using the expression
\begin{equation}
    \Omega_{\phi}(z, \alpha) = \frac{8 \pi \rho_{\phi}}{3 m^2_p H^2_0}.
\end{equation}
In the spatially-flat $\phi$CDM model we choose to constrain $\Omega_{\rm m0}$, $\alpha$, and $H_0$ while in the non-flat $\phi$CDM model we constrain $\Omega_{\rm m0}$, $\Omega_{\rm k0}$, $\alpha$, and $H_0$. For analyses which involve $H(z)$ + BAO data, instead of $\Omega_{\rm m0}$ we constrain $\Omega_b h^2$ and $\Omega_c h^2$. When $\alpha=0$ the $\phi$CDM models reduce to the $\Lambda$CDM models.

\section{Data}
\label{sec:data}

In this paper, we analyze four different GRB data sets as well as some combinations of these data sets. We also use a joint $H(z)$ + BAO data set. These data sets are summarized in Table \ref{tab:data} and described in what follows.\footnote{In this table and elsewhere, for compactness, we sometimes use ML, MS, and GL as abbreviations for the MD-LGRB, MD-SGRB, and GW-LGRB data sets compiled by \cite{Wangetal_2021} and \cite{Huetal_2021}.}

\begin{itemize}

\item[]{\bf MD-LGRB sample}. This includes 31 long GRBs, with burst duration longer than 2 seconds, listed in Table 1 of \cite{Wangetal_2021}. For this data set, measured quantities for a GRB are redshift $z$, X-ray flux $F_0$, characteristic time scale $t_b$, and spectral index during the plateau phase $\beta^{\prime}$.\footnote{ML, MS, and GL data error bars on $F_0$ and $t_b$ are mostly asymmetric. We symmetrize these error bars using the method applied in \cite{Wangetal_2021} and \cite{Huetal_2021}, with the symmetrized error bar $\sigma = \sqrt{(\sigma_u^2 + \sigma_d^2)/2}$, where $\sigma_u$ and $\sigma_d$ are the asymmetric upper and lower error bars.} This sample probes the redshift range $1.45 \leq z \leq 5.91$.

\item[]{\bf MD-SGRB sample}. This includes 5 short GRBs, with burst duration shorter than 2 seconds, listed in Table 1 of \cite{Huetal_2021}. For this data set, measured quantities for a GRB are the same as those for the MD-LGRB sample. This data set probes the redshift range $0.35 \leq z \leq 2.6$.

\item[]{\bf GW-LGRB sample}. This includes 24 long GRBs listed in Table 1 of \cite{Huetal_2021}. For this data set, measured quantities for a GRB are the same as those for the MD-LGRB sample. This sample probes the redshift range $0.55 \leq z \leq 4.81$.

\item[]{\bf A118 sample}. This sample include 118 long GRBs listed in Table 7 of \cite{Khadkaetal_2021a}. For this data set, measured quantities for a GRB are $z$, rest-frame spectral peak energy $E_{\rm p}$, and measured bolometric fluence $S_{\rm bolo}$, computed in the standard rest-frame energy band $1-10^4$ keV. This sample probes the redshift range $0.3399 \leq z \leq 8.2$. 

The A118 data and the MD-LGRB data sets have 3 common GRBs, GRB060526, GRB081008, and GRB090516. We exclude these common GRBs from the A118 data set to form the A115 data set for joint analyses with the MD-LGRB data set. There are also 3 common GRBs between the A118 data set and the GW-LGRB data set, GRB060206, GRB091029, and GRB131105A. We exclude these common GRBs from the A118 data set to form the A115$^{\prime}$ data set for joint analyses with the GW-LGRB data set.

\item[]{$\textbf{ \emph{H(z)}}$ \bf and BAO data}. In addition to the GRB data, we also use 31 $H(z)$ and 11 BAO measurements. These $H(z)$ and BAO measurements probe the redshift range $0.07 \leq z \leq 1.965$ and $0.0106 \leq z \leq 2.33$, respectively. The $H(z)$ data are in Table 2 of \cite{Ryan_1} and the BAO data are in Table 1 of \cite{KhadkaRatra2020c}. We use cosmological constraints from the better-established $H(z)$ + BAO data to compare with those obtained from the GRB data sets.

\end{itemize}

\section{Data Analysis Methodology}
\label{sec:analysis}

\begin{table}
\centering
\resizebox{\columnwidth}{!}{%
\begin{threeparttable}
\caption{Summary of data sets used.}
\label{tab:data}
\begin{tabular}{lcc}
\toprule
Data set & $N$ (Number of points) & Redshift range\\
\midrule
ML & 31 & $1.45 \leq z \leq 5.91$ \\
MS & 5 & $0.35 \leq z \leq 2.6$ \\
GL & 24 & $0.55 \leq z \leq 4.81$ \\
MS + GL & 29 & $0.35 \leq z \leq 4.81$ \\
A118 & 118 & $0.3399 \leq z \leq 8.2$ \\
A115\tnote{a} & 115 & $0.3399 \leq z \leq 8.2$ \\
A115$^{\prime}$\tnote{b} & 115 & $0.3399 \leq z \leq 8.2$ \\
\midrule
$H(z)$ & 31 & $0.070 \leq z \leq 1.965$ \\
BAO & 11 & $0.38 \leq z \leq 2.334$ \\
\bottomrule
\end{tabular}
\begin{tablenotes}[flushleft]
\item [a] Excluding from A118 those GRBs in common with MD-LGRB (GRB060526, GRB081008, and GRB090516).
\item [b] Excluding from A118 those GRBs in common with GW-LGRB (GRB060206, GRB091029, and GRB131105A).
\end{tablenotes}
\end{threeparttable}%
}
\end{table}

\begin{table}
\centering
\resizebox{\columnwidth}{!}{%
\begin{threeparttable}
\caption{Flat priors of the constrained parameters.}
\label{tab:priors}
\begin{tabular}{lcc}
\toprule
Parameter & & Prior\\
\midrule
 & Cosmological Parameters & \\
\midrule
$H_0$\,\tnote{a} &  & [None, None]\\
\obhs\,\tnote{b} &  & [0, 1]\\
\ochs\,\tnote{c} &  & [0, 1]\\
\ok &  & [-2, 2]\\
$\alpha$ &  & [0, 10]\\
\wx &  & [-5, 0.33]\\
\midrule
 & GRB Nuisance Parameters\tnote{d} & \\
\midrule
$k$ &  & [-10, 10]\\
$b$\,\tnote{e} &  & [0, 10]\\
$\sigma_{\rm int}$ &  & [0, 5]\\
$\beta$ &  & [0, 5]\\
$\gamma$ &  & [0, 300]\\
\bottomrule
\end{tabular}
\begin{tablenotes}[flushleft]
\item [a] \hunit. In the GRB alone cases, $H_0$ is set to be 70 \hunit, while in the $H(z)$ + BAO case, the prior range is irrelevant (unbounded).
\item [b] In the GRB alone cases, \obhs\ is set to be 0.0245, i.e. $\Omega_{b}=0.05$.
\item [c] In the GRB alone cases, $\Omega_{c}\in[-0.05,0.95]$ to ensure $\Om\in[0,1]$.
\item [d] Note that $k$, $b$, and $\sigma_{\rm int}$ of MD-LGRBs are different from those of MD-SGRBs/GW-LGRBs, but with the same prior ranges.
\item [e] $b<0$ values are possible for MD-SGRBs (due to fewer data points) but, as discussed below, requiring $b\geq 0$ does not have significant consequences.
\end{tablenotes}
\end{threeparttable}%
}
\end{table}

For GRBs which obey the Dainotti correlation the luminosity of the plateau phase is \citep{Dianotti_2008, Dainotti_2010, Dainotti_2011}
\be
\label{eq:L0}
    L_0=\frac{4\pi D_L^2F_0}{(1+z)^{1-\beta^{\prime}}},
\ee
where $F_0$ is the GRB X-ray flux, $\beta^{\prime}$ is the spectral index in the plateau phase, and $D_L$ is the luminosity distance. 

$D_L$, as a function of redshift $z$ and cosmological parameters $\textbf{\emph{p}}$, is given by
\begin{equation}
  \label{eq:DL}
  \frac{H_0\sqrt{\left|\Omega_{\rm k0}\right|}D_L(z, \textbf{\emph{p}})}{c(1+z)} = 
    \begin{cases}
    {\rm sinh}\left[g(z, \textbf{\emph{p}})\right] & \text{if}\ \Omega_{\rm k0} > 0, \\
    \vspace{1mm}
    g(z, \textbf{\emph{p}}) & \text{if}\ \Omega_{\rm k0} = 0,\\
    \vspace{1mm}
    {\rm sin}\left[g(z, \textbf{\emph{p}})\right] & \text{if}\ \Omega_{\rm k0} < 0,
    \end{cases}   
\end{equation}
where
\begin{equation}
\label{eq:gz}
   g(z, \textbf{\emph{p}}) = H_0\sqrt{\left|\Omega_{\rm k0}\right|}\int^z_0 \frac{dz'}{H(z', \textbf{\emph{p}})},
\end{equation}
$c$ is the speed of light, and $H(z, \textbf{\emph{p}})$ is the Hubble parameter that is described in Sec. \ref{sec:model} for each cosmological model.

For these GRBs the luminosity of the plateau phase $L_0$ and the characteristic time scale $t_b$ are correlated through the Dainotti or luminosity-time relation
\begin{equation}
    \label{eq:dcorr}
    y\equiv\log \left(\frac{L_0}{10^{47}\ \mathrm{erg/s}}\right) = k\log \frac{t_b}{10^3(1+z)\ \mathrm{s}} + b\equiv kx + b,
\end{equation}
where $\log=\log_{10}$ and the slope $k$ and the intercept $b$ are free parameters to be determined from the data.

We predict $L_0$ as a function of cosmological parameters $\textbf{\emph{p}}$ at the redshift of each GRB by using eqs.\ \eqref{eq:L0}, \eqref{eq:DL}, and \eqref{eq:dcorr}. We then compare predicted and measured values of $L_0$ by using the natural log of the likelihood function \citep{D'Agostini_2005}
\be
\label{eq:LH_MD-LGRB}
    \ln\mathcal{L}_{\rm GRB}= -\frac{1}{2}\Bigg[\chi^2_{\rm GRB}+\sum^{N}_{i=1}\ln\left(2\pi(\sigma_{\rm int}^2+\sigma_{{y_i}}^2+k^2\sigma_{{x_i}}^2)\right)\Bigg],
\ee
where
\be
\label{eq:chi2_MD-LGRB}
    \chi^2_{\rm GRB} = \sum^{N}_{i=1}\bigg[\frac{(y_i-k x_i-b)^2}{(\sigma_{\rm int}^2+\sigma_{{y_i}}^2+k^2\sigma_{{x_i}}^2)}\bigg].
\ee
Here $N$ is the number of data points (e.g., for MD-LGRB $N=31$), and $\sigma_{\rm int}$ is the intrinsic scatter parameter (which also contains the unknown systematic uncertainty).

For GRBs which obey the Amati correlation the rest frame isotropic radiated energy $E_{\rm iso}$ is
\be
\label{eq:Eiso}
    E_{\rm iso}=\frac{4\pi D_L^2}{1+z}S_{\rm bolo},
\ee
where $S_{\rm bolo}$ is the bolometric fluence. 

For these GRBs the rest frame peak photon energy $E_{\rm p}$ and $E_{\rm iso}$ are correlated through the Amati (or $E_{\rm p}-E_{\rm iso}$) relation \citep{Amati2008, Amati2009} 
\begin{equation}
    \label{eq:Amati}
    \log E_{\rm iso} = \gamma  + \beta\log E_{\rm p},
\end{equation}
where the intercept $\gamma$ and the slope $\beta$ are free parameters to be determined from the data. Note that the peak energy $E_{\rm p} = (1+z)E_{\rm p, obs}$ where $E_{\rm p, obs}$ is the observed peak energy.

We predict $E_{\rm iso}$ as a function of cosmological parameters $\textbf{\emph{p}}$ at the redshift of each GRB by using eqs.\ \eqref{eq:DL}, \eqref{eq:Eiso}, and \eqref{eq:Amati}. We then compare predicted and measured values of $E_{\rm iso}$ by using the natural log of the likelihood function \citep{D'Agostini_2005}
\be
\label{eq:LH_GRB}
    \ln\mathcal{L}_{\rm A118}= -\frac{1}{2}\Bigg[\chi^2_{\rm A118}+\sum^{N}_{i=1}\ln\left(2\pi(\sigma_{\rm int}^2+\sigma_{{y^{\prime}_i}}^2+\beta^2\sigma_{{x^{\prime}_i}}^2)\right)\Bigg],
\ee
where
\be
\label{eq:chi2_GRB}
    \chi^2_{\rm A118} = \sum^{N}_{i=1}\bigg[\frac{(y^{\prime}_i-\beta x^{\prime}_i-\gamma)^2}{(\sigma_{\rm int}^2+\sigma_{{y^{\prime}_i}}^2+\beta^2\sigma_{{x^{\prime}_i}}^2)}\bigg].
\ee
Here $x^{\prime}=\log(E_{\rm p}/{\rm keV})$, $\sigma_{x^{\prime}}=\sigma_{E_{\rm p}}/(E_{\rm p}\ln 10)$, $y^{\prime}=\log(E_{\rm iso}/{\rm erg})$, and $\sigma_{\rm int}$ is the intrinsic scatter parameter, which also contains the unknown systematic uncertainty.

The $H(z)$ + BAO data analyses follow the method described in Sec.\ 4 of \cite{KhadkaRatra2021a}.

We maximize the likelihood function using the Markov chain Monte Carlo (MCMC) method as implemented in the \textsc{MontePython} code \citep{Brinckmann2019} and determine the best-fitting and posterior mean values and the corresponding uncertainties for all free parameters. We assure convergence of the MCMC chains for each free parameter from the Gelman-Rubin criterion ($R-1 < 0.05$). Flat priors used for the free parameters are given in Table \ref{tab:priors}.

The Akaike Information Criterion ($AIC$) and the Bayesian Information Criterion ($BIC$) are used to compare the goodness of fit of models with different numbers of parameters. These are
\be
\label{AIC}
    AIC=-2\ln \mathcal{L}_{\rm max} + 2n,
\ee
and
\be
\label{BIC}
    BIC=-2\ln \mathcal{L}_{\rm max} + n\ln N.
\ee
In these equations, $\mathcal{L}_{\rm max}$ is the maximum value of the relevant likelihood function and $n$ is the number of free parameters of the model under consideration.

\section{Results}
\label{sec:results}

\subsection{Constraints from ML, MS, and GL data}
 \label{subsec:MLSGL}
 
\begin{table}
\centering
\begin{threeparttable}
\caption{One-dimensional marginalized posterior means and 68.27\% limits of the Dainotti correlation parameters for the ML, GL, and MS data sets using the flat \lcdm\ model with $\Omega_{\rm m0} = 0.3$ and $H_0 = 70$ \hunit, and comparison with the results given in \protect\cite{Wangetal_2021} and \protect\cite{Huetal_2021}.}
\label{tab:fix}
\setlength{\tabcolsep}{3.5pt}
\begin{tabular}{lcccc}
\toprule
Data set & Source & $k$ & $b$ & $\sigma_{\rm int}$\\
\midrule
 & {}\tnote{a} & $-1.02^{+0.09}_{-0.08}$ & $1.72^{+0.07}_{-0.07}$ & --\\
ML & {}\tnote{b} & $-1.026\pm0.085$ & $1.726\pm0.074$ & $0.303^{+0.032}_{-0.050}$\\
 & {}\tnote{c} & $-1.026\pm0.086$ & $1.726\pm0.074$ & $0.303^{+0.032}_{-0.050}$\\
\midrule 
 & {}\tnote{d} & $-1.77^{+0.20}_{-0.20}$ & $0.66^{+0.01}_{-0.01}$ & $0.42^{+0.08}_{-0.06}$\\
GL & {}\tnote{b} & $-1.753^{+0.187}_{-0.208}$ & $0.642^{+0.100}_{-0.071}$ & $0.428^{+0.053}_{-0.086}$\\
 & {}\tnote{c} & $-1.769\pm0.205$ & $0.656\pm0.094$ & $0.431^{+0.054}_{-0.088}$\\
\midrule
 & {}\tnote{d} & $-1.38^{+0.17}_{-0.19}$ & $0.33^{+0.17}_{-0.16}$ & $0.35^{+0.20}_{-0.12}$\\
MS & {}\tnote{b} & $-1.381^{+0.209}_{-0.213}$ & $0.327^{+0.195}_{-0.189}$ & $0.420^{+0.086}_{-0.242}$\\
 & {}\tnote{c} & $-1.397^{+0.247}_{-0.241}$ & $0.354^{+0.195}_{-0.242}$ & $0.525^{+0.044}_{-0.358}$\\
\bottomrule
\end{tabular}
\begin{tablenotes}[flushleft]
\item [a] Results from \cite{Wangetal_2021} with the prior ranges of the parameters being $k\in(-1.3,-0.75)$, $b\in(1.4,2.0)$, and $\sigma_{\rm int}\in(0.1,0.6)$ for ML.
\item [b] Our results with the same prior ranges of the parameters as \cite{Wangetal_2021} or \cite{Huetal_2021}.
\item [c] Our results with wider prior ranges of the parameters $k\in[-10,10]$, $b\in[0,10]$ ($b\in[-0.5,10]$), and $\sigma_{\rm int}\in[0,3]$ for ML and GL (MS).
\item [d] Results from \cite{Huetal_2021} with the prior ranges of the parameters being $k\in(-2.2,-1)$, $b\in(0.1,0.8)$, and $\sigma_{\rm int}\in(0.01,0.8)$ for GL and being $k\in(-2.1,-0.55)$, $b\in(-0.5,1.0)$, and $\sigma_{\rm int}\in(0.01,1)$ for MS.
\end{tablenotes}
\end{threeparttable}%
\end{table}

In Table \ref{tab:fix} we list Dainotti correlation parameters computed using the ML, GL, and MS data sets. These are computed in the flat \lcdm\ model with $\Omega_{\rm m0} = 0.3$ and $H_0 = 70$ \hunit, the same model and parameter values used in \cite{Wangetal_2021} and \cite{Huetal_2021}. The first line of parameter values in each of the three subpanels of Table \ref{tab:fix} are taken from these papers.\footnote{\cite{Wangetal_2021} do not list a value for $\sigma_{\rm int}$ in the ML case.} To compare to these results, we used \textsc{emcee} (\citealp{emcee}) to compute the values listed in the second and third lines of each subpanel. Comparing the first and second lines in each subpanel, we find that they are consistent, except: i) for the GL case our $b$ uncertainties are larger than those of \cite{Huetal_2021}; and, ii) for the MS case we have larger $b$ and $k$ error bars and a larger central value of $\sigma_{\rm int}$ than those of \cite{Huetal_2021}, but they agree within 1$\sigma$. In the third line of each subpanel we list results obtained assuming wider prior ranges of the parameters. We find that the ML results do not change, the GL results are shifted closer to those of \cite{Huetal_2021}, except for the values of $\sigma_{\rm int}$, and the MS results are shifted away from those of \cite{Huetal_2021} with larger error bars, especially for $\sigma_{\rm int}$.

\begin{figure*}
\centering
 \subfloat[MD-LGRB]{%
    \includegraphics[width=3.45in,height=2.5in]{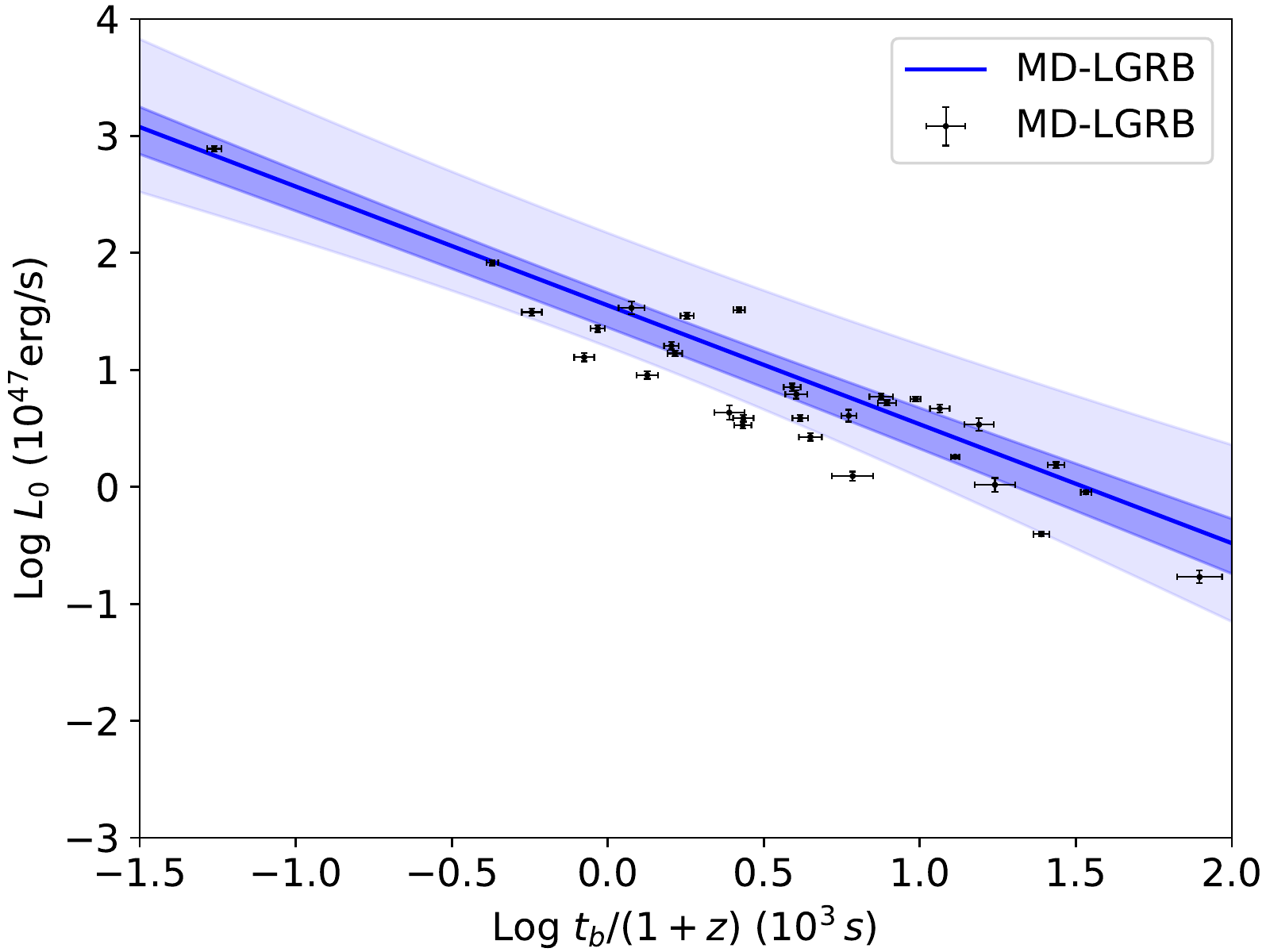}}
 \subfloat[MD-SGRB]{%
    \includegraphics[width=3.45in,height=2.5in]{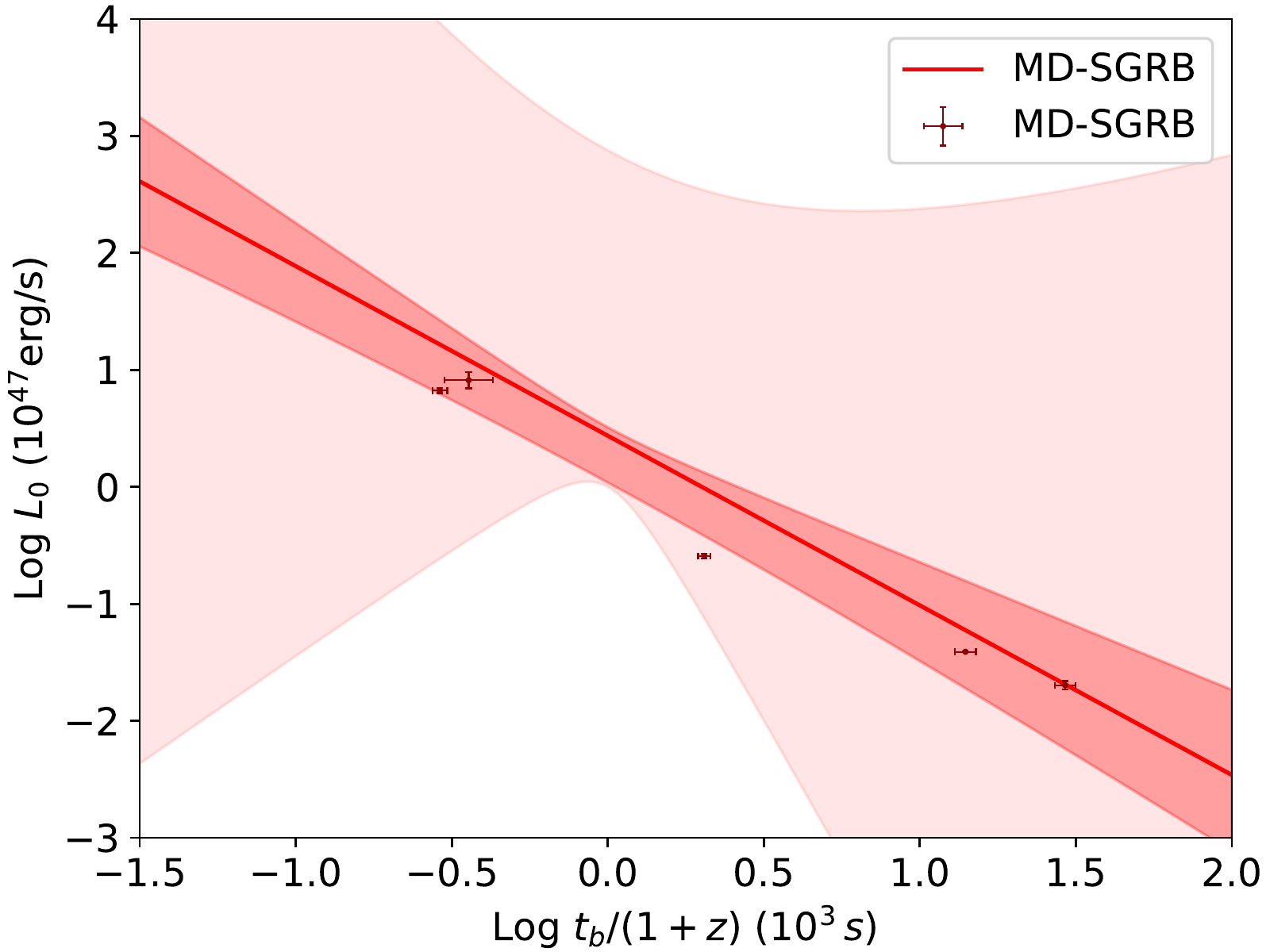}}\\
 \subfloat[GW-LGRB and MD-SGRB]{%
    \includegraphics[width=3.45in,height=2.5in]{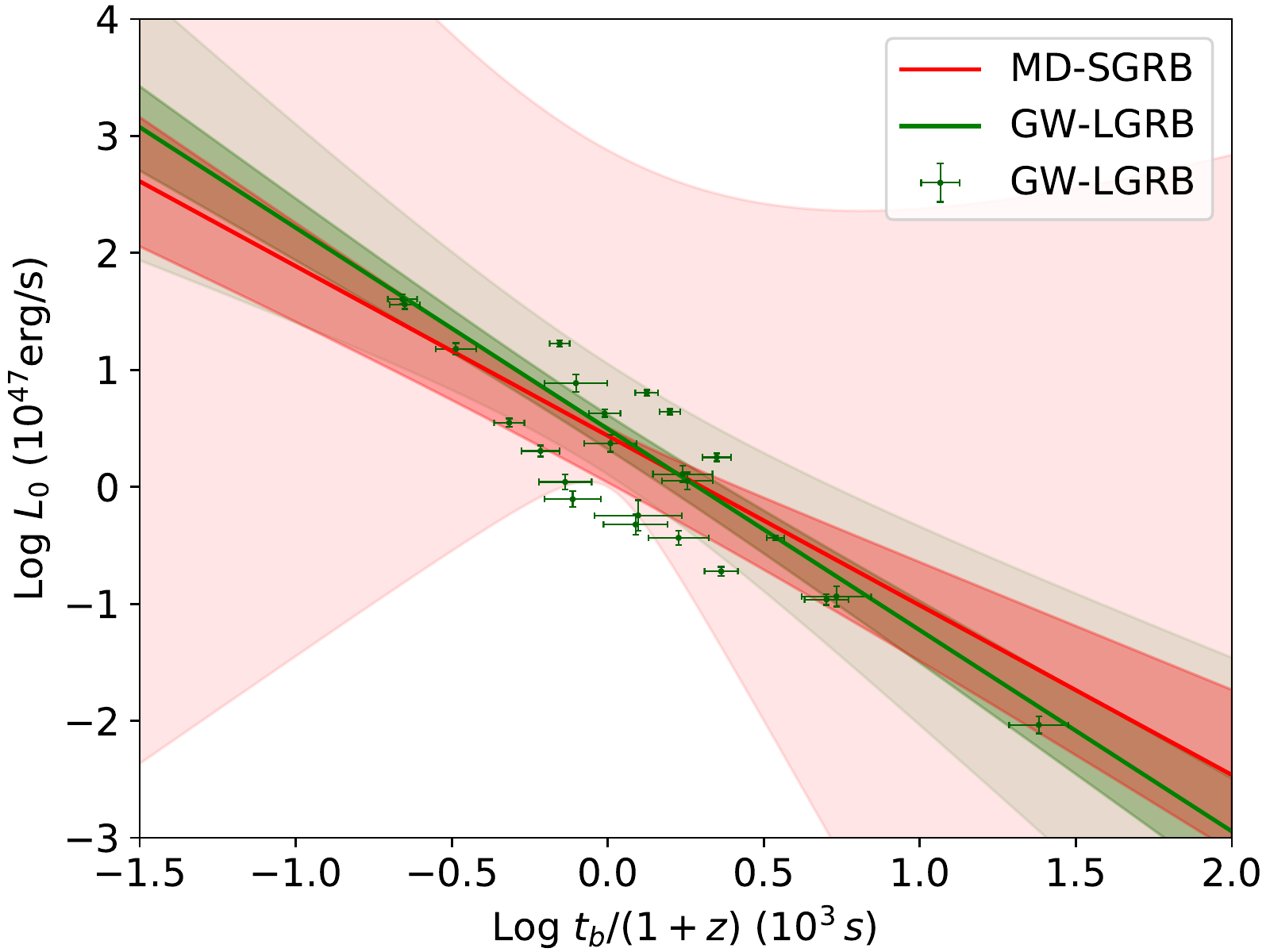}}
 \subfloat[MD-SGRB + GW-LGRB]{%
    \includegraphics[width=3.45in,height=2.5in]{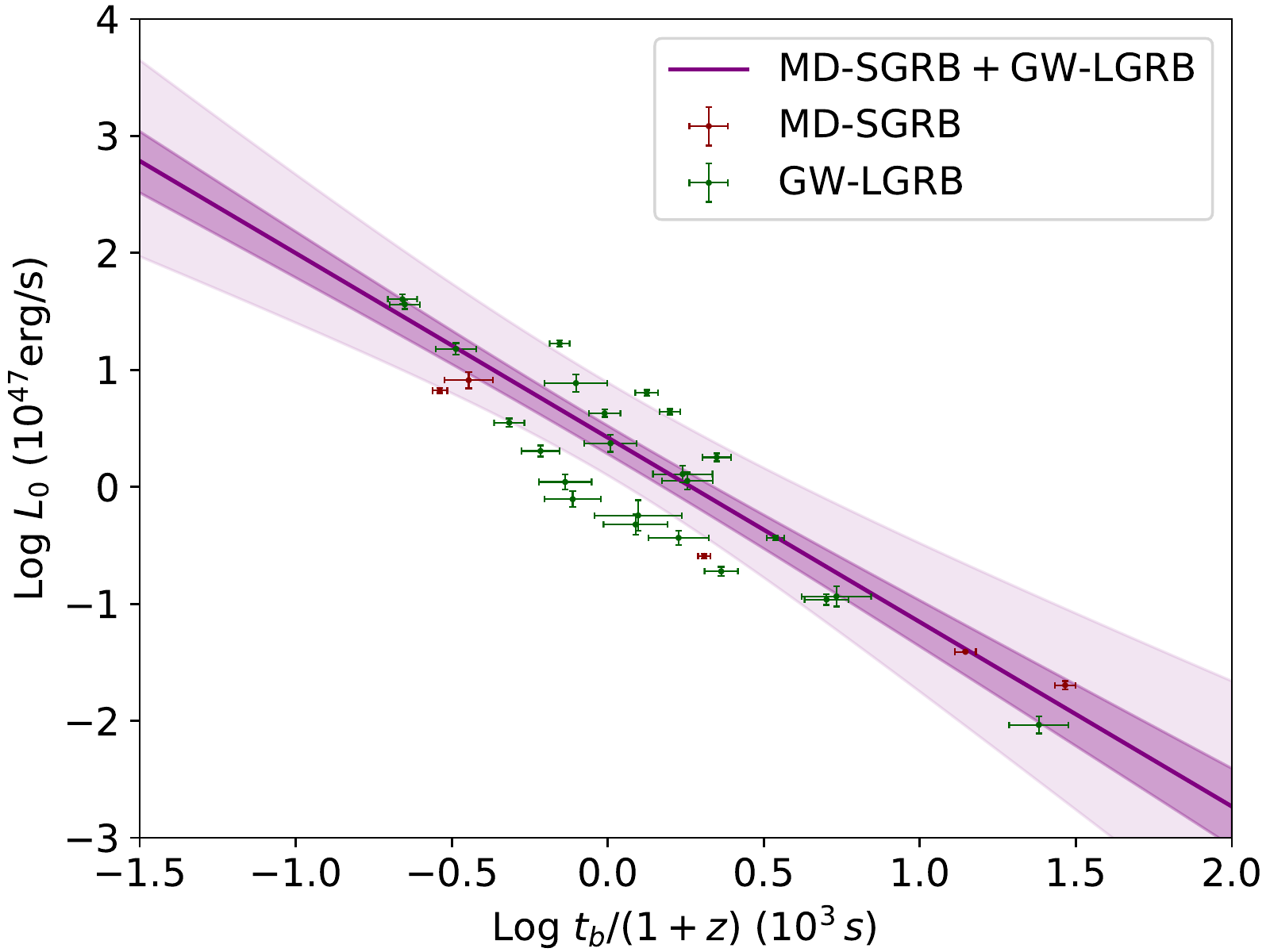}}\\
\caption{$L_0-t_b$ correlations for MD-LGRB, MD-SGRB, GW-LGRB, and MD-SGRB + GW-LGRB data using the flat \lcdm\ model. The MD-LGRB, MD-SGRB, and GW-LGRB data with error bars are shown in black, dark red, and dark green, respectively. The solid lines are the $L_0-t_b$ correlations with posterior mean values for the slopes and intercepts listed in Table \ref{tab:1d_BFP}, for MD-LGRB (blue), MD-SGRB (red), GW-LGRB (green), and MD-SGRB + GW-LGRB (purple) data. The $1\sigma$ and $3\sigma$ confidence regions are the dark and light colored shaded regions with the uncertainties propagated from those of $k$ and $b$ (without considering $\sigma_{\rm int}$).}
\label{fig00}
\end{figure*}

\begin{figure*}
\centering
 \subfloat[Flat \lcdm]{%
    \includegraphics[width=3.45in,height=2.5in]{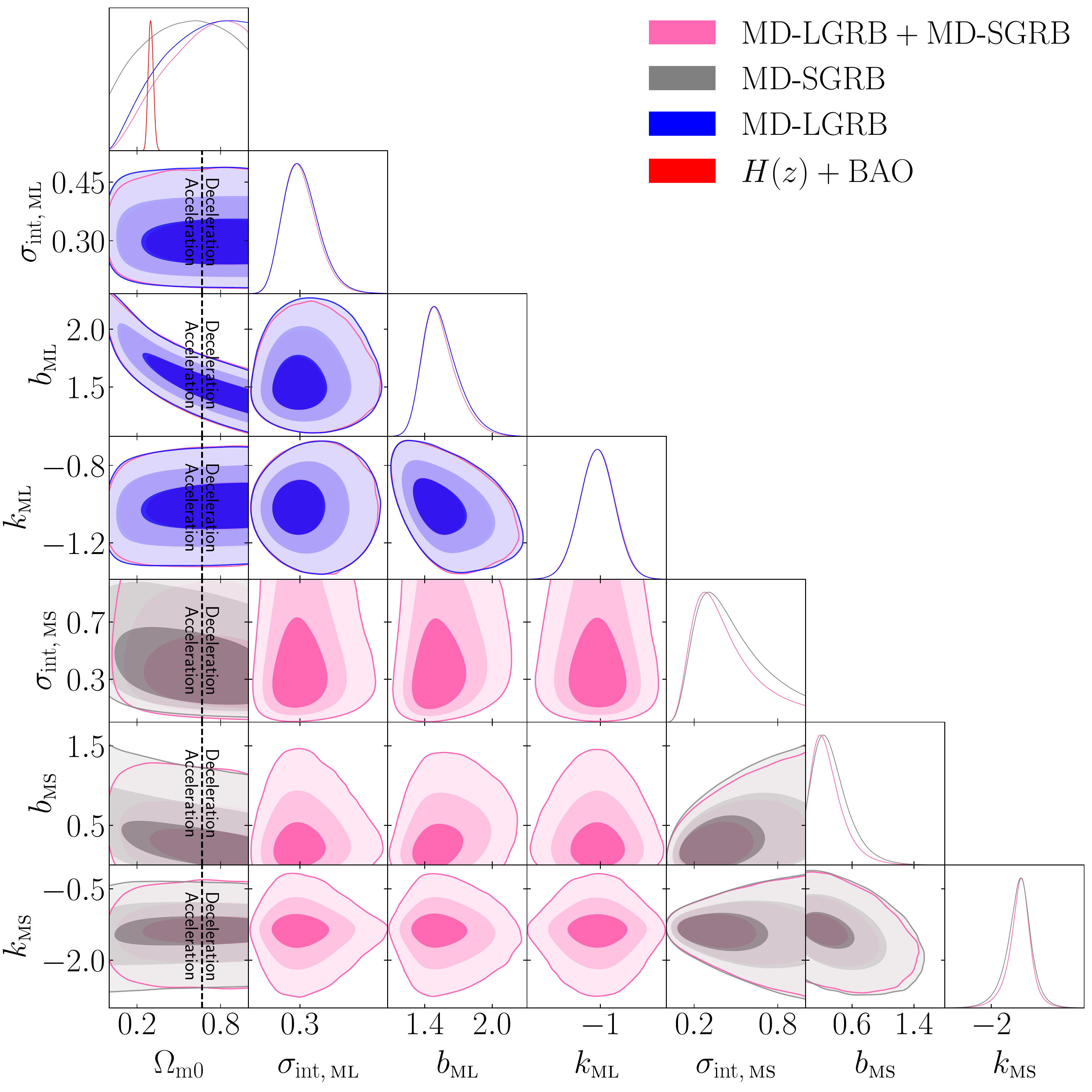}}
 \subfloat[Non-flat \lcdm]{%
    \includegraphics[width=3.45in,height=2.5in]{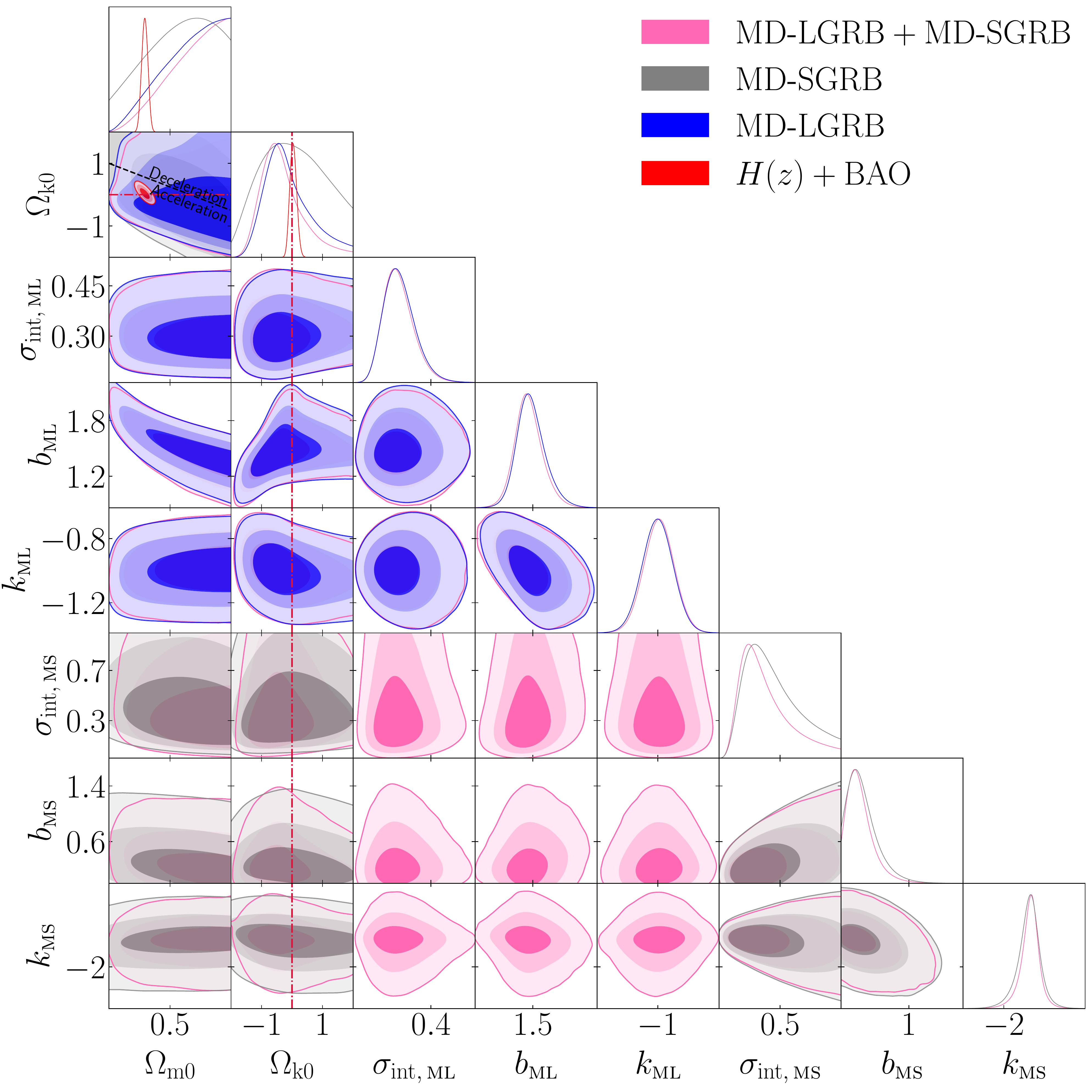}}\\
 \subfloat[Flat XCDM]{%
    \includegraphics[width=3.45in,height=2.5in]{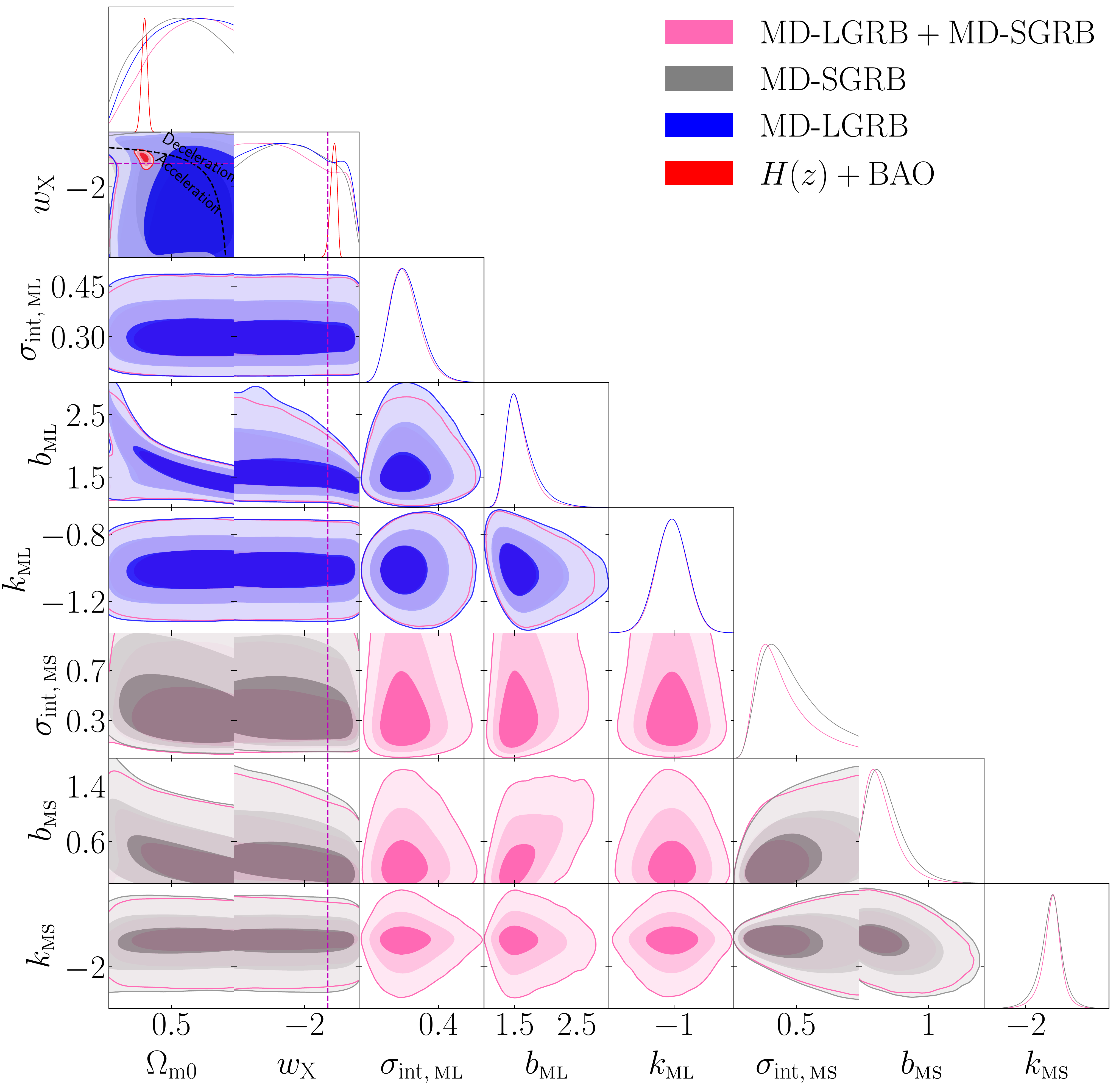}}
 \subfloat[Non-flat XCDM]{%
    \includegraphics[width=3.45in,height=2.5in]{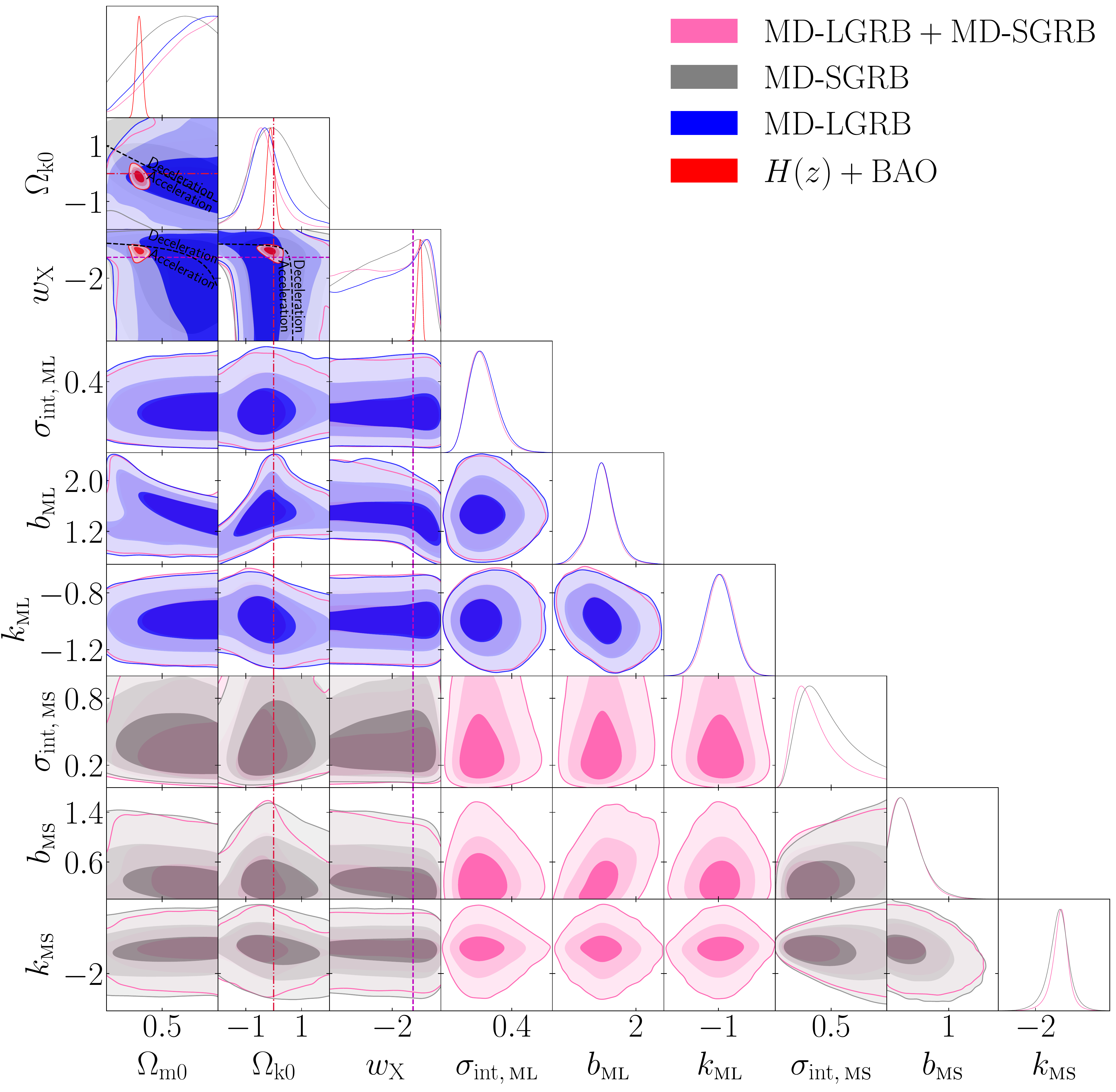}}\\
 \subfloat[Flat \pcdm]{%
    \includegraphics[width=3.45in,height=2.5in]{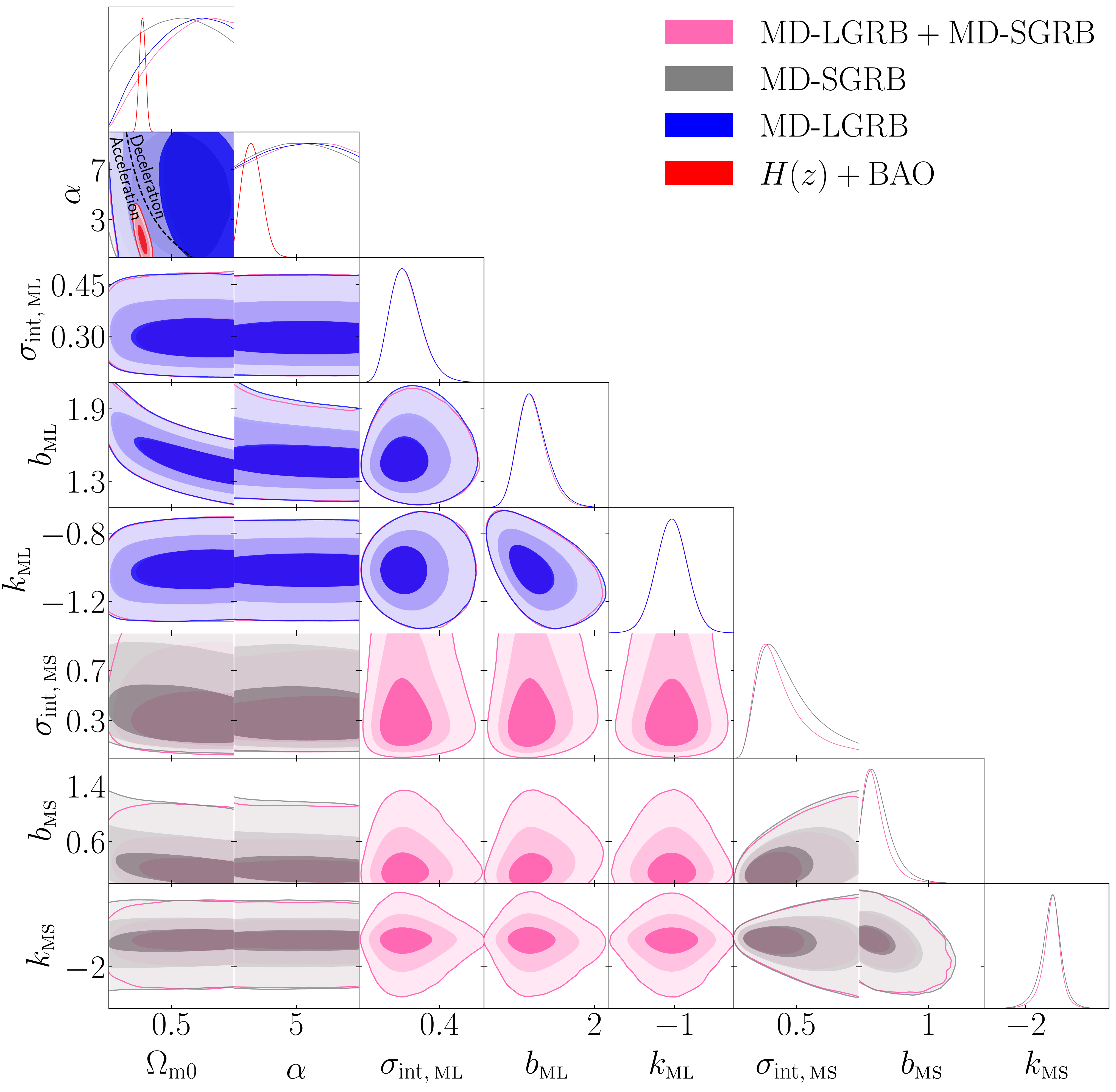}}
  \subfloat[Non-flat \pcdm]{%
     \includegraphics[width=3.45in,height=2.5in]{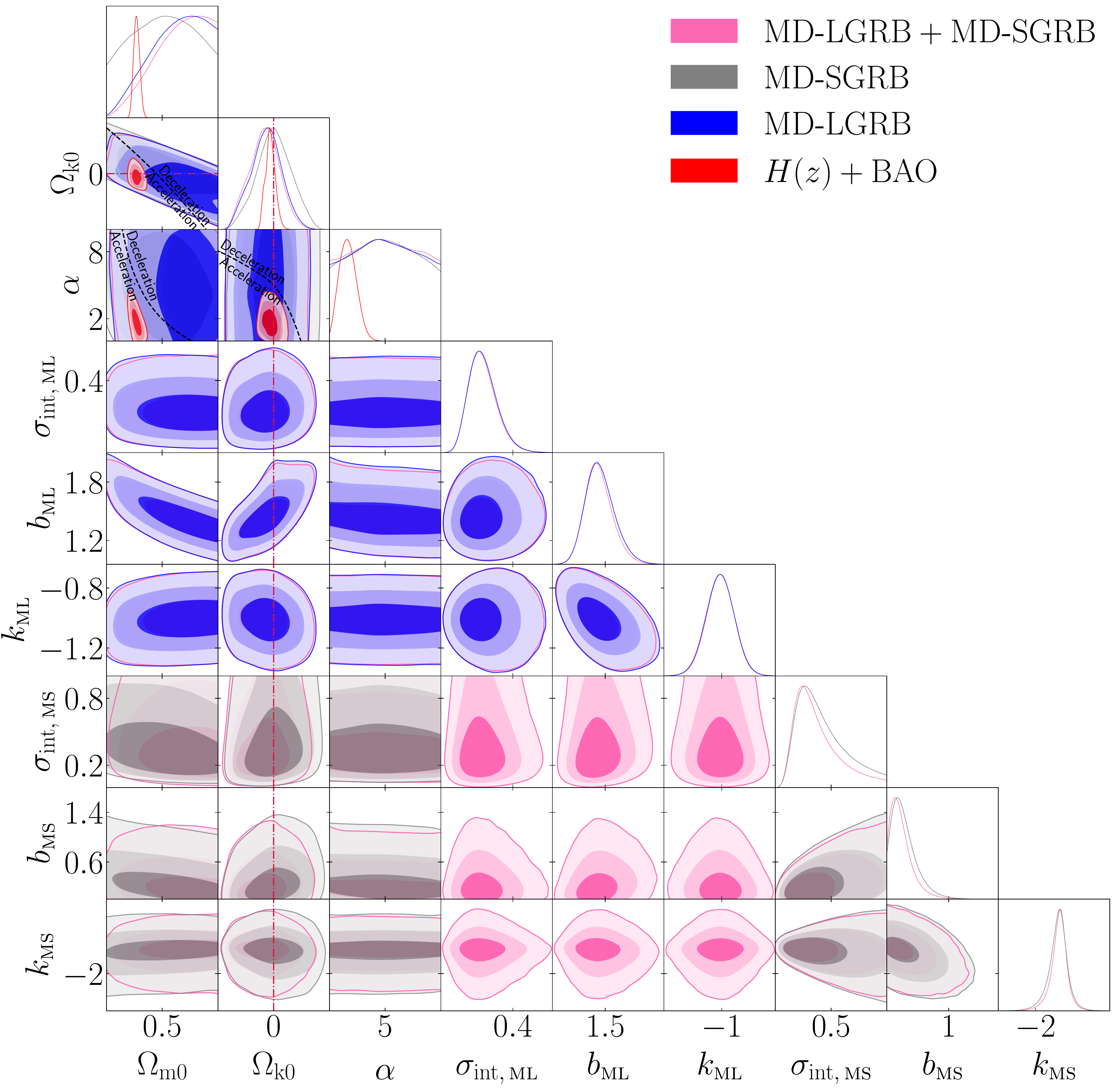}}\\
\caption{One-dimensional likelihoods and 1$\sigma$, 2$\sigma$, and 3$\sigma$ two-dimensional likelihood confidence contours from MD-LGRB (blue), MD-SGRB (gray), MD-LGRB + MD-SGRB (pink), and $H(z)$ + BAO (red) data for all six models. The zero-acceleration lines are shown as black dashed lines, which divide the parameter space into regions associated with currently-accelerating and currently-decelerating cosmological expansion. In the non-flat XCDM and non-flat \pcdm\ cases, the zero-acceleration lines are computed for the third cosmological parameter set to the $H(z)$ + BAO data best-fitting values listed in Table \ref{tab:BFP}. The crimson dash-dot lines represent flat hypersurfaces, with closed spatial hypersurfaces either below or to the left. The magenta lines represent $w_{\rm X}=-1$, i.e.\ flat or non-flat \lcdm\ models. The $\alpha = 0$ axes correspond to flat and non-flat \lcdm\ models in panels (e) and (f), respectively.}
\label{fig1}
\end{figure*}

\begin{figure*}
\centering
 \subfloat[Flat \lcdm]{%
    \includegraphics[width=3.45in,height=2.5in]{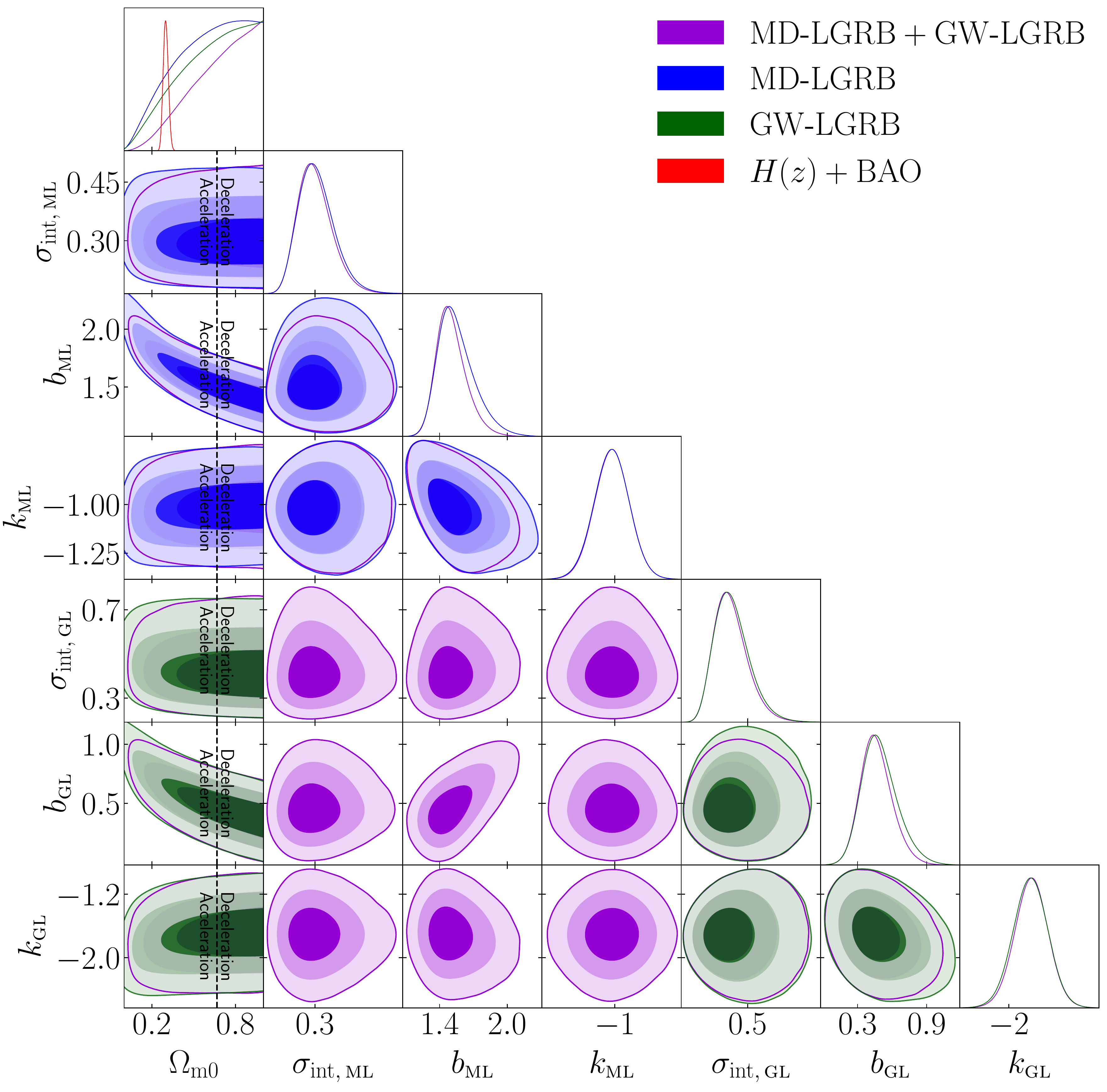}}
 \subfloat[Non-flat \lcdm]{%
    \includegraphics[width=3.45in,height=2.5in]{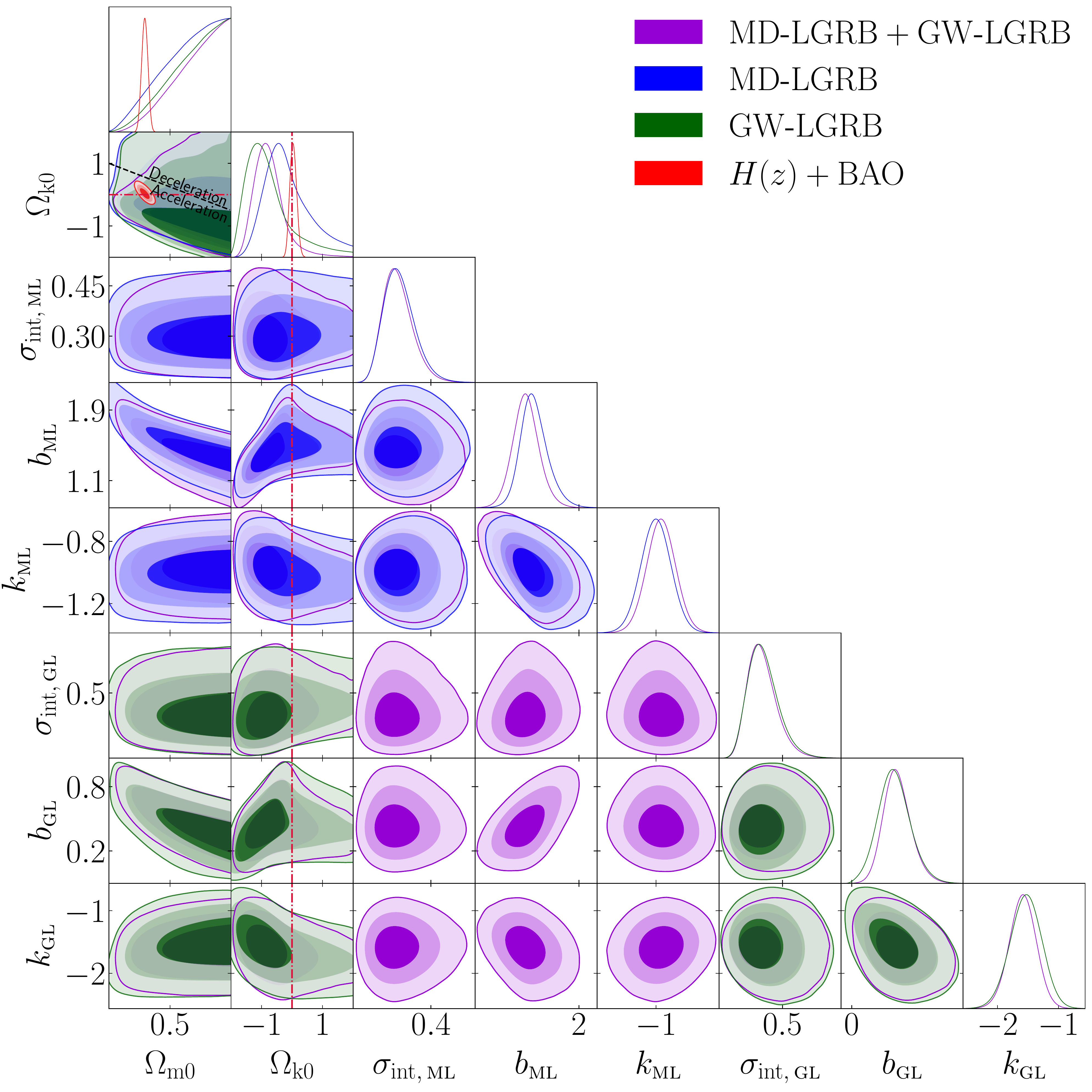}}\\
 \subfloat[Flat XCDM]{%
    \includegraphics[width=3.45in,height=2.5in]{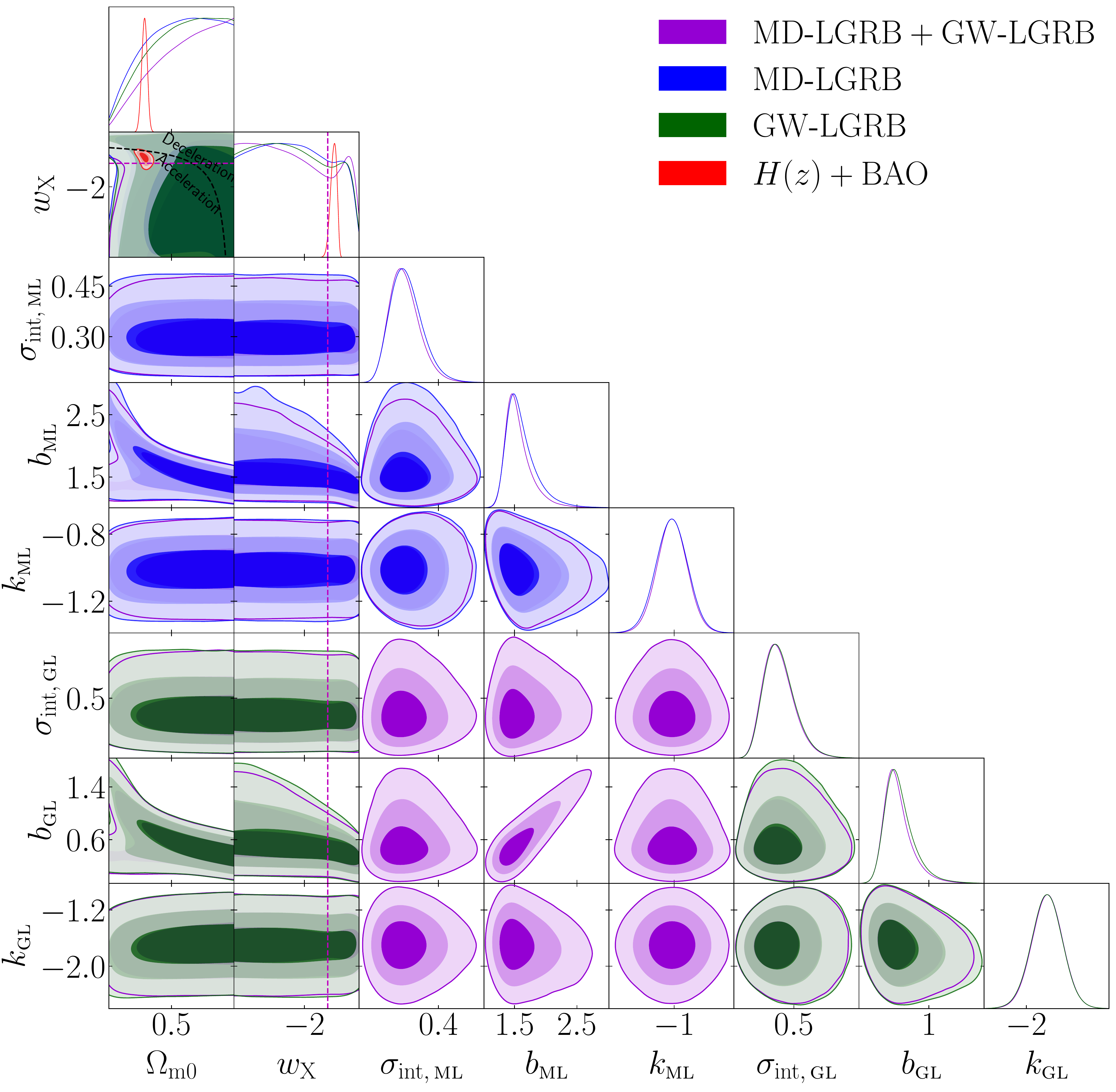}}
 \subfloat[Non-flat XCDM]{%
    \includegraphics[width=3.45in,height=2.5in]{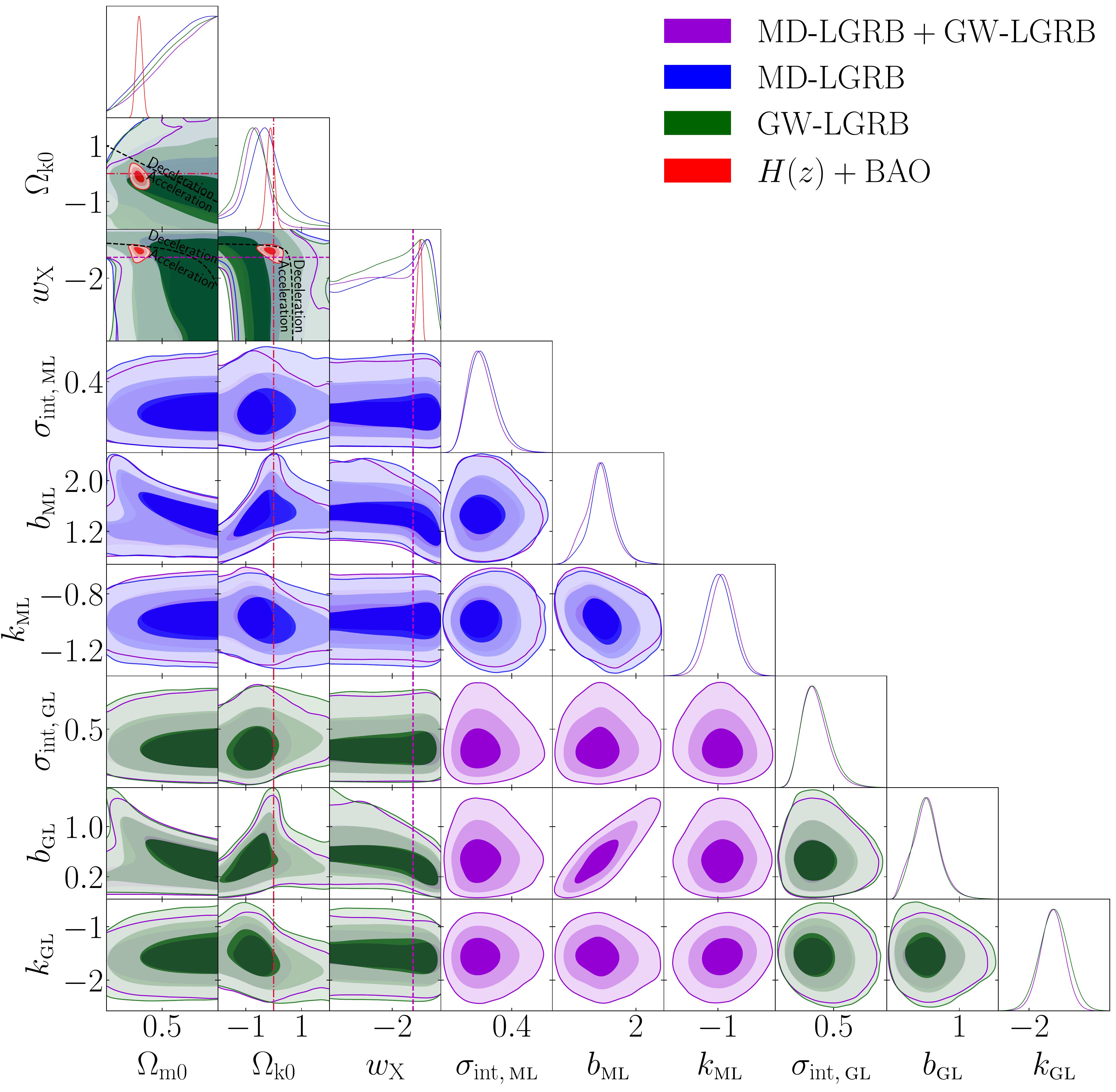}}\\
 \subfloat[Flat \pcdm]{%
    \includegraphics[width=3.45in,height=2.5in]{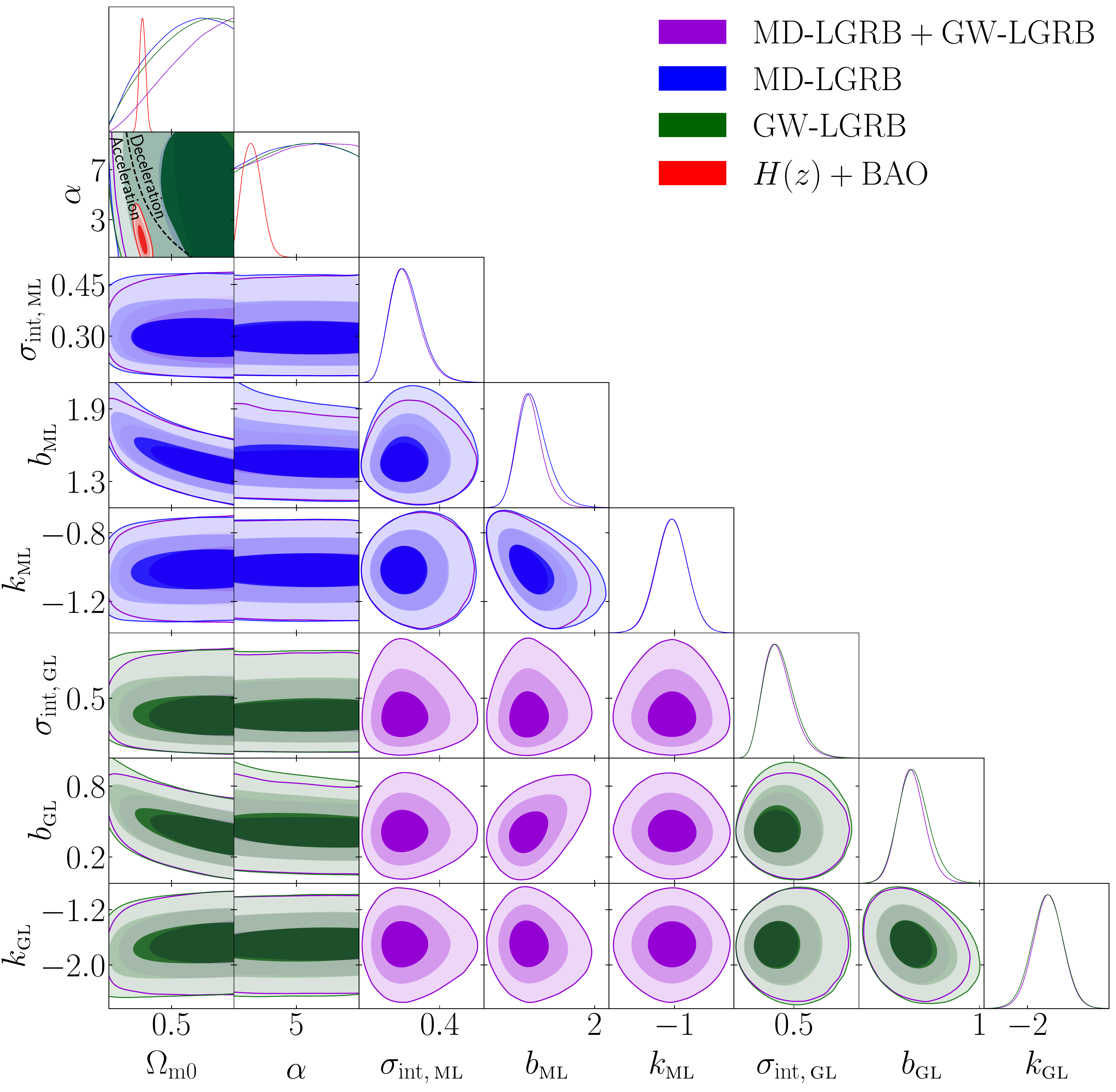}}
  \subfloat[Non-flat \pcdm]{%
     \includegraphics[width=3.45in,height=2.5in]{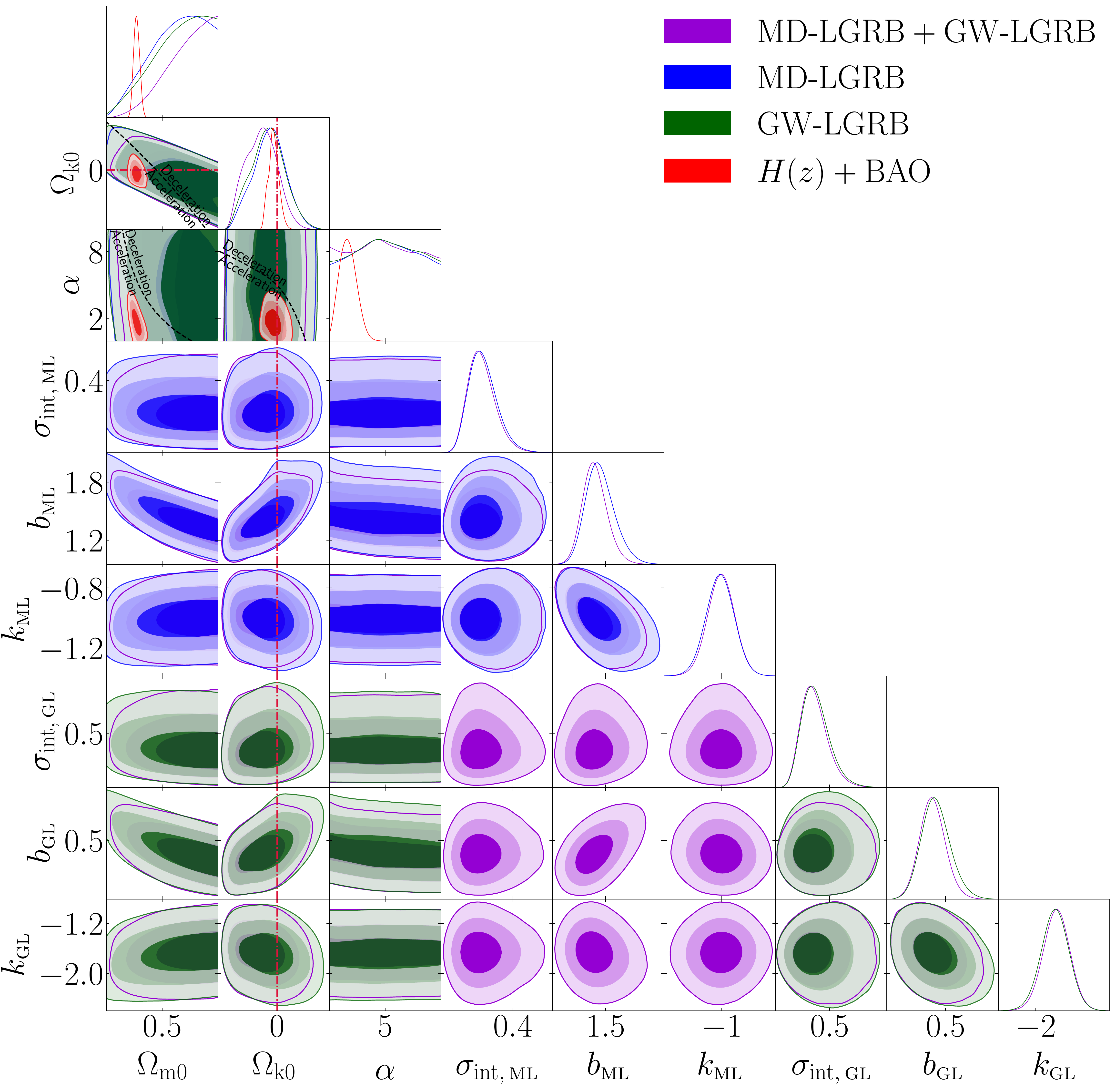}}\\
\caption{One-dimensional likelihoods and 1$\sigma$, 2$\sigma$, and 3$\sigma$ two-dimensional likelihood confidence contours from MD-LGRB (blue), GW-LGRB (green), MD-LGRB + GW-LGRB (violet), and $H(z)$ + BAO (red) data for all six models. The zero-acceleration lines are shown as black dashed lines, which divide the parameter space into regions associated with currently-accelerating and currently-decelerating cosmological expansion. In the non-flat XCDM and non-flat \pcdm\ cases, the zero-acceleration lines are computed for the third cosmological parameter set to the $H(z)$ + BAO data best-fitting values listed in Table \ref{tab:BFP}. The crimson dash-dot lines represent flat hypersurfaces, with closed spatial hypersurfaces either below or to the left. The magenta lines represent $w_{\rm X}=-1$, i.e.\ flat or non-flat \lcdm\ models. The $\alpha = 0$ axes correspond to flat and non-flat \lcdm\ models in panels (e) and (f), respectively.}
\label{fig2}
\end{figure*}

\begin{figure*}
\centering
 \subfloat[Flat \lcdm]{%
    \includegraphics[width=3.45in,height=2.5in]{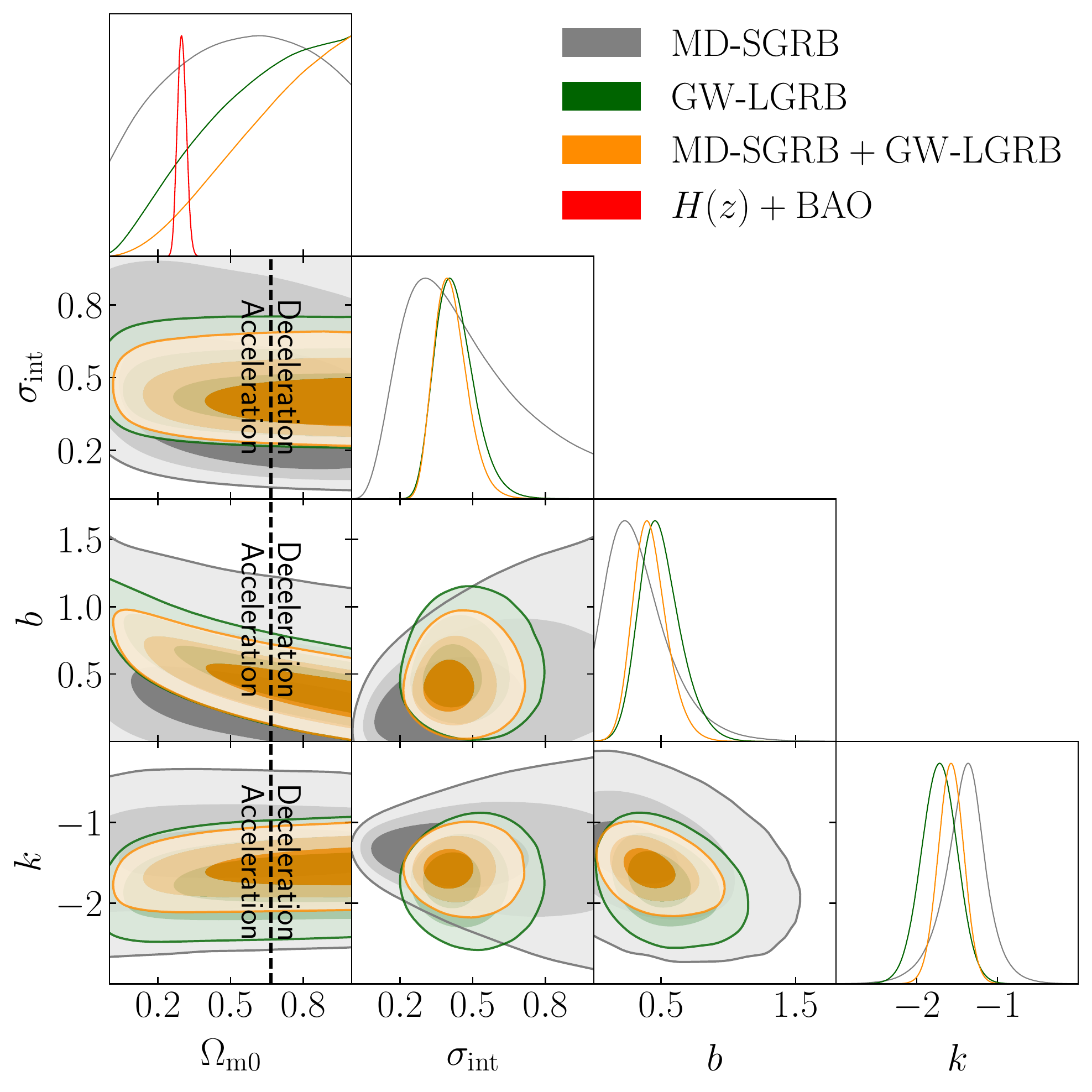}}
 \subfloat[Non-flat \lcdm]{%
    \includegraphics[width=3.45in,height=2.5in]{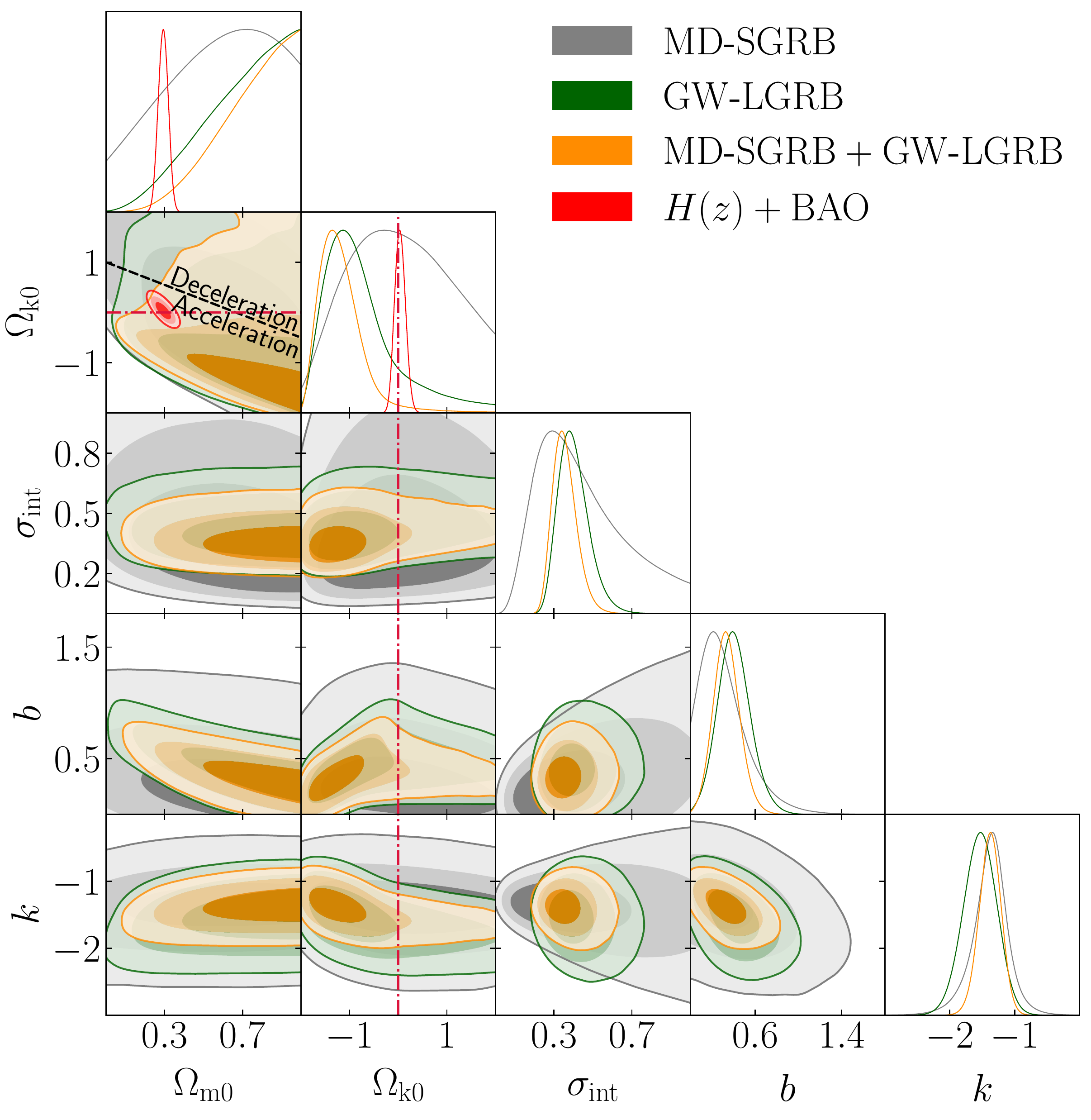}}\\
 \subfloat[Flat XCDM]{%
    \includegraphics[width=3.45in,height=2.5in]{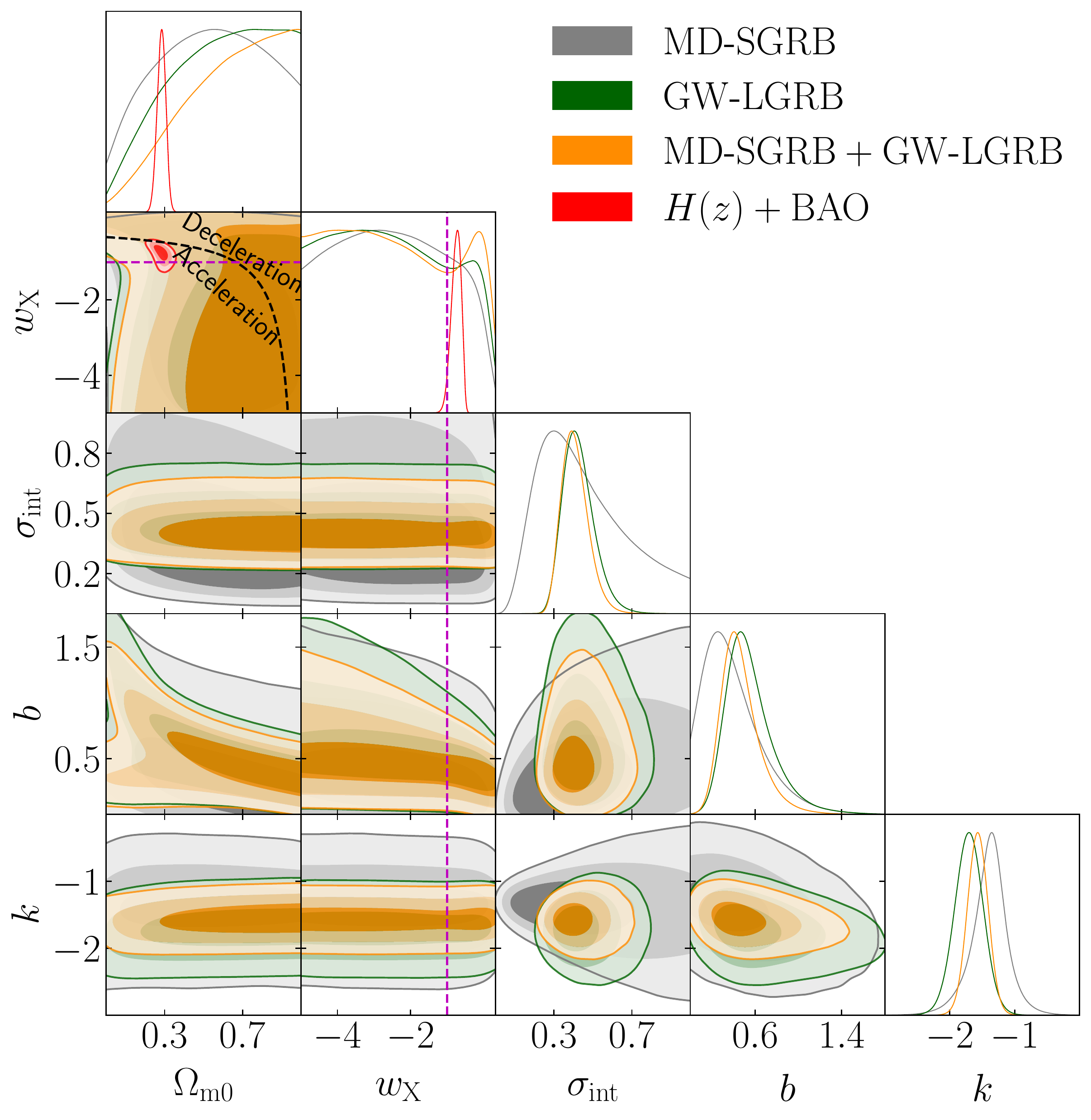}}
 \subfloat[Non-flat XCDM]{%
    \includegraphics[width=3.45in,height=2.5in]{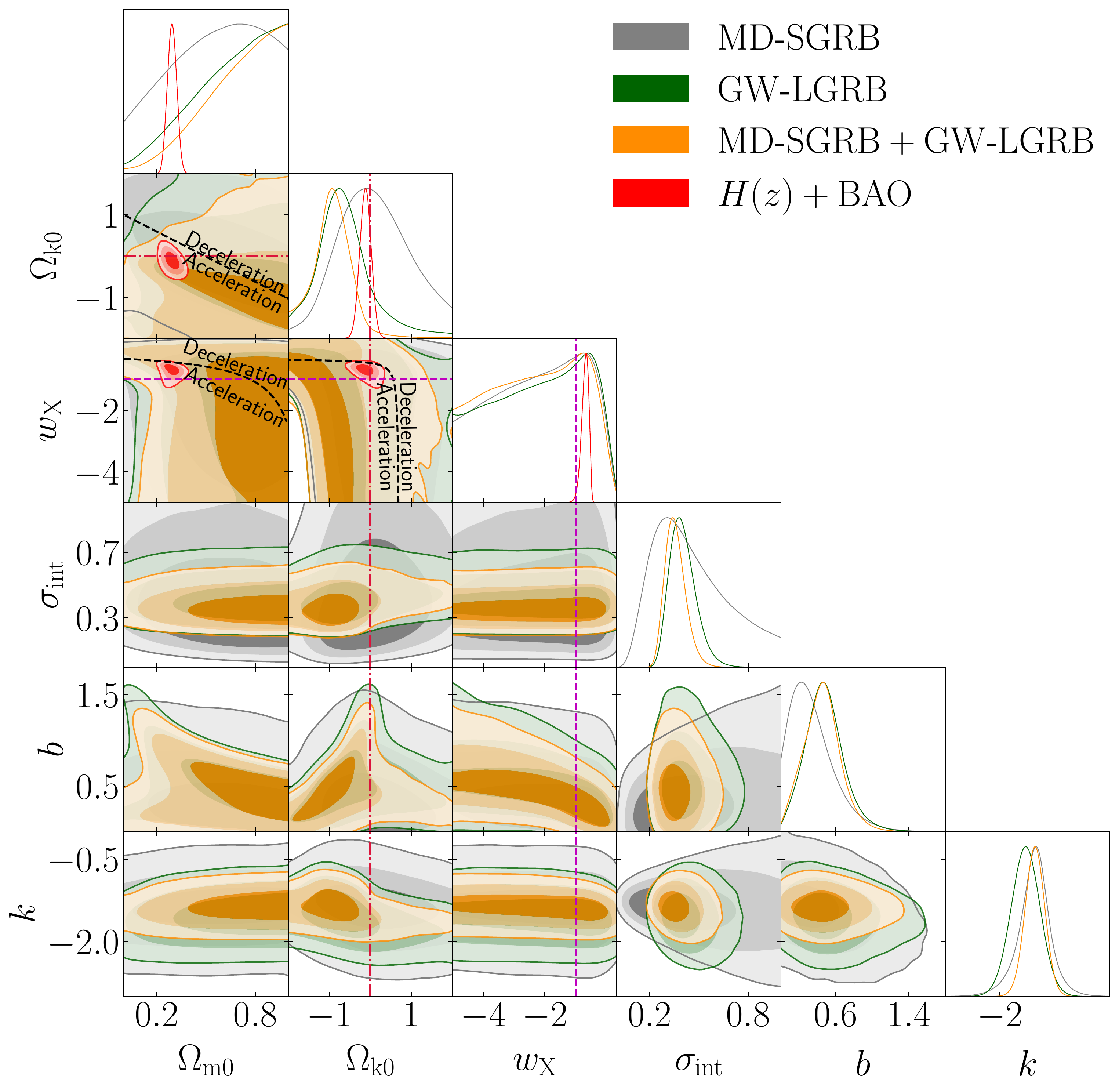}}\\
 \subfloat[Flat \pcdm]{%
    \includegraphics[width=3.45in,height=2.5in]{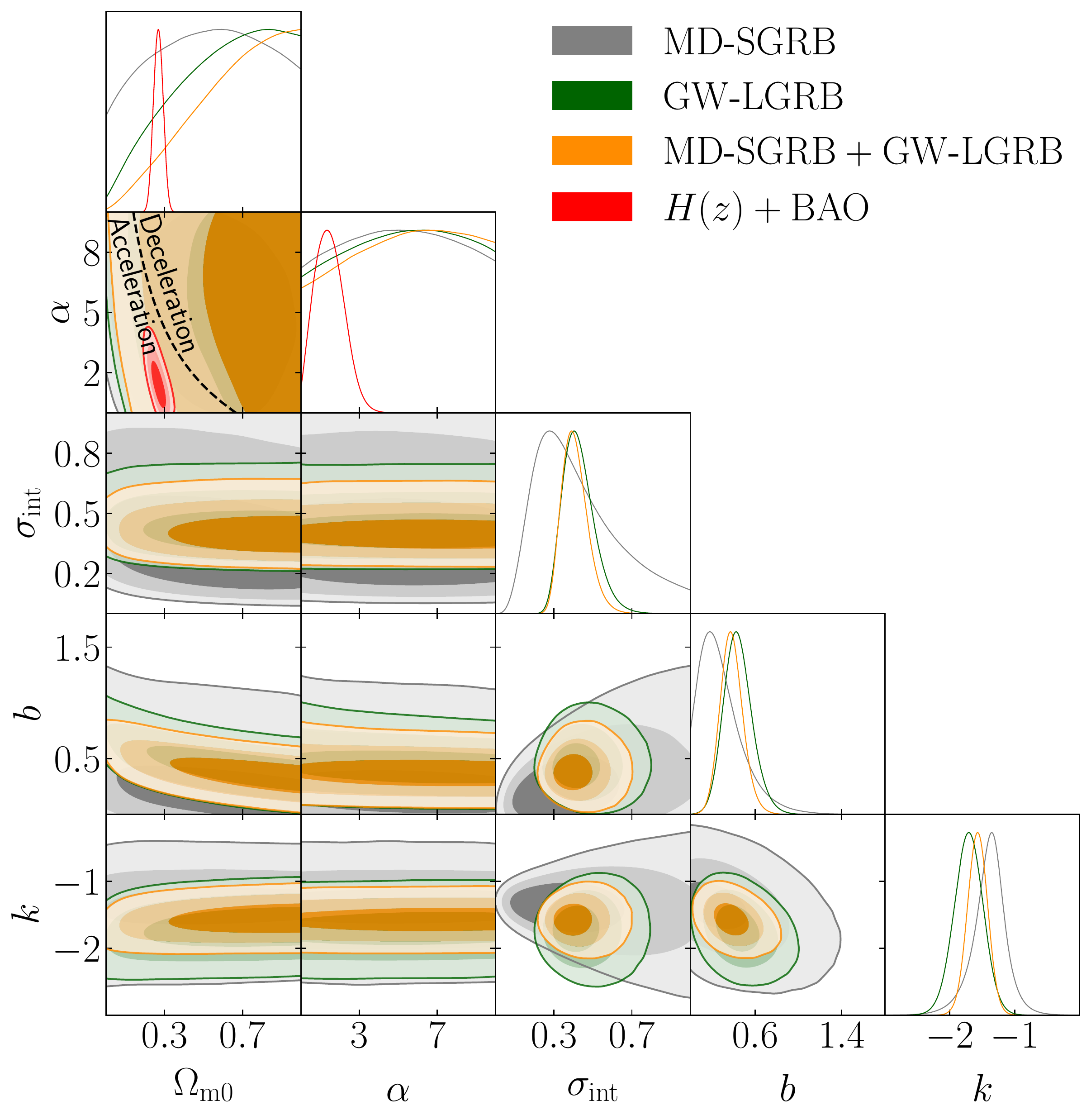}}
 \subfloat[Non-flat \pcdm]{%
    \includegraphics[width=3.45in,height=2.5in]{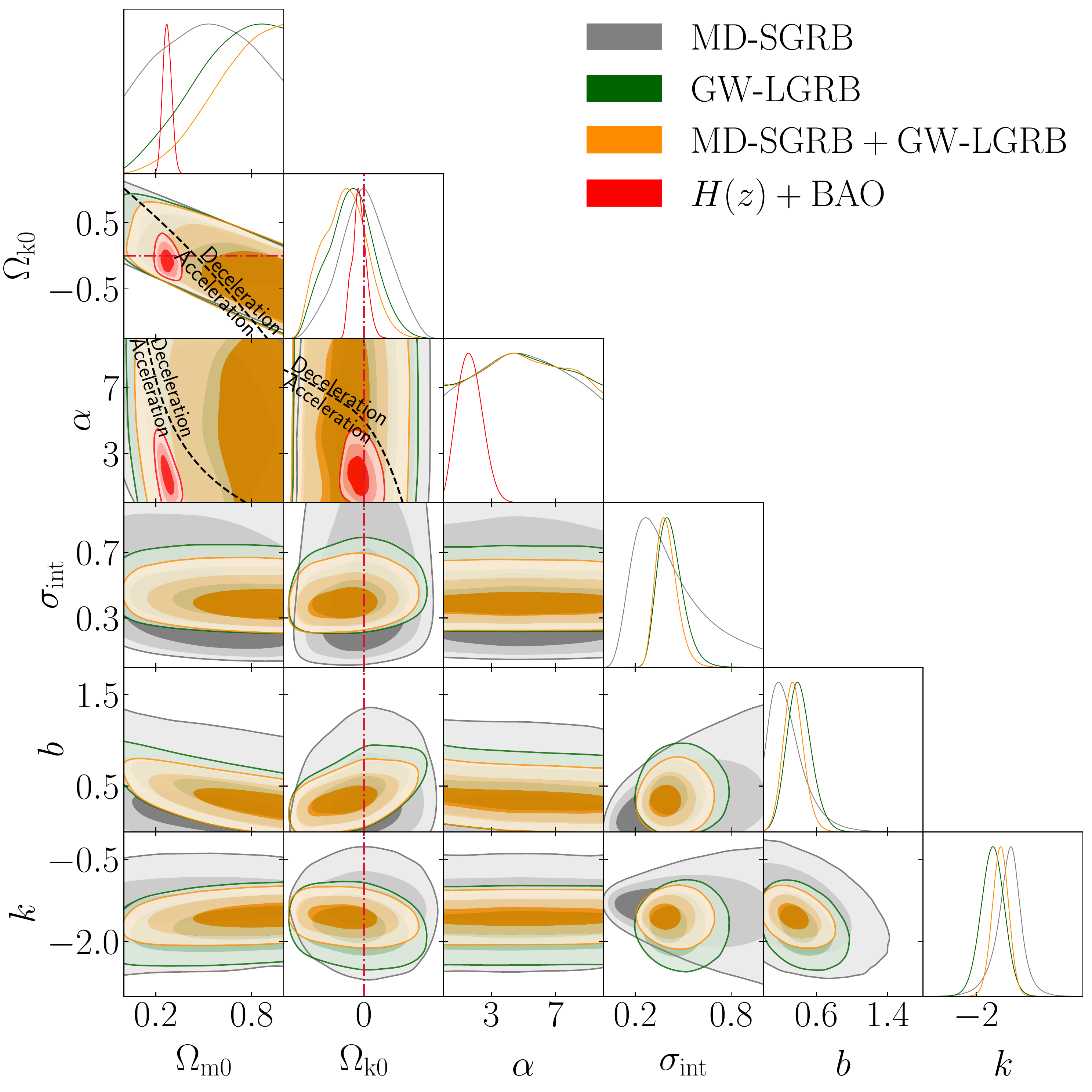}}\\
\caption{One-dimensional likelihoods and 1$\sigma$, 2$\sigma$, and 3$\sigma$ two-dimensional likelihood confidence contours from MD-SGRB (gray), GW-LGRB (green), MD-SGRB + GW-LGRB (orange), and $H(z)$ + BAO (red) data for all six models, without subscripts on $\sigma_{\rm int}$, $k$, and $b$. The zero-acceleration lines are shown as black dashed lines, which divide the parameter space into regions associated with currently-accelerating and currently-decelerating cosmological expansion. In the non-flat XCDM and non-flat \pcdm\ cases, the zero-acceleration lines are computed for the third cosmological parameter set to the $H(z)$ + BAO data best-fitting values listed in Table \ref{tab:BFP}. The crimson dash-dot lines represent flat hypersurfaces, with closed spatial hypersurfaces either below or to the left. The magenta lines represent $w_{\rm X}=-1$, i.e.\ flat or non-flat \lcdm\ models. The $\alpha = 0$ axes correspond to flat and non-flat \lcdm\ models in panels (e) and (f), respectively.}
\label{fig3}
\end{figure*}

We now record and discuss results when these data sets are used to jointly constrain the Dainotti parameters and the cosmological parameters of the six spatially-flat and non-flat dark energy cosmological models. Figure \ref{fig00} shows the flat $\Lambda$CDM Dainotti correlations for the ML, MS, GL, and MS + GL data sets. The unmarginalized best-fitting results and the one-dimensional (1D) posterior mean values and uncertainties are reported in Tables \ref{tab:BFP} and \ref{tab:1d_BFP}, respectively. The corresponding posterior 1D probability distributions and two-dimensional (2D) confidence regions of these parameters are shown in Figs.\ \ref{fig1}--\ref{fig3}, in blue (ML), gray (MS), green (GL), pink (ML + MS), violet (ML + GL), orange (MS + GL), and red ($H(z)$ + BAO, as a baseline). Note that $H_0=70$ \hunit\ and $\Omega_{b}=0.05$ are applied in the GRB cases.

\begin{sidewaystable*}
\centering
\resizebox*{\columnwidth}{0.65\columnwidth}{%
\begin{threeparttable}
\caption{Unmarginalized best-fitting parameter values for all models from various combinations of data.}\label{tab:BFP}
\begin{tabular}{lccccccccccccccccccccccccc}
\toprule
Model & Data set & $\Omega_{b}h^2$ & $\Omega_{c}h^2$ & $\Omega_{\mathrm{m0}}$ & $\Omega_{\mathrm{k0}}$ & $w_{\mathrm{X}}$ & $\alpha$ & $H_0$\tnote{a} & $\sigma_{\mathrm{int,\,\textsc{ml}}}$ & $b_{\mathrm{\textsc{ml}}}$ & $k_{\mathrm{\textsc{ml}}}$ & $\sigma_{\mathrm{int,\,\textsc{ms}}}$ & $b_{\mathrm{\textsc{ms}}}$ & $k_{\mathrm{\textsc{ms}}}$ & $\sigma_{\mathrm{int,\,\textsc{gl}}}$ & $b_{\mathrm{\textsc{gl}}}$ & $k_{\mathrm{\textsc{gl}}}$ & $\sigma_{\mathrm{int,\,\textsc{ms+gl}}}$ & $b_{\mathrm{\textsc{ms+gl}}}$ & $k_{\mathrm{\textsc{ms+gl}}}$ & $-2\ln\mathcal{L}_{\mathrm{max}}$ & $AIC$ & $BIC$ & $\Delta AIC$ & $\Delta BIC$ \\
\midrule
 & $H(z)$ + BAO & 0.0239 & 0.1187 & 0.298 & -- & -- & -- & 69.13 & -- & -- & -- & -- & -- & -- & -- & -- & -- & -- & -- & -- & 23.66 & 29.66 & 34.87 & 0.00 & 0.00\\
 & ML & 0.0245 & 0.4645 & 0.998 & -- & -- & -- & 70 & 0.275 & 1.383 & $-1.010$ & -- & -- & -- & -- & -- & -- & -- & -- & -- & 8.68 & 16.68 & 22.41 & 0.00 & 0.00\\
 & MS & 0.0245 & 0.4651 & 0.999 & -- & -- & -- & 70 & -- & -- & -- & 0.159 & 0.105 & $-1.310$ & -- & -- & -- & -- & -- & -- & $-3.49$ & 4.51 & 2.95 & 0.00 & 0.00\\
Flat \lcdm & GL & 0.0245 & 0.4641 & 0.997 & -- & -- & -- & 70 & -- & -- & -- & -- & -- & -- & 0.370 & 0.359 & $-1.675$ & -- & -- & -- & 22.94 & 30.94 & 35.65 & 0.00 & 0.00\\
 & MS + GL & 0.0245 & 0.4652 & 0.999 & -- & -- & -- & 70 & -- & -- & -- & -- & -- & -- & -- & -- & -- & 0.361 & 0.313 & $-1.545$ & 25.59 & 33.59 & 38.30 & 0.00 & 0.00\\
 & ML + GL & 0.0245 & 0.4642 & 0.997 & -- & -- & -- & 70 & 0.274 & 1.378 & $-1.008$ & -- & -- & -- & 0.362 & 0.358 & $-1.708$ & -- & -- & -- & 31.66 & 45.66 & 59.71 & 0.00 & 0.00\\
 & ML + MS & 0.0245 & 0.463 & 0.995 & -- & -- & -- & 70 & 0.274 & 1.388 & $-1.013$ & 0.148 & 0.080 & $-1.322$ & -- & -- & -- & -- & -- & -- & 5.34 & 19.34 & 30.42 & 0.00 & 0.00\\
\\
 & $H(z)$ + BAO & 0.0247 & 0.1140 & 0.294 & 0.029 & -- & -- & 68.68 & -- & -- & -- & -- & -- & -- & -- & -- & -- & -- & -- & -- & 23.60 & 31.60 & 38.55 & 1.94 & 3.68\\
 & ML & 0.0245 & 0.4410 & 0.950 & $-0.973$ & -- & -- & 70 & 0.268 & 1.316 & $-0.967$ & -- & -- & -- & -- & -- & -- & -- & -- & -- & 7.48 & 17.48 & 24.65 & 0.80 & 2.24\\
 & MS & 0.0245 & 0.2727 & 0.607 & $-1.730$ & -- & -- & 70 & -- & -- & -- & 0.007 & 0.023 & $-0.919$ & -- & -- & -- & -- & -- & -- & $-15.81$ & $-5.81$ & $-7.76$ & $-10.32$ & $-10.71$\\
Non-flat \lcdm & GL & 0.0245 & 0.4640 & 0.997 & $-1.703$ & -- & -- & 70 & -- & -- & -- & -- & -- & -- & 0.329 & 0.238 & $-1.377$ & -- & -- & -- & 17.00 & 27.00 & 32.89 & $-3.94$ & $-2.76$\\
 & MS + GL & 0.0245 & 0.4649 & 0.999 & $-1.738$ & -- & -- & 70 & -- & -- & -- & -- & -- & -- & -- & -- & -- & 0.304 & 0.220 & $-1.303$ & 16.02 & 26.02 & 31.91 & $-7.57$ & $-6.39$\\
 & ML + GL & 0.0245 & 0.4330 & 0.934 & $-1.288$ & -- & -- & 70 & 0.269 & 1.248 & $-0.953$ & -- & -- & -- & 0.336 & 0.330 & $-1.529$ & -- & -- & -- & 26.79 & 42.79 & 58.84 & $-2.87$ & $-0.87$\\
 & ML + MS & 0.0245 & 0.4625 & 0.994 & $-1.130$ & -- & -- & 70 & 0.259 & 1.278 & $-0.966$ & 0.136 & 0.159 & $-1.249$ & -- & -- & -- & -- & -- & -- & 2.91 & 18.91 & 31.58 & $-0.43$ & 1.14\\
\\
 & $H(z)$ + BAO & 0.0304 & 0.0891 & 0.281 & -- & $-0.701$ & -- & 65.18 & -- & -- & -- & -- & -- & -- & -- & -- & -- & -- & -- & -- & 19.65 & 27.65 & 34.60 & $-2.01$ & $-0.27$\\
 & ML & 0.0245 & 0.0327 & 0.117 & -- & 0.133 & -- & 70 & 0.275 & 1.288 & $-0.997$ & -- & -- & -- & -- & -- & -- & -- & -- & -- & 8.14 & 18.14 & 25.31 & 1.46 & 2.90\\
 & MS & 0.0245 & 0.0939 & 0.242 & -- & 0.141 & -- & 70 & -- & -- & -- & 0.160 & 0.054 & $-1.285$ & -- & -- & -- & -- & -- & -- & $-4.23$ & 5.77 & 3.81 & 1.26 & 0.86\\
Flat XCDM & GL & 0.0245 & 0.0035 & 0.057 & -- & 0.139 & -- & 70 & -- & -- & -- & -- & -- & -- & 0.364 & 0.259 & $-1.651$ & -- & -- & -- & 21.97 & 31.97 & 37.86 & 1.03 & 2.21\\ 
 & MS + GL & 0.0245 & 0.0058 & 0.062 & -- & 0.141 & -- & 70 & -- & -- & -- & -- & -- & -- & -- & -- & -- & 0.346 & 0.248 & $-1.518$ & 23.92 & 33.92 & 39.81 & 0.33 & 1.51\\
 & ML + GL & 0.0245 & 0.0735 & 0.200 & -- & 0.143 & -- & 70 & 0.273 & 1.300 & $-1.015$ & -- & -- & -- & 0.359 & 0.273 & $-1.636$ & -- & -- & -- & 30.20 & 46.20 & 62.26 & 0.54 & 6.55\\
 & ML + MS & 0.0245 & $-0.0207$ & 0.008 & -- & 0.137 & -- & 70 & 0.278 & 1.269 & $-0.984$ & 0.175 & 0.031 & $-1.282$ & -- & -- & -- & -- & -- & -- & 3.99 & 19.99 & 32.65 & 0.65 & 2.23\\
\\
 & $H(z)$ + BAO & 0.0290 & 0.0980 & 0.295 & $-0.152$ & $-0.655$ & -- & 65.59 & -- & -- & -- & -- & -- & -- & -- & -- & -- & -- & -- & -- & 18.31 & 28.31 & 37.00 & $-1.35$ & 2.13\\
 & ML & 0.0245 & 0.1525 & 0.361 & $-1.893$ & 0.036 & -- & 70 & 0.269 & 0.949 & $-0.976$ & -- & -- & -- & -- & -- & -- & -- & -- & -- & 7.39 & 19.39 & 27.99 & 2.71 & 5.58\\
 & MS & 0.0245 & 0.3596 & 0.784 & $-1.915$ & $-1.108$ & -- & 70 & -- & -- & -- & 0.057 & 0.098 & $-0.995$ & -- & -- & -- & -- & -- & -- & $-13.61$ & $-1.61$ & $-3.95$ & $-6.12$ & $-6.90$\\
Non-flat XCDM & GL & 0.0245 & 0.0378 & 0.127 & $-0.174$ & $-4.518$ & -- & 70 & -- & -- & -- & -- & -- & -- & 0.327 & 1.237 & $-1.299$ & -- & -- & -- & 16.61 & 28.61 & 35.68 & $-2.33$ & 0.03\\
 & MS + GL & 0.0245 & 0.4212 & 0.910 & $-1.218$ & $-2.308$ & -- & 70 & -- & -- & -- & -- & -- & -- & -- & -- & -- & 0.298 & 0.403 & $-1.248$ & 15.65 & 27.65 & 34.72 & $-5.94$ & $-3.58$\\
 & ML + GL & 0.0245 & 0.3594 & 0.783 & $-0.789$ & $-4.432$ & -- & 70 & 0.269 & 1.449 & $-0.927$ & -- & -- & -- & 0.338 & 0.576 & $-1.429$ & -- & -- & -- & 26.04 & 44.04 & 62.11 & $-1.62$ & 2.40\\
 & ML + MS & 0.0245 & 0.1499 & 0.306 & $-1.993$ & $0.130$ & -- & 70 & 0.281 & 0.881 & $-0.978$ & 0.095 & $-0.164$ & $-1.206$ & -- & -- & -- & -- & -- & -- & $-0.04$ & 17.96 & 32.21 & $-1.38$ & 1.79\\
\\
 & $H(z)$ + BAO & 0.0333 & 0.0788 & 0.264 & -- & -- & 1.504 & 65.20 & -- & -- & -- & -- & -- & -- & -- & -- & -- & -- & -- & -- & 19.49 & 27.49 & 34.44 & $-2.17$ & $-0.43$\\
 & ML & 0.0245 & 0.4651 & 0.999 & -- & -- & 5.225 & 70 & 0.275 & 1.383 & $-1.011$ & -- & -- & -- & -- & -- & -- & -- & -- & -- & 8.68 & 18.68 & 25.85 & 2.00 & 3.44\\
 & MS & 0.0245 & 0.4649 & 0.999 & -- & -- & 8.046 & 70 & -- & -- & -- & 0.160 & 0.099 & $-1.306$ & -- & -- & -- & -- & -- & -- & $-3.49$ & 6.51 & 4.56 & 2.00 & 1.61\\
Flat $\phi$CDM & GL & 0.0245 & 0.4641 & 0.997 & -- & -- & 4.299 & 70 & -- & -- & -- & -- & -- & -- & 0.372 & 0.360 & $-1.674$ & -- & -- & -- & 22.94 & 32.94 & 38.83 & 2.00 & 3.18\\
 & MS + GL & 0.0245 & 0.4653 & 1.000 & -- & -- & 6.323 & 70 & -- & -- & -- & -- & -- & -- & -- & -- & -- & 0.359 & 0.314 & $-1.545$ & 25.59 & 35.59 & 41.48 & 2.00 & 3.18\\
 & ML + GL & 0.0245 & 0.4648 & 0.999 & -- & -- & 7.886 & 70 & 0.274 & 1.375 & $-1.013$ & -- & -- & -- & 0.371 & 0.333 & $-1.647$ & -- & -- & -- & 31.72 & 47.72 & 63.78 & 2.06 & 4.07\\
 & ML + MS & 0.0245 & 0.4611 & 0.991 & -- & -- & 6.029 & 70 & 0.268 & 1.371 & $-0.997$ & 0.148 & 0.113 & $-1.317$ & -- & -- & -- & -- & -- & -- & 5.32 & 21.32 & 33.99 & 1.98 & 3.57\\
\\
 & $H(z)$ + BAO & 0.0334 & 0.0816 & 0.266 & $-0.147$ & -- & 1.915 & 65.70 & -- & -- & -- & -- & -- & -- & -- & -- & -- & -- & -- & -- & 18.15 & 28.15 & 36.84 & $-1.51$ & 1.97\\
 & ML & 0.0245 & 0.4558 & 0.980 & $-0.980$ & -- & 0.423 & 70 & 0.266 & 1.296 & $-0.973$ & -- & -- & -- & -- & -- & -- & -- & -- & -- & 7.48 & 19.48 & 28.09 & 2.80 & 5.68\\
 & MS & 0.0245 & 0.4482 & 0.965 & $-0.964$ & -- & 8.262 & 70 & -- & -- & -- & 0.132 & 0.003 & $-1.261$ & -- & -- & -- & -- & -- & -- & $-5.25$ & 6.75 & 4.40 & 2.24 & 1.45\\
Non-flat $\phi$CDM & GL & 0.0245 & 0.4644 & 0.998 & $-0.993$ & -- & 0.173 & 70 & -- & -- & -- & -- & -- & -- & 0.340 & 0.337 & $-1.547$ & -- & -- & -- & 20.15 & 32.15 & 39.22 & 1.21 & 3.57\\
 & MS + GL & 0.0245 & 0.4601 & 0.989 & $-0.982$ & -- & 0.011 & 70 & -- & -- & -- & -- & -- & -- & -- & -- & -- & 0.315 & 0.310 & $-1.484$ & 21.46 & 33.46 & 40.53 & $-0.13$ & 2.23\\
 & ML + GL & 0.0245 & 0.4465 & 0.961 & $-0.950$ & -- & 0.232 & 70 & 0.255 & 1.290 & $-0.969$ & -- & -- & -- & 0.348 & 0.356 & $-1.600$ & -- & -- & -- & 28.04 & 46.04 & 64.11 & 0.38 & 4.40\\
 & ML + MS & 0.0245 & 0.4628 & 0.995 & $-0.936$ & -- & 8.517 & 70 & 0.281 & 1.202 & $-0.994$ & 0.152 & 0.027 & $-1.291$ & -- & -- & -- & -- & -- & -- & 2.71 & 20.71 & 34.96 & 1.36 & 4.53\\
\bottomrule
\end{tabular}
\begin{tablenotes}[flushleft]
\item [a] \hunit. In the GRB only cases, $H_0$ is set to be 70 \hunit.
\end{tablenotes}
\end{threeparttable}%
}
\end{sidewaystable*}

\begin{sidewaystable*}
\centering
\resizebox*{\columnwidth}{0.65\columnwidth}{%
\begin{threeparttable}
\caption{One-dimensional marginalized posterior mean values and uncertainties ($\pm 1\sigma$ error bars or $2\sigma$ limits) of the parameters for all models from various combinations of data.}\label{tab:1d_BFP}
\begin{tabular}{lcccccccccccccccccccc}
\toprule
Model & Data set & $\Omega_{b}h^2$ & $\Omega_{c}h^2$ & $\Omega_{\mathrm{m0}}$ & $\Omega_{\mathrm{k0}}$ & $w_{\mathrm{X}}$ & $\alpha$ & $H_0$\tnote{a} & $\sigma_{\mathrm{int,\,\textsc{ml}}}$ & $b_{\mathrm{\textsc{ml}}}$ & $k_{\mathrm{\textsc{ml}}}$ & $\sigma_{\mathrm{int,\,\textsc{ms}}}$ & $b_{\mathrm{\textsc{ms}}}$ & $k_{\mathrm{\textsc{ms}}}$ & $\sigma_{\mathrm{int,\,\textsc{gl}}}$ & $b_{\mathrm{\textsc{gl}}}$ & $k_{\mathrm{\textsc{gl}}}$ & $\sigma_{\mathrm{int,\,\textsc{ms+gl}}}$ & $b_{\mathrm{\textsc{ms+gl}}}$ & $k_{\mathrm{\textsc{ms+gl}}}$ \\
\midrule
 & $H(z)$ + BAO & $0.0241\pm0.0029$ & $0.1193^{+0.0082}_{-0.0090}$ & $0.299^{+0.017}_{-0.019}$ & -- & -- & -- & $69.30\pm1.84$ & -- & -- & -- & -- & -- & -- & -- & -- & -- & -- & -- & -- \\
 & ML & -- & -- & $>0.188$ & -- & -- & -- & -- & $0.305^{+0.035}_{-0.053}$ & $1.552^{+0.108}_{-0.189}$ & $-1.017\pm0.090$ & -- & -- & -- & -- & -- & -- & -- & -- & -- \\
 & MS & -- & -- & $0.520^{+0.379}_{-0.253}$ & -- & -- & -- & -- & -- & -- & -- & $0.695^{+0.044}_{-0.550}$ & $0.437^{+0.073}_{-0.400}$ & $-1.450^{+0.362}_{-0.258}$ & -- & -- & -- & -- & -- & -- \\
Flat \lcdm & GL & -- & -- & $>0.202$ & -- & -- & -- & -- & -- & -- & -- & -- & -- & -- & $0.429^{+0.059}_{-0.094}$ & $0.495^{+0.120}_{-0.173}$ & $-1.720\pm0.219$ & -- & -- & -- \\
 & MS + GL & -- & -- & $>0.293$ & -- & -- & -- & -- & -- & -- & -- & -- & -- & -- & -- & -- & -- & $0.412^{+0.052}_{-0.079}$ & $0.421^{+0.101}_{-0.141}$ & $-1.577\pm0.155$ \\
 & ML + GL & -- & -- & $>0.294$ & -- & -- & -- & -- & $0.301^{+0.033}_{-0.051}$ & $1.507^{+0.095}_{-0.151}$ & $-1.015\pm0.089$ & -- & -- & -- & $0.424^{+0.056}_{-0.090}$ & $0.465^{+0.113}_{-0.149}$ & $-1.708\pm0.210$ & -- & -- & -- \\
 & ML + MS & -- & -- & $>0.206$ & -- & -- & -- & -- & $0.302^{+0.034}_{-0.052}$ & $1.542^{+0.104}_{-0.184}$ & $-1.016\pm0.091$ & $0.613^{+0.010}_{-0.479}$ & $0.379^{+0.049}_{-0.350}$ & $-1.426^{+0.312}_{-0.210}$ & -- & -- & -- & -- & -- & -- \\
\\
 & $H(z)$ + BAO & $0.0253^{+0.0041}_{-0.0050}$ & $0.1135^{+0.0196}_{-0.0197}$ & $0.293\pm0.025$ & $0.039^{+0.102}_{-0.115}$ & -- & -- & $68.75^{+2.37}_{-2.36}$ & -- & -- & -- & -- & -- & -- & -- & -- & -- & -- & -- & -- \\
 & ML & -- & -- & $>0.241$ &  $-0.131^{+0.450}_{-0.919}$ & -- & -- & -- & $0.304^{+0.035}_{-0.053}$ & $1.478^{+0.123}_{-0.166}$ & $-1.000\pm0.096$ & -- & -- & -- & -- & -- & -- & -- & -- & -- \\
 & MS & -- & -- & $0.564^{+0.426}_{-0.149}$ &  $0.066^{+1.002}_{-1.199}$ & -- & -- & -- & -- & -- & -- & $0.670^{+0.024}_{-0.533}$ & $0.413^{+0.047}_{-0.393}$ & $-1.430^{+0.360}_{-0.239}$ & -- & -- & -- & -- & -- & -- \\
Non-flat \lcdm & GL & -- & -- & $>0.290$ &  $-0.762^{+0.271}_{-0.888}$ & -- & -- & -- & -- & -- & -- & -- & -- & -- & $0.402^{+0.057}_{-0.090}$ & $0.407^{+0.136}_{-0.160}$ & $-1.536\pm0.252$ & -- & -- & -- \\
 & MS + GL & -- & -- & $>0.391$ &  $-1.165^{+0.225}_{-0.519}$ & -- & -- & -- & -- & -- & -- & -- & -- & -- & -- & -- & -- & $0.357^{+0.046}_{-0.070}$ & $0.337^{+0.110}_{-0.127}$ & $-1.382^{+0.164}_{-0.163}$ \\
 & ML + GL & -- & -- & $>0.338$ &  $-0.737^{+0.299}_{-0.547}$ & -- & -- & -- & $0.300^{+0.033}_{-0.051}$ & $1.386^{+0.138}_{-0.154}$ & $-0.966\pm0.093$ & -- & -- & -- & $0.397^{+0.053}_{-0.086}$ & $0.437^{+0.110}_{-0.142}$ & $-1.588^{+0.215}_{-0.214}$ & -- & -- & -- \\
 & ML + MS & -- & -- & $>0.270$ &  $-0.300^{+0.380}_{-0.836}$ & -- & -- & -- & $0.302^{+0.034}_{-0.052}$ & $1.457^{+0.128}_{-0.163}$ & $-0.993\pm0.095$ & $0.518^{+0.009}_{-0.393}$ & $0.339^{+0.056}_{-0.301}$ & $-1.385^{+0.271}_{-0.186}$ & -- & -- & -- & -- & -- & -- \\
\\
 & $H(z)$ + BAO & $0.0296^{+0.0046}_{-0.0052}$ & $0.0939^{+0.0194}_{-0.0171}$ & $0.284^{+0.023}_{-0.021}$ & -- & $-0.754^{+0.155}_{-0.107}$ & -- & $65.89^{+2.41}_{-2.71}$ & -- & -- & -- & -- & -- & -- & -- & -- & -- & -- & -- & -- \\
 & ML & -- & -- & $>0.123$ & -- & $-2.456^{+2.567}_{-2.180}$ & -- & -- & $0.306^{+0.036}_{-0.054}$ & $1.611^{+0.113}_{-0.277}$ & $-1.014\pm0.092$ & -- & -- & -- & -- & -- & -- & -- & -- & -- \\
 & MS & -- & -- & $0.520^{+0.340}_{-0.276}$ & -- & $-2.494^{+1.264}_{-2.050}$ & -- & -- & -- & -- & -- & $0.704^{+0.035}_{-0.564}$ & $0.497^{+0.086}_{-0.458}$ & $-1.441^{+0.366}_{-0.259}$ & -- & -- & -- & -- & -- & -- \\
Flat XCDM & GL & -- & -- & $>0.141$ & -- & $<-0.046$ & -- & -- & -- & -- & -- & -- & -- & -- & $0.428^{+0.058}_{-0.092}$ & $0.556^{+0.127}_{-0.256}$ & $-1.706\pm0.215$ & -- & -- & -- \\
 & MS + GL & -- & -- & $>0.192$ & -- & $<0.028$ & -- & -- & -- & -- & -- & -- & -- & -- & -- & -- & -- & $0.409^{+0.052}_{-0.078}$ & $0.470^{+0.108}_{-0.206}$ & $-1.570\pm0.155$ \\
 & ML + GL & -- & -- & $>0.164$ & -- & $<0.022$ & -- & -- & $0.300^{+0.033}_{-0.051}$ & $1.566^{+0.099}_{-0.239}$ & $-1.012\pm0.087$ & -- & -- & -- & $0.424^{+0.057}_{-0.092}$ & $0.531^{+0.116}_{-0.240}$ & $-1.700\pm0.213$ & -- & -- & -- \\
 & ML + MS & -- & -- & $>0.142$ & -- & $<-0.072$ & -- & -- & $0.303^{+0.034}_{-0.052}$ & $1.597^{+0.108}_{-0.255}$ & $-1.013\pm0.089$ & $0.562^{+0.010}_{-0.431}$ & $0.409^{+0.066}_{-0.377}$ & $-1.408^{+0.288}_{-0.196}$ & -- & -- & -- & -- & -- & -- \\
\\
 & $H(z)$ + BAO & $0.0290^{+0.0052}_{-0.0055}$ & $0.0990^{+0.0214}_{-0.0215}$ & $0.293\pm0.028$ & $-0.116\pm0.134$ & $-0.700^{+0.138}_{-0.083}$ & -- & $65.96^{+2.32}_{-2.55}$ & -- & -- & -- & -- & -- & -- & -- & -- & -- & -- & -- & -- \\
 & ML & -- & -- & $>0.174$ & $-0.262^{+0.580}_{-0.724}$ & $-2.000^{+2.117}_{-1.264}$ & -- & -- & $0.305^{+0.036}_{-0.054}$ & $1.462^{+0.194}_{-0.196}$ & $-0.996\pm0.097$ & -- & -- & -- & -- & -- & -- & -- & -- & -- \\
 & MS & -- & -- & $0.552^{+0.442}_{-0.152}$ & $0.134^{+0.793}_{-0.987}$ & $-2.234^{+2.159}_{-0.969}$ & -- & -- & -- & -- & -- & $0.733^{+0.031}_{-0.596}$ & $0.464^{+0.052}_{-0.444}$ & $-1.433^{+0.394}_{-0.270}$ & -- & -- & -- & -- & -- & -- \\
Non-flat XCDM & GL & -- & -- & $>0.194$ & $-0.615^{+0.470}_{-0.685}$ & $-2.212^{+2.186}_{-0.962}$ & -- & -- & -- & -- & -- & -- & -- & -- & $0.403^{+0.058}_{-0.092}$ & $0.480^{+0.177}_{-0.223}$ & $-1.532^{+0.259}_{-0.260}$ & -- & -- & -- \\
 & MS + GL & -- & -- & $>0.268$ & $-0.920^{+0.460}_{-0.386}$ & $-2.323^{+2.085}_{-1.095}$ & -- & -- & -- & -- & -- & -- & -- & -- & -- & -- & -- & $0.358^{+0.047}_{-0.072}$ & $0.452^{+0.185}_{-0.203}$ & $-1.363\pm0.175$ \\
 & ML + GL & -- & -- & $>0.196$ & $-0.696^{+0.484}_{-0.408}$ & $-2.158^{+2.254}_{-1.424}$ & -- & -- & $0.298^{+0.033}_{-0.051}$ & $1.422^{+0.201}_{-0.221}$ & $-0.968\pm0.092$ & -- & -- & -- & $0.397^{+0.054}_{-0.087}$ & $0.480^{+0.205}_{-0.224}$ & $-1.561^{+0.220}_{-0.219}$ & -- & -- & -- \\
 & ML + MS & -- & -- & $>0.198$ & $-0.375^{+0.458}_{-0.611}$ & $-2.235^{+2.289}_{-2.007}$ & -- & -- & $0.301^{+0.034}_{-0.052}$ & $1.470^{+0.184}_{-0.183}$ & $-0.988^{+0.095}_{-0.094}$ & $0.525^{+0.007}_{-0.409}$ & $0.397^{+0.079}_{-0.364}$ & $-1.367^{+0.275}_{-0.193}$ & -- & -- & -- & -- & -- & -- \\
\\
 & $H(z)$ + BAO & $0.0321^{+0.0056}_{-0.0039}$ & $0.0823^{+0.0186}_{-0.0183}$ & $0.268\pm0.024$ & -- & -- & $1.467^{+0.637}_{-0.866}$ & $65.24^{+2.15}_{-2.35}$ & -- & -- & -- & -- & -- & -- & -- & -- & -- & -- & -- & -- \\
 & ML & -- & -- & $>0.148$ & -- & -- & -- & -- & $0.304^{+0.035}_{-0.053}$ & $1.493^{+0.093}_{-0.143}$ & $-1.017\pm0.089$ & -- & -- & -- & -- & -- & -- & -- & -- & -- \\
 & MS & -- & -- & $0.514^{+0.365}_{-0.275}$ & -- & -- & -- & -- & -- & -- & -- & $0.597^{+0.029}_{-0.457}$ & $0.355^{+0.052}_{-0.331}$ & $-1.425^{+0.311}_{-0.221}$ & -- & -- & -- & -- & -- & -- \\
Flat $\phi$CDM & GL & -- & -- & $>0.148$ & -- & -- & -- & -- & -- & -- & -- & -- & -- & -- & $0.428^{+0.059}_{-0.094}$ & $0.444^{+0.112}_{-0.141}$ & $-1.710\pm0.218$ & -- & -- & -- \\
 & MS + GL & -- & -- & $>0.222$ & -- & -- & -- & -- & -- & -- & -- & -- & -- & -- & -- & -- & -- & $0.408^{+0.050}_{-0.076}$ & $0.384^{+0.092}_{-0.111}$ & $-1.573\pm0.151$ \\
 & ML + GL & -- & -- & $>0.235$ & -- & -- & -- & -- & $0.301^{+0.033}_{-0.051}$ & $1.463^{+0.086}_{-0.119}$ & $-1.015\pm0.088$ & -- & -- & -- & $0.423^{+0.056}_{-0.091}$ & $0.423^{+0.104}_{-0.121}$ & $-1.701\pm0.209$ & -- & -- & -- \\
 & ML + MS & -- & -- & $>0.166$ & -- & -- & -- & -- & $0.303^{+0.034}_{-0.053}$ & $1.485^{+0.092}_{-0.137}$ & $-1.017\pm0.090$ & $0.517^{+0.001}_{-0.385}$ & $0.304^{+0.033}_{-0.283}$ & $-1.410^{+0.268}_{-0.173}$ & -- & -- & -- & -- & -- & -- \\
\\
 & $H(z)$ + BAO & $0.0319^{+0.0062}_{-0.0037}$ & $0.0848^{+0.0181}_{-0.0220}$ & $0.271^{+0.025}_{-0.028}$ & $-0.074^{+0.104}_{-0.110}$ & -- & $1.653^{+0.685}_{-0.856}$ & $65.46^{+2.31}_{-2.29}$ & -- & -- & -- & -- & -- & -- & -- & -- & -- & -- & -- & -- \\
 & ML & -- & -- & $>0.207$ & $-0.163^{+0.355}_{-0.317}$ & -- & -- & -- & $0.303^{+0.035}_{-0.053}$ & $1.448^{+0.120}_{-0.165}$ & $-1.011\pm0.091$ & -- & -- & -- & -- & -- & -- & -- & -- & -- \\
 & MS & -- & -- & $0.505^{+0.310}_{-0.313}$ & $0.017^{+0.387}_{-0.375}$ & -- & -- & -- & -- & -- & -- & $0.589^{+0.023}_{-0.455}$ & $0.352^{+0.048}_{-0.331}$ & $-1.432^{+0.314}_{-0.207}$ & -- & -- & -- & -- & -- & -- \\
Non-flat $\phi$CDM & GL & -- & -- & $>0.207$ & $-0.193^{+0.364}_{-0.347}$ & -- & -- & -- & -- & -- & -- & -- & -- & -- & $0.422^{+0.057}_{-0.092}$ & $0.408^{+0.121}_{-0.149}$ & $-1.693^{+0.215}_{-0.214}$ & -- & -- & -- \\
 & MS + GL & -- & -- & $>0.313$ & $-0.293^{+0.330}_{-0.359}$ & -- & -- & -- & -- & -- & -- & -- & -- & -- & -- & -- & -- & $0.397^{+0.050}_{-0.076}$ & $0.343^{+0.101}_{-0.120}$ & $-1.546\pm0.150$ \\
 & ML + GL & -- & -- & $>0.340$ & $-0.327^{+0.321}_{-0.345}$ & -- & -- & -- & $0.299^{+0.033}_{-0.051}$ & $1.391^{+0.107}_{-0.136}$ & $-1.004\pm0.088$ & -- & -- & -- & $0.414^{+0.055}_{-0.088}$ & $0.374^{+0.113}_{-0.128}$ & $-1.668^{+0.207}_{-0.208}$ & -- & -- & -- \\
 & ML + MS & -- & -- & $>0.238$ & $-0.198^{+0.351}_{-0.321}$ & -- & -- & -- & $0.301^{+0.034}_{-0.052}$ & $1.435^{+0.114}_{-0.156}$ & $-1.010\pm0.089$ & $0.507^{+0.003}_{-0.379}$ & $0.292^{+0.025}_{-0.281}$ & $-1.393^{+0.258}_{-0.173}$ & -- & -- & -- & -- & -- & -- \\
\bottomrule
\end{tabular}
\begin{tablenotes}[flushleft]
\item [a] \hunit. In the GRB only cases, $H_0$ is set to be 70 \hunit.
\end{tablenotes}
\end{threeparttable}%
}
\end{sidewaystable*}

ML, MS, and GL GRB data have almost cosmological-model independent Dainotti parameters. This means that it is not unreasonable to treat the ML, MS, and GL GRBs as standardizable candles, as was assumed in \cite{Wangetal_2021} and \cite{Huetal_2021}. 

In the ML case (with subscript ``ML'' in the first line of Tables \ref{tab:BFP}, \ref{tab:1d_BFP}, \ref{tab:BFP2}, and \ref{tab:1d_BFP2}), the slope $k$ ranges from a high of $-0.996\pm0.097$ (non-flat XCDM) to a low of $-1.017\pm0.090$ (flat \lcdm), the intercept $b$ ranges from a high of $1.611^{+0.113}_{-0.277}$ (flat XCDM) to a low of $1.448^{+0.120}_{-0.165}$ (non-flat \pcdm), and the intrinsic scatter $\sigma_{\rm int}$ ranges from a high of $0.306^{+0.036}_{-0.054}$ (flat XCDM) to a low of $0.303^{+0.035}_{-0.053}$ (non-flat \pcdm), with central values of each pair being $0.16\sigma$, $0.54\sigma$, and $0.05\sigma$ away from each other, respectively. 

In the MS case (with subscript ``MS'' in the first line of Tables \ref{tab:BFP} and \ref{tab:1d_BFP}) with prior range of $b\in[0,10]$, the slope $k$ ranges from a high of $-1.425^{+0.311}_{-0.221}$ (flat \pcdm) to a low of $-1.450^{+0.362}_{-0.258}$ (flat \lcdm), the intercept $b$ ranges from a high of $0.497^{+0.086}_{-0.458}$ (flat XCDM) to a low of $0.352^{+0.048}_{-0.331}$ (non-flat \pcdm), and the intrinsic scatter $\sigma_{\rm int}$ ranges from a high of $0.733^{+0.031}_{-0.596}$ (non-flat XCDM) to a low of $0.589^{+0.023}_{-0.455}$ (non-flat \pcdm), with central values of each pair being $0.06\sigma$, $0.54\sigma$, and $0.05\sigma$ away from each other, respectively.\footnote{Note, however, that the lower error bars of $b$ and $\sigma_{\rm int}$ are considerably larger than the upper ones due to cut-off prior ranges of the former and skewed distributions of the latter. Therefore here we also consider the MS case with wider prior range of $b\in[-10,10]$, which are not listed in the tables due to their insignificant differences. Because the lowest and highest values of $k$, $b$, and $\sigma_{\rm int}$ from these two MS cases differ from each other at only $0.22\sigma$, $0.56\sigma$, and $0.37\sigma$, respectively, and the constraints of the cosmological parameters are also within $1\sigma$ range, the prior range of $b\in[0,10]$ is an acceptable choice.}

In the GL case (with subscript ``GL'' in the first line of Tables \ref{tab:BFP}, \ref{tab:1d_BFP}, \ref{tab:BFP2}, and \ref{tab:1d_BFP2}), the slope $k$ ranges from a high of $-1.532^{+0.259}_{-0.260}$ (non-flat XCDM) to a low of $-1.720\pm0.219$ (flat \lcdm), the intercept $b$ ranges from a high of $0.556^{+0.127}_{-0.256}$ (flat XCDM) to a low of $0.407^{+0.136}_{-0.160}$ (non-flat \lcdm), and the intrinsic scatter $\sigma_{\rm int}$ ranges from a high of $0.429^{+0.059}_{-0.094}$ (flat \lcdm) to a low of $0.402^{+0.057}_{-0.090}$ (non-flat \pcdm), with central values of each pair being $0.55\sigma$, $0.51\sigma$, and $0.25\sigma$ away from each other, respectively.

Figure \ref{fig00}, panel (c), shows that the GL and MS GRBs obey the same Dainotti correlation in the flat \lcdm\ model, within the uncertainties.\footnote{It is unclear if this is more than just a coincidence, as the plateau phases in the two cases are dominated by GW emission (GL) and MD radiation (MS), respectively.} Table \ref{tab:comp} shows that the differences between the GL and MS Dainotti parameters in all six cosmological models are within $1\sigma$. GL and MS GRBs however follow a different Dainotti correlation than the ML GRBs. Given the similarity of the GL and MS Dainotti correlation parameters, it is not unreasonable to use just three (not six) correlation parameters in joint analyses of MS and GL data (with subscript ``MS+GL'' in the first line of Tables \ref{tab:BFP} and \ref{tab:1d_BFP}). In this case, the slope $k$ ranges from a high of $-1.363\pm0.175$ (non-flat XCDM) to a low of $-1.577\pm0.155$ (flat \lcdm), the intercept $b$ ranges from a high of $0.470^{+0.108}_{-0.206}$ (flat XCDM) to a low of $0.337^{+0.110}_{-0.127}$ (non-flat \lcdm), and the intrinsic scatter $\sigma_{\rm int}$ ranges from a high of $0.412^{+0.052}_{-0.079}$ (flat \lcdm) to a low of $0.357^{+0.046}_{-0.070}$ (non-flat \lcdm), with central values of each pair being $0.92\sigma$, $0.57\sigma$, and $0.60\sigma$ away from each other, respectively. In contrast to the GL case, the MS + GL case tightens the constraints a little bit, with smaller error bars, and prefers lower values of $b$ and $\sigma_{\rm int}$, and higher values of $k$. When we jointly analyze ML + GL and ML + MS, the constraints on the Dainotti parameters follow the same pattern as that of MS + GL against GL.

The constraints on the cosmological parameters are very loose for all of these cases. In the flat \lcdm\ model, the highest $2\sigma$ lower limit of \om\ among these cases is $\Om>0.294$ of the ML + GL data. In the non-flat \lcdm\ model, the highest $2\sigma$ lower limit of \om\ is $\Om>0.391$ of the MS + GL case, which is inconsistent with that of the $H(z)$ + BAO case. The MS data favor open hypersufaces while all other cases favor closed hypersufaces, with the favored spatial geometries for GL, ML + GL, and MS + GL data being more than $1\sigma$ (or even $2\sigma$) away from flat geometry. In the flat XCDM parametrization, the highest $2\sigma$ lower limit of \om\ is $\Om>0.192$ for the MS + GL data, and the constraints on the X-fluid equation of state parameter \wx\ are very loose, with the highest $1\sigma$ upper limit being $0.111$ for the ML case. In the non-flat XCDM parametrization, the highest $2\sigma$ lower limit of \om\ is $\Om>0.268$ for the MS + GL data, and the constraints on \wx\ are very loose, with the highest $1\sigma$ upper limit being $0.238$ for the MS + GL data. The favored spatial geometries for these cases follow the same pattern as that for non-flat \lcdm, but with larger upper limits of \ok\ except for the MS data. In the flat \pcdm\ model, the highest $2\sigma$ lower limit of \om\ is $\Om>0.235$ for the ML + GL data. In the non-flat \pcdm\ model, the highest $2\sigma$ lower limit of \om\ is $\Om>0.340$ for the ML + GL case, which is inconsistent with that of the $H(z)$ + BAO data. Except for the MS case, closed spatial hypersurfaces are favored, but only in the ML + GL case is flat geometry slightly more than $1\sigma$ away. There are no constraints on $\alpha$ from these GRB data.

In the \lcdm\ and XCDM cases, all GRB data combinations more favor currently accelerating cosmological expansion. They however more favor currently decelerating cosmological expansion in the \pcdm\ models, in the $\Om-\alpha$ and $\Om-\Ok$ parameter subspaces.

From the $AIC$ and $BIC$ values we compute $\Delta AIC$ and $\Delta BIC$ values with respect to the flat \lcdm\ model. These are listed in the last two columns of Table \ref{tab:BFP}. In the ML case, flat \lcdm\ is the most favored model but there is only weak or positive evidence against any other model. In the MS case non-flat \lcdm\ model is the most favored model and, except for non-flat XCDM (with positive evidence against it), the other models are very strongly disfavored. In the GL case, non-flat \lcdm\ is again the most favored model, while the evidence against the others are mostly positive, except for non-flat \pcdm\ (with strong $BIC$ evidence against it). In the MS + GL case, similar to the MS case, non-flat \lcdm\ is the most favored model and, except for non-flat XCDM (with weak $AIC$ and positive $BIC$ evidence against it), the others are strongly disfavored. In the ML + GL case, non-flat \lcdm\ is the most favored model but, except for flat XCDM (with strong $BIC$ evidence against it), the evidence against the other models is either weak or positive. In the ML + MS case, the best candidates are non-flat XCDM based on $AIC$ and flat \lcdm\ based on $BIC$, while the evidence against the other models is either weak or positive.

\begin{table}
\begin{threeparttable}
\caption{MD-SGRB and GW-LGRB data $L_0-t_b$ correlation parameters (and $\sigma_{\rm int}$) differences.}
\label{tab:comp}
\setlength{\tabcolsep}{14.5pt}
\begin{tabular}{lccc}
\toprule
Model & $\Delta \sigma_{\rm int}$  & $\Delta k$  & $\Delta b$\\
\midrule
Flat \lcdm\  & $0.48\sigma$ & $0.31\sigma$ & $0.80\sigma$\\
Non-flat \lcdm\ & $0.50\sigma$ & $0.31\sigma$ & $0.01\sigma$ \\
Flat XCDM  & $0.49\sigma$ & $0.79\sigma$ & $0.22\sigma$\\
Non-flat XCDM  & $0.55\sigma$ & $0.26\sigma$ & $0.07\sigma$\\
Flat $\phi$CDM  & $0.37\sigma$ & $0.92\sigma$ & $0.59\sigma$\\
Non-flat $\phi$CDM & $0.36\sigma$ & $0.87\sigma$ & $0.36\sigma$\\
\bottomrule
\end{tabular}
\end{threeparttable}
\end{table}

\subsection{Constraints from A118, A115 (and jointly with ML), and A115$^{\prime}$ (and jointly with GL) data}
 \label{subsec:GRB-A}

\begin{sidewaystable*}
\centering
\resizebox*{\columnwidth}{0.65\columnwidth}{%
\begin{threeparttable}
\caption{Unmarginalized best-fitting parameter values for all models from various combinations of data.\tnote{a}}\label{tab:BFP2}
\begin{tabular}{lcccccccccccccccccccc}
\toprule
Model & Data set & $\Omega_{c}h^2$ & $\Omega_{\mathrm{m0}}$ & $\Omega_{\mathrm{k0}}$ & $w_{\mathrm{X}}$ & $\alpha$ & $\sigma_{\mathrm{int,\,\textsc{ml}}}$ & $b_{\mathrm{\textsc{ml}}}$ & $k_{\mathrm{\textsc{ml}}}$ & $\sigma_{\mathrm{int}}$ & $\gamma$ & $\beta$ & $\sigma_{\mathrm{int,\,\textsc{gl}}}$ & $b_{\mathrm{\textsc{gl}}}$ & $k_{\mathrm{\textsc{gl}}}$ & $-2\ln\mathcal{L}_{\mathrm{max}}$ & $AIC$ & $BIC$ & $\Delta AIC$ & $\Delta BIC$ \\
\midrule
 & A118 & 0.4089 & 0.884 & -- & -- & -- & -- & -- & -- & 0.401 & 50.02 & 1.099 & -- & -- & -- & 128.72 & 136.72 & 147.81 & 0.00 & 0.00\\
 & ML & 0.4645 & 0.998 & -- & -- & -- & 0.275 & 1.383 & $-1.010$ & -- & -- & -- & -- & -- & -- & 8.68 & 16.68 & 22.41 & 0.00 & 0.00\\
 & A115 & 0.4172 & 0.901 & -- & -- & -- & -- & -- & -- & 0.405 & 50.01 & 1.099 & -- & -- & -- & 127.97 & 135.97 & 146.95 & 0.00 & 0.00\\
Flat \lcdm & ML + A115 & 0.4540 & 0.977 & -- & -- & -- & 0.274 & 1.400 & $-1.019$ & 0.407 & 50.00 & 1.097 & -- & -- & -- & 136.70 & 150.70 & 171.59 & 0.00 & 0.00\\
 & GL & 0.4641 & 0.997 & -- & -- & -- & -- & -- & -- & -- & -- & -- & 0.370 & 0.359 & $-1.675$ & 22.94 & 30.94 & 35.65 & 0.00 & 0.00\\
 & A115$^{\prime}$ & 0.4629 & 0.995 & -- & -- & -- & -- & -- & -- & 0.403 & 50.01 & 1.091 & -- & -- & -- & 126.34 & 134.34 & 145.32 & 0.00 & 0.00\\
 & GL + A115$^{\prime}$ & 0.4652 & 0.999 & -- & -- & -- & -- & -- & -- & 0.402 & 49.96 & 1.110 & 0.370 & 0.363 & $-1.666$ & 149.34 & 163.34 & 183.88 & 0.00 & 0.00\\
\\
 & A118 & 0.4622 & 0.993 & 0.907 & -- & -- & -- & -- & -- & 0.400 & 49.92 & 1.115 & -- & -- & -- & 127.96 & 137.96 & 151.82 & 1.24 & 4.01\\
 & ML & 0.4410 & 0.950 & $-0.973$ & -- & -- & 0.268 & 1.316 & $-0.967$ & -- & -- & -- & -- & -- & -- & 7.48 & 17.48 & 24.65 & 0.80 & 2.24\\
 & A115 & 0.4631 & 0.995 & 1.014 & -- & -- & -- & -- & -- & 0.403 & 49.90 & 1.118 & -- & -- & -- & 127.18 & 137.18 & 150.90 & 1.21 & 3.95\\
Non-flat \lcdm & ML + A115 & 0.4637 & 0.996 & 0.062 & -- & -- & 0.283 & 1.383 & $-1.006$ & 0.410 & 50.01 & 1.088 & -- & -- & -- & 136.74 & 152.74 & 176.61 & 2.04 & 5.02\\
 & GL & 0.4640 & 0.997 & $-1.703$ & -- & -- & -- & -- & -- & -- & -- & -- & 0.329 & 0.238 & $-1.377$ & 17.00 & 27.00 & 32.89 & $-3.94$ & $-2.76$\\
 & A115$^{\prime}$ & 0.4647 & 0.998 & 0.729 & -- & -- & -- & -- & -- & 0.404 & 49.94 & 1.110 & -- & -- & -- & 125.93 & 135.93 & 149.65 & 1.59 & 4.33\\
 & GL + A115$^{\prime}$ & 0.4544 & 0.977 & $-0.244$ & -- & -- & -- & -- & -- & 0.402 & 50.00 & 1.100 & 0.362 & 0.364 & $-1.674$ & 149.29 & 165.29 & 188.77 & 1.95 & 4.89\\
\\
 & A118 & $-0.0115$ & 0.027 & -- & $-0.098$ & -- & -- & -- & -- & 0.399 & 50.04 & 1.102 & -- & -- & -- & 128.43 & 138.43 & 152.29 & 1.71 & 4.48\\
 & ML & 0.0327 & 0.117 & -- & 0.133 & -- & 0.275 & 1.288 & $-0.997$ & -- & -- & -- & -- & -- & -- & 8.14 & 18.14 & 25.31 & 1.46 & 2.90\\
 & A115 & $-0.0197$ & 0.010 & -- & $-0.102$ & -- & -- & -- & -- & 0.407 & 50.01 & 1.115 & -- & -- & -- & 127.66 & 137.66 & 151.38 & 1.69 & 4.43\\
Flat XCDM & ML + A115 & 0.4065 & 0.880 & -- & $-4.386$ & -- & 0.273 & 1.431 & $-1.008$ & 0.403 & 50.05 & 1.093 & -- & -- & -- & 136.70 & 152.70 & 176.57 & 2.00 & 4.98\\
 & GL & 0.0035 & 0.057 & -- & 0.139 & -- & -- & -- & -- & -- & -- & -- & 0.364 & 0.259 & $-1.651$ & 21.97 & 31.97 & 37.86 & 1.03 & 2.21\\ 
 & A115$^{\prime}$ & 0.0119 & 0.074 & -- & $-0.082$ & -- & -- & -- & -- & 0.402 & 50.03 & 1.100 & -- & -- & -- & 126.22 & 136.22 & 149.94 & 1.88 & 4.62\\
 & GL + A115$^{\prime}$ & 0.2817 & 0.625 & -- & 0.122 & -- & -- & -- & -- & 0.401 & 49.98 & 1.084 & 0.376 & 0.330 & $-1.672$ & 149.21 & 165.21 & 188.69 & 1.87 & 4.81\\
\\
 & A118 & 0.4603 & 0.989 & 0.841 & $-1.048$ & -- & -- & -- & -- & 0.399 & 49.93 & 1.112 & -- & -- & -- & 127.99 & 139.99 & 156.61 & 3.27 & 8.80\\
 & ML & 0.1525 & 0.361 & $-1.893$ & 0.036 & -- & 0.269 & 0.949 & $-0.976$ & -- & -- & -- & -- & -- & -- & 7.39 & 19.39 & 27.99 & 2.71 & 5.58\\
 & A115 & 0.4602 & 0.989 & 0.955 & $-1.097$ & -- & -- & -- & -- & 0.404 & 49.91 & 1.115 & -- & -- & -- & 127.18 & 139.18 & 155.65 & 3.21 & 8.70\\
Non-flat XCDM & ML + A115 & 0.3647 & 0.794 & 0.002 & $-4.000$ & -- & 0.269 & 1.478 & $-1.015$ & 0.404 & 50.05 & 1.107 & -- & -- & -- & 136.75 & 154.75 & 181.60 & 4.05 & 10.01\\
 & GL & 0.0378 & 0.127 & $-0.174$ & $-4.518$ & -- & -- & -- & -- & -- & -- & -- & 0.327 & 1.237 & $-1.299$ & 16.61 & 28.61 & 35.68 & $-2.33$ & 0.03\\
 & A115$^{\prime}$ & 0.4643 & 0.997 & 0.783 & $-0.956$ & -- & -- & -- & -- & 0.403 & 49.94 & 1.112 & -- & -- & -- & 125.92 & 137.92 & 154.39 & 3.58 & 9.07\\
 & GL + A115$^{\prime}$ & 0.4349 & 0.938 & $-0.255$ & $-0.043$ & -- & -- & -- & -- & 0.410 & 50.03 & 1.063 & 0.358 & 0.296 & $-1.616$ & 149.16 & 167.16 & 193.57 & 3.82 & 9.69\\
\\
 & A118 & 0.2301 & 0.520 & -- & -- & 9.936 & -- & -- & -- & 0.402 & 50.02 & 1.109 & -- & -- & -- & 128.56 & 138.56 & 152.41 & 1.84 & 4.60\\
 & ML & 0.4651 & 0.999 & -- & -- & 5.225 & 0.275 & 1.383 & $-1.011$ & -- & -- & -- & -- & -- & -- & 8.68 & 18.68 & 25.85 & 2.00 & 3.44\\
 & A115 & 0.2065 & 0.471 & -- & -- & 9.932 & -- & -- & -- & 0.405 & 50.04 & 1.106 & -- & -- & -- & 127.79 & 137.79 & 151.52 & 1.82 & 4.57\\
Flat $\phi$CDM & ML + A115 & 0.4563 & 0.981 & -- & -- & 2.762 & 0.273 & 1.391 & $-1.021$ & 0.405 & 49.99 & 1.099 & -- & -- & -- & 136.70 & 152.70 & 176.57 & 2.00 & 4.98\\
 & GL & 0.4641 & 0.997 & -- & -- & 4.299 & -- & -- & -- & -- & -- & -- & 0.372 & 0.360 & $-1.674$ & 22.94 & 32.94 & 38.83 & 2.00 & 3.18\\
 & A115$^{\prime}$ & 0.3334 & 0.730 & -- & -- & 9.652 & -- & -- & -- & 0.402 & 50.02 & 1.096 & -- & -- & -- & 126.29 & 136.29 & 150.01 & 1.95 & 4.69\\
 & GL + A115$^{\prime}$ & 0.4484 & 0.965 & -- & -- & 8.745 & -- & -- & -- & 0.403 & 50.05 & 1.077 & 0.372 & 0.348 & $-1.663$ & 149.36 & 165.36 & 188.83 & 2.02 & 4.95\\
\\
 & A118 & 0.3245 & 0.712 & 0.245 & -- & 8.862 & -- & -- & -- & 0.400 & 50.01 & 1.116 & -- & -- & -- & 128.42 & 140.42 & 157.04 & 3.70 & 9.23\\
 & ML & 0.4558 & 0.980 & $-0.980$ & -- & 0.423 & 0.266 & 1.296 & $-0.973$ & -- & -- & -- & -- & -- & -- & 7.48 & 19.48 & 28.09 & 2.80 & 5.68\\
 & A115 & 0.3217 & 0.706 & 0.290 & -- & 3.150 & -- & -- & -- & 0.406 & 50.05 & 1.108 & -- & -- & -- & 127.64 & 139.64 & 156.11 & 3.67 & 9.16\\
Non-flat $\phi$CDM & ML + A115 & 0.3044 & 0.671 & $-0.076$ & -- & 8.893 & 0.274 & 1.406 & $-1.001$ & 0.404 & 50.03 & 1.095 & -- & -- & -- & 136.77 & 154.77 & 181.62 & 4.07 & 10.03\\
 & GL & 0.4644 & 0.998 & $-0.993$ & -- & 0.173 & -- & -- & -- & -- & -- & -- & 0.340 & 0.337 & $-1.547$ & 20.15 & 32.15 & 39.22 & 1.21 & 3.57\\
 & A115$^{\prime}$ & 0.3787 & 0.823 & 0.161 & -- & 7.940 & -- & -- & -- & 0.404 & 50.02 & 1.105 & -- & -- & -- & 126.21 & 138.21 & 154.68 & 3.87 & 9.36\\
 & GL + A115$^{\prime}$ & 0.4420 & 0.952 & $-0.266$ & -- & 8.343 & -- & -- & -- & 0.400 & 50.00 & 1.077 & 0.372 & 0.310 & $-1.622$ & 149.18 & 167.18 & 193.59 & 3.84 & 9.71\\
\bottomrule
\end{tabular}
\begin{tablenotes}[flushleft]
\item [a] In these GRB cases, $\Omega_b$ and $H_0$ are set to be 0.05 and 70 \hunit, respectively.
\end{tablenotes}
\end{threeparttable}%
}
\end{sidewaystable*}

\begin{sidewaystable*}
\centering
\resizebox*{\columnwidth}{0.65\columnwidth}{%
\begin{threeparttable}
\caption{One-dimensional marginalized posterior mean values and uncertainties ($\pm 1\sigma$ error bars or $2\sigma$ limits) of the parameters for all models from various combinations of data.\tnote{a}}\label{tab:1d_BFP2}
\begin{tabular}{lcccccccccccccc}
\toprule
Model & Data set & $\Omega_{\mathrm{m0}}$ & $\Omega_{\mathrm{k0}}$ & $w_{\mathrm{X}}$ & $\alpha$ & $\sigma_{\mathrm{int,\,\textsc{ml}}}$ & $b_{\mathrm{\textsc{ml}}}$ & $k_{\mathrm{\textsc{ml}}}$ & $\sigma_{\mathrm{int}}$ & $\gamma$ & $\beta$ & $\sigma_{\mathrm{int,\,\textsc{gl}}}$ & $b_{\mathrm{\textsc{gl}}}$ & $k_{\mathrm{\textsc{gl}}}$ \\
\midrule
 & A118 & $>0.247$ & -- & -- & -- & -- & -- & -- & $0.412^{+0.027}_{-0.033}$ & $50.09\pm0.26$ & $1.110\pm0.090$ & -- & -- & -- \\
 & ML & $>0.188$ & -- & -- & -- & $0.305^{+0.035}_{-0.053}$ & $1.552^{+0.108}_{-0.189}$ & $-1.017\pm0.090$ & -- & -- & -- & -- & -- & -- \\
 & A115 & $0.630^{+0.352}_{-0.135}$ & -- & -- & -- & -- & -- & -- & $0.417^{+0.028}_{-0.035}$ & $50.09\pm0.26$ & $1.112\pm0.093$ & -- & -- & -- \\
Flat \lcdm & ML + A115 & $>0.298$ & -- & -- & -- & $0.301^{+0.033}_{-0.051}$ & $1.515^{+0.101}_{-0.151}$ & $-1.015\pm0.088$ & $0.416^{+0.027}_{-0.034}$ & $50.07\pm0.25$ & $1.111\pm0.089$ & -- & -- & -- \\
 & GL & $>0.202$ & -- & -- & -- & -- & -- & -- & -- & -- & -- & $0.429^{+0.059}_{-0.094}$ & $0.495^{+0.120}_{-0.173}$ & $-1.720\pm0.219$ \\
 & A115$^{\prime}$ & $>0.264$ & -- & -- & -- & -- & -- & -- & $0.414^{+0.028}_{-0.034}$ & $50.10\pm0.26$ & $1.107\pm0.090$ & -- & -- & -- \\%
 & GL + A115$^{\prime}$ & $>0.339$ & -- & -- & -- & -- & -- & -- & $0.413^{+0.027}_{-0.034}$ & $50.08\pm0.25$ & $1.104\pm0.089$ & $0.423^{+0.056}_{-0.090}$ & $0.458^{+0.112}_{-0.139}$ & $-1.705\pm0.210$ \\
\\
 & A118 & $>0.287$ &  $0.694^{+0.626}_{-0.848}$ & -- & -- & -- & -- & -- & $0.412^{+0.027}_{-0.034}$ & $50.01\pm0.26$ & $1.121\pm0.090$ & -- & -- & -- \\
 & ML & $>0.241$ &  $-0.131^{+0.450}_{-0.919}$ & -- & -- & $0.304^{+0.035}_{-0.053}$ & $1.478^{+0.123}_{-0.166}$ & $-1.000\pm0.096$ & -- & -- & -- & -- & -- & -- \\
 & A115 & $>0.275$ &  $0.698^{+0.639}_{-0.857}$ & -- & -- & -- & -- & -- & $0.417^{+0.028}_{-0.034}$ & $50.00\pm0.27$ & $1.124\pm0.092$ & -- & -- & -- \\
Non-flat \lcdm & ML + A115 & $>0.346$ &  $0.352^{+0.427}_{-0.830}$ & -- & -- & $0.304^{+0.034}_{-0.052}$ & $1.486^{+0.096}_{-0.136}$ & $-1.019\pm0.091$ & $0.416^{+0.028}_{-0.034}$ & $50.03\pm0.26$ & $1.113\pm0.091$ & -- & -- & -- \\
 & GL & $>0.290$ &  $-0.762^{+0.271}_{-0.888}$ & -- & -- & -- & -- & -- & -- & -- & -- & $0.402^{+0.057}_{-0.090}$ & $0.407^{+0.136}_{-0.160}$ & $-1.536\pm0.252$ \\
 & A115$^{\prime}$ & $>0.299$ & $0.599^{+0.582}_{-0.887}$ & -- & -- & -- & -- & -- & $0.414^{+0.028}_{-0.034}$ & $50.02\pm0.26$ & $1.117\pm0.090$ & -- & -- & -- \\
 & GL + A115$^{\prime}$ & $>0.381$ &  $0.214^{+0.428}_{-0.855}$ & -- & -- & -- & -- & -- & $0.414^{+0.027}_{-0.034}$ & $50.05\pm0.26$ & $1.103\pm0.090$ & $0.423^{+0.056}_{-0.090}$ & $0.432^{+0.111}_{-0.132}$ & $-1.701\pm0.212$ \\
\\
 & A118 & $0.599^{+0.350}_{-0.175}$ & -- & $-2.440^{+1.779}_{-1.715}$ & -- & -- & -- & -- & $0.412^{+0.028}_{-0.034}$ & $50.15^{+0.26}_{-0.30}$ & $1.106\pm0.089$ & -- & -- & -- \\
 & ML & $>0.123$ & -- & $-2.456^{+2.567}_{-2.180}$ & -- & $0.306^{+0.036}_{-0.054}$ & $1.611^{+0.113}_{-0.277}$ & $-1.014\pm0.092$ & -- & -- & -- & -- & -- & -- \\
 & A115 & $0.589^{+0.357}_{-0.184}$ & -- & $-2.411^{+1.797}_{-1.729}$ & -- & -- & -- & -- & $0.417^{+0.028}_{-0.035}$ & $50.14^{+0.27}_{-0.31}$ & $1.109\pm0.092$ & -- & -- & -- \\
Flat XCDM & ML + A115 & $>0.191$ & -- & $<-0.041$ & -- & $0.300^{+0.033}_{-0.051}$ & $1.562^{+0.103}_{-0.220}$ & $-1.012\pm0.088$ & $0.417^{+0.028}_{-0.034}$ & $50.13^{+0.27}_{-0.30}$ & $1.108\pm0.090$ & -- & -- & -- \\
 & GL & $>0.141$ & -- & $<-0.046$ & -- & -- & -- & -- & -- & -- & -- & $0.428^{+0.058}_{-0.092}$ & $0.556^{+0.127}_{-0.256}$ & $-1.706\pm0.215$ \\
 & A115$^{\prime}$ & $0.605^{+0.394}_{-0.126}$ & -- & $-2.391^{+1.826}_{-1.758}$ & -- & -- & -- & -- & $0.414^{+0.028}_{-0.034}$ & $50.15^{+0.27}_{-0.30}$ & $1.103\pm0.092$ & -- & -- & -- \\
 & GL + A115$^{\prime}$ & $>0.205$ & -- & $<-0.017$ & -- & -- & -- & -- & $0.413^{+0.027}_{-0.034}$ & $50.14^{+0.26}_{-0.30}$ & $1.101\pm0.088$ & $0.421^{+0.055}_{-0.090}$ & $0.512^{+0.113}_{-0.210}$ & $-1.698\pm0.208$ \\
\\
 & A118 & $>0.246$ & $0.590^{+0.476}_{-0.796}$ & $-2.358^{+2.032}_{-1.154}$ & -- & -- & -- & -- & $0.412^{+0.027}_{-0.033}$ & $50.01\pm0.28$ & $1.121\pm0.091$ & -- & -- & -- \\
 & ML & $>0.174$ & $-0.262^{+0.580}_{-0.724}$ & $-2.000^{+2.117}_{-1.264}$ & -- & $0.305^{+0.036}_{-0.054}$ & $1.462^{+0.194}_{-0.196}$ & $-0.996\pm0.097$ & -- & -- & -- & -- & -- & -- \\
 & A115 & $>0.240$ & $0.563^{+0.498}_{-0.796}$ & $-2.290^{+2.146}_{-1.032}$ & -- & -- & -- & -- & $0.418^{+0.028}_{-0.034}$ & $50.01\pm0.28$ & $1.122\pm0.093$ & -- & -- & -- \\
Non-flat XCDM & ML + A115 & $>0.231$ & $0.202^{+0.376}_{-0.635}$ & $-2.155^{+2.224}_{-1.156}$ & -- & $0.302^{+0.033}_{-0.051}$ & $1.489^{+0.122}_{-0.167}$ & $-1.016\pm0.089$ & $0.417^{+0.027}_{-0.034}$ & $50.05\pm0.28$ & $1.109\pm0.090$ & -- & -- & -- \\
 & GL & $>0.194$ & $-0.615^{+0.470}_{-0.685}$ & $-2.212^{+2.186}_{-0.962}$ & -- & -- & -- & -- & -- & -- & -- & $0.403^{+0.058}_{-0.092}$ & $0.480^{+0.177}_{-0.223}$ & $-1.532^{+0.259}_{-0.260}$ \\
 & A115$^{\prime}$ & $>0.236$ & $0.451^{+0.469}_{-0.789}$ & $-2.210^{+2.208}_{-0.946}$ & -- & -- & -- & -- & $0.415^{+0.028}_{-0.034}$ & $50.03\pm0.28$ & $1.114\pm0.091$ & -- & -- & -- \\
 & GL + A115$^{\prime}$ & $>0.226$ & $0.014^{+0.408}_{-0.604}$ & $-2.080^{+2.201}_{-1.138}$ & -- & -- & -- & -- & $0.415^{+0.027}_{-0.033}$ & $50.09^{+0.27}_{-0.30}$ & $1.096\pm0.090$ & $0.416^{+0.055}_{-0.088}$ & $0.446^{+0.142}_{-0.183}$ & $-1.682\pm0.208$ \\
\\
 & A118 & $0.568^{+0.332}_{-0.230}$ & -- & -- & -- & -- & -- & -- & $0.411^{+0.027}_{-0.033}$ & $50.05\pm0.25$ & $1.110\pm0.089$ & -- & -- & -- \\
 & ML & $>0.148$ & -- & -- & -- & $0.304^{+0.035}_{-0.053}$ & $1.493^{+0.093}_{-0.143}$ & $-1.017\pm0.089$ & -- & -- & -- & -- & -- & -- \\
 & A115 & $0.565^{+0.339}_{-0.228}$ & -- & -- & -- & -- & -- & -- & $0.416^{+0.028}_{-0.034}$ & $50.04\pm0.25$ & $1.113\pm0.091$ & -- & -- & -- \\
Flat $\phi$CDM & ML + A115 & $>0.198$ & -- & -- & -- & $0.302^{+0.033}_{-0.051}$ & $1.477^{+0.091}_{-0.126}$ & $-1.016\pm0.088$ & $0.416^{+0.027}_{-0.034}$ & $50.03\pm0.25$ & $1.110\pm0.089$ & -- & -- & -- \\
 & GL & $>0.148$ & -- & -- & -- & -- & -- & -- & -- & -- & -- & $0.428^{+0.059}_{-0.094}$ & $0.444^{+0.112}_{-0.141}$ & $-1.710\pm0.218$ \\
 & A115$^{\prime}$ & $0.586^{+0.391}_{-0.156}$ & -- & -- & -- & -- & -- & -- & $0.414^{+0.028}_{-0.034}$ & $50.06\pm0.25$ & $1.106\pm0.089$ & -- & -- & -- \\
 & GL + A115$^{\prime}$ & $>0.231$ & -- & -- & -- & -- & -- & -- & $0.413^{+0.027}_{-0.033}$ & $50.05\pm0.24$ & $1.102\pm0.088$ & $0.423^{+0.055}_{-0.090}$ & $0.426^{+0.105}_{-0.119}$ & $-1.703\pm0.210$ \\
\\
 & A118 & $0.560^{+0.256}_{-0.247}$ & $-0.002^{+0.294}_{-0.286}$ & -- & $5.203^{+3.808}_{-2.497}$ & -- & -- & -- & $0.412^{+0.027}_{-0.033}$ & $50.05\pm0.25$ & $1.111^{+0.089}_{-0.090}$ & -- & -- & -- \\
 & ML & $>0.207$ & $-0.163^{+0.355}_{-0.317}$ & -- & -- & $0.303^{+0.035}_{-0.053}$ & $1.448^{+0.120}_{-0.165}$ & $-1.011\pm0.091$ & -- & -- & -- & -- & -- & -- \\
 & A115 & $0.546^{+0.260}_{-0.253}$ & $0.011^{+0.299}_{-0.291}$ & -- & -- & -- & -- & -- & $0.417^{+0.028}_{-0.034}$ & $50.04\pm0.26$ & $1.115\pm0.092$ & -- & -- & -- \\
Non-flat $\phi$CDM & ML + A115 & $0.630^{+0.286}_{-0.181}$ & $-0.108^{+0.296}_{-0.263}$ & -- & -- & $0.301^{+0.033}_{-0.051}$ & $1.453^{+0.115}_{-0.139}$ & $-1.013\pm0.089$ & $0.416^{+0.027}_{-0.034}$ & $50.03\pm0.25$ & $1.106\pm0.090$ & -- & -- & -- \\
 & GL & $>0.207$ & $-0.193^{+0.364}_{-0.347}$ & -- & -- & -- & -- & -- & -- & -- & -- & $0.422^{+0.057}_{-0.092}$ & $0.408^{+0.121}_{-0.149}$ & $-1.693^{+0.215}_{-0.214}$ \\
 & A115$^{\prime}$ & $0.581^{+0.254}_{-0.247}$ & $-0.033^{+0.296}_{-0.290}$ & -- & $5.215^{+3.853}_{-2.429}$ & -- & -- & -- & $0.414^{+0.028}_{-0.034}$ & $50.05\pm0.25$ & $1.106\pm0.091$ & -- & -- & -- \\
 & GL + A115$^{\prime}$ & $0.666^{+0.327}_{-0.106}$ & $-0.169^{+0.317}_{-0.270}$ & -- & -- & -- & -- & -- &  $0.413^{+0.027}_{-0.034}$ & $50.04\pm0.25$ & $1.096\pm0.089$ & $0.419^{+0.055}_{-0.089}$ & $0.401^{+0.115}_{-0.128}$ & $-1.686\pm0.209$ \\
\bottomrule
\end{tabular}
\begin{tablenotes}[flushleft]
\item [a] In these GRB cases, $\Omega_b$ and $H_0$ are set to be 0.05 and 70 \hunit, respectively.
\end{tablenotes}
\end{threeparttable}%
}
\end{sidewaystable*}

\begin{figure*}
\centering
 \subfloat[Flat \lcdm]{%
    \includegraphics[width=3.45in,height=2.5in]{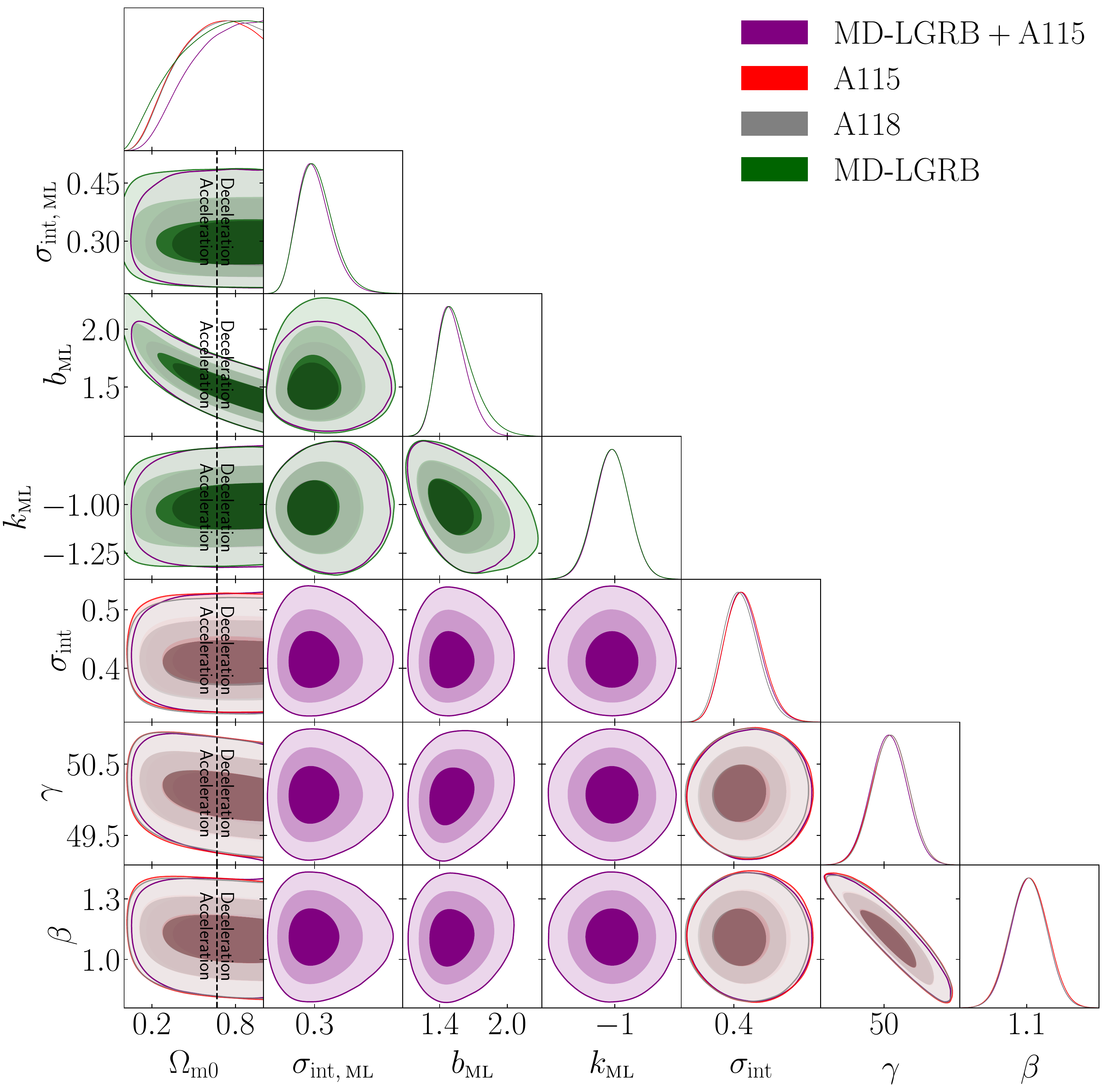}}
 \subfloat[Non-flat \lcdm]{%
    \includegraphics[width=3.45in,height=2.5in]{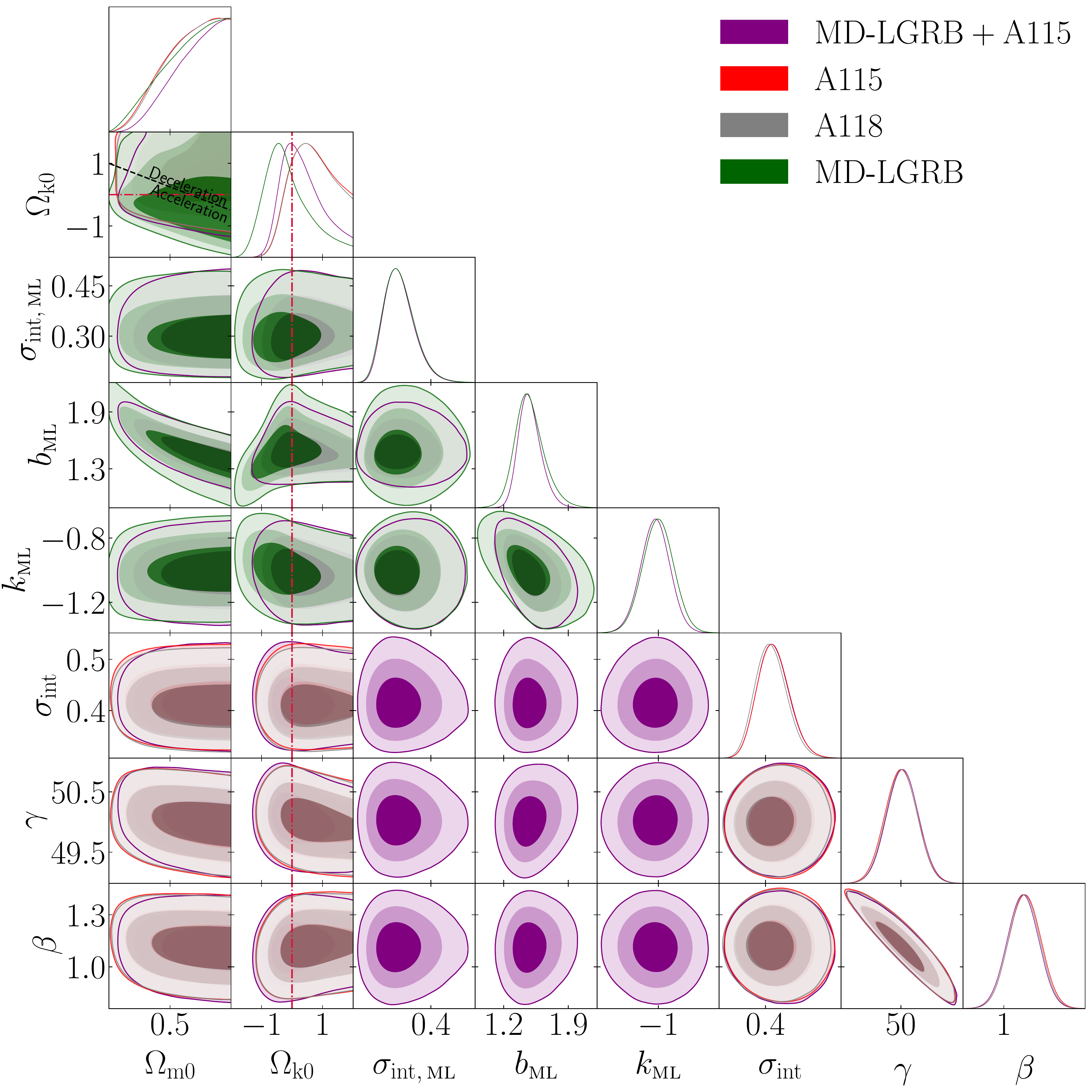}}\\
 \subfloat[Flat XCDM]{%
    \includegraphics[width=3.45in,height=2.5in]{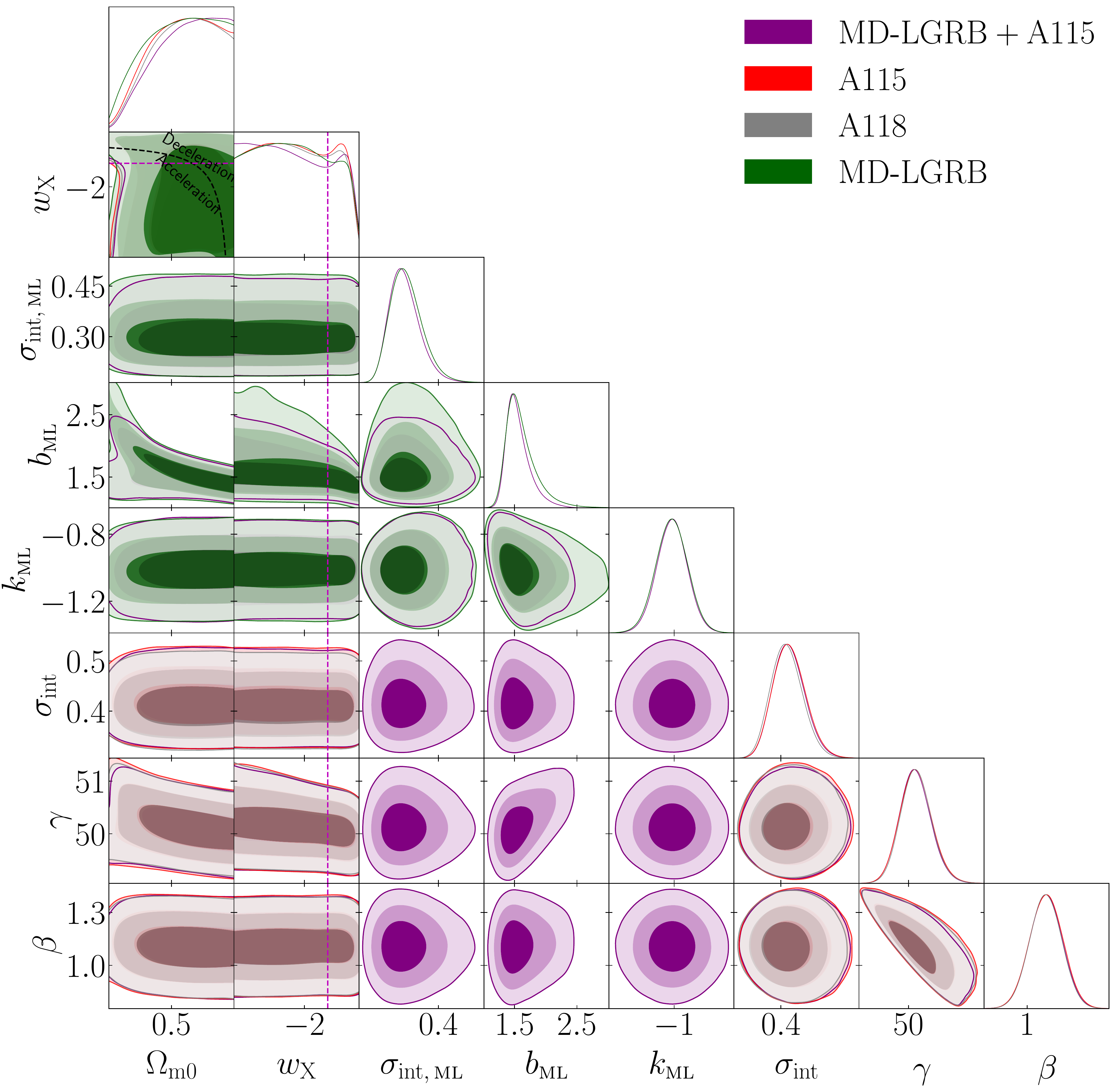}}
 \subfloat[Non-flat XCDM]{%
    \includegraphics[width=3.45in,height=2.5in]{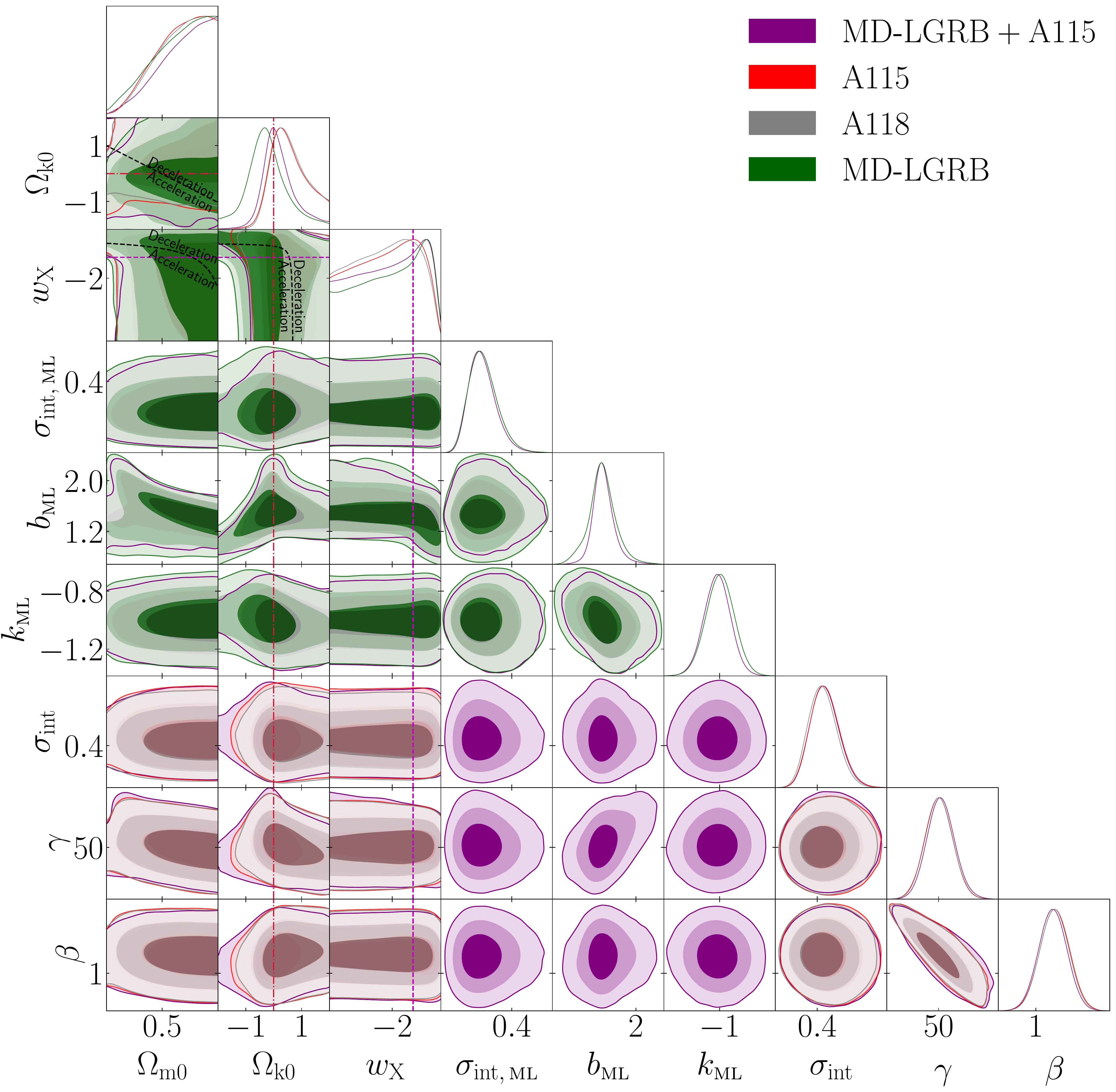}}\\
 \subfloat[Flat \pcdm]{%
    \includegraphics[width=3.45in,height=2.5in]{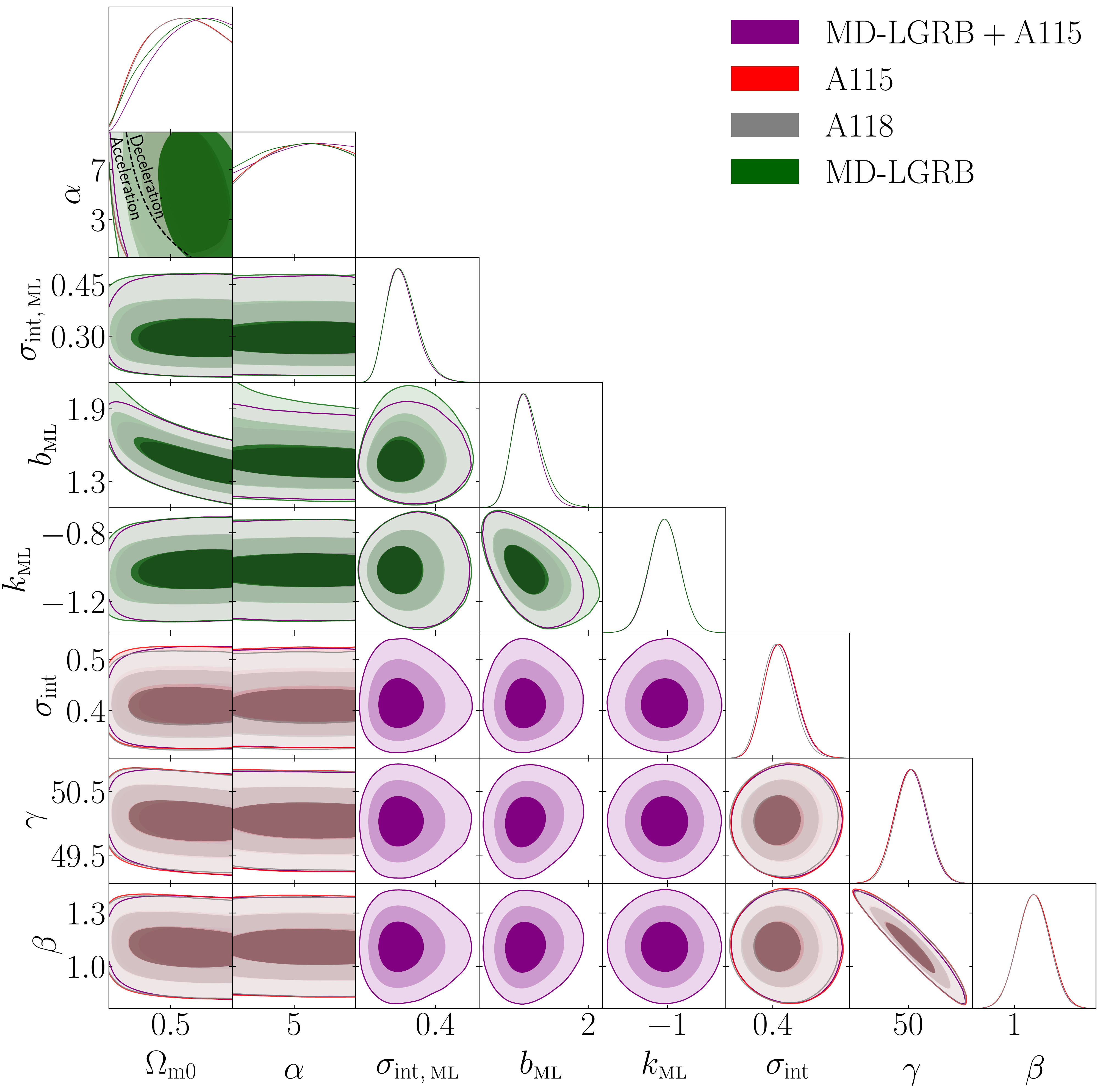}}
  \subfloat[Non-flat \pcdm]{%
     \includegraphics[width=3.45in,height=2.5in]{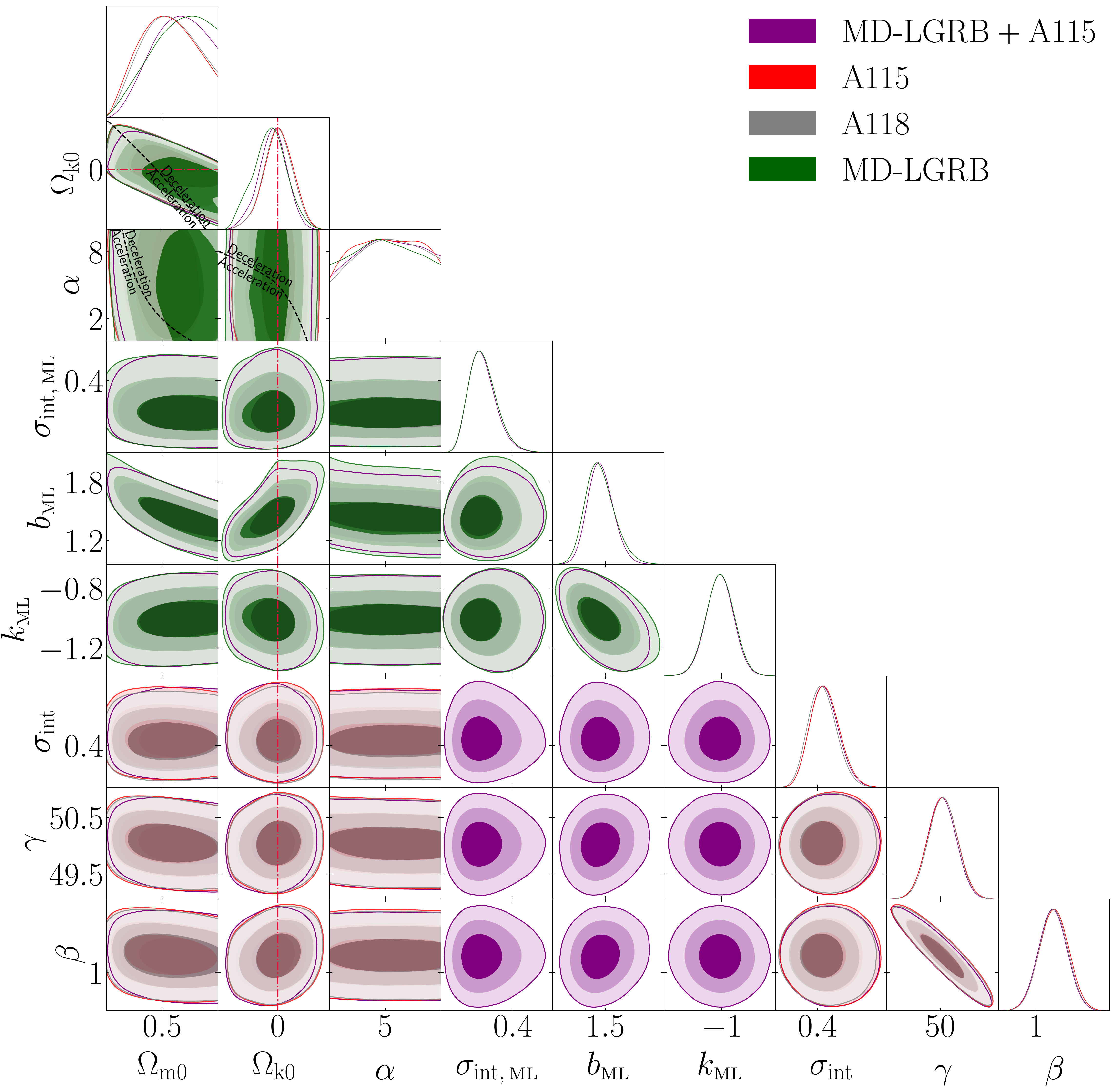}}\\
\caption{One-dimensional likelihoods and 1$\sigma$, 2$\sigma$, and 3$\sigma$ two-dimensional likelihood confidence contours from MD-LGRB (green), A118 (gray), A115 (red), and MD-LGRB + A115 (purple) data for all six models. The zero-acceleration lines are shown as black dashed lines, which divide the parameter space into regions associated with currently-accelerating and currently-decelerating cosmological expansion. In the non-flat XCDM and non-flat \pcdm\ cases, the zero-acceleration lines are computed for the third cosmological parameter set to the $H(z)$ + BAO data best-fitting values listed in Table \ref{tab:BFP2}. The crimson dash-dot lines represent flat hypersurfaces, with closed spatial hypersurfaces either below or to the left. The magenta lines represent $w_{\rm X}=-1$, i.e.\ flat or non-flat \lcdm\ models. The $\alpha = 0$ axes correspond to flat and non-flat \lcdm\ models in panels (e) and (f), respectively.}
\label{fig5}
\end{figure*}

\begin{figure*}
\centering
 \subfloat[Flat \lcdm]{%
    \includegraphics[width=3.45in,height=2.5in]{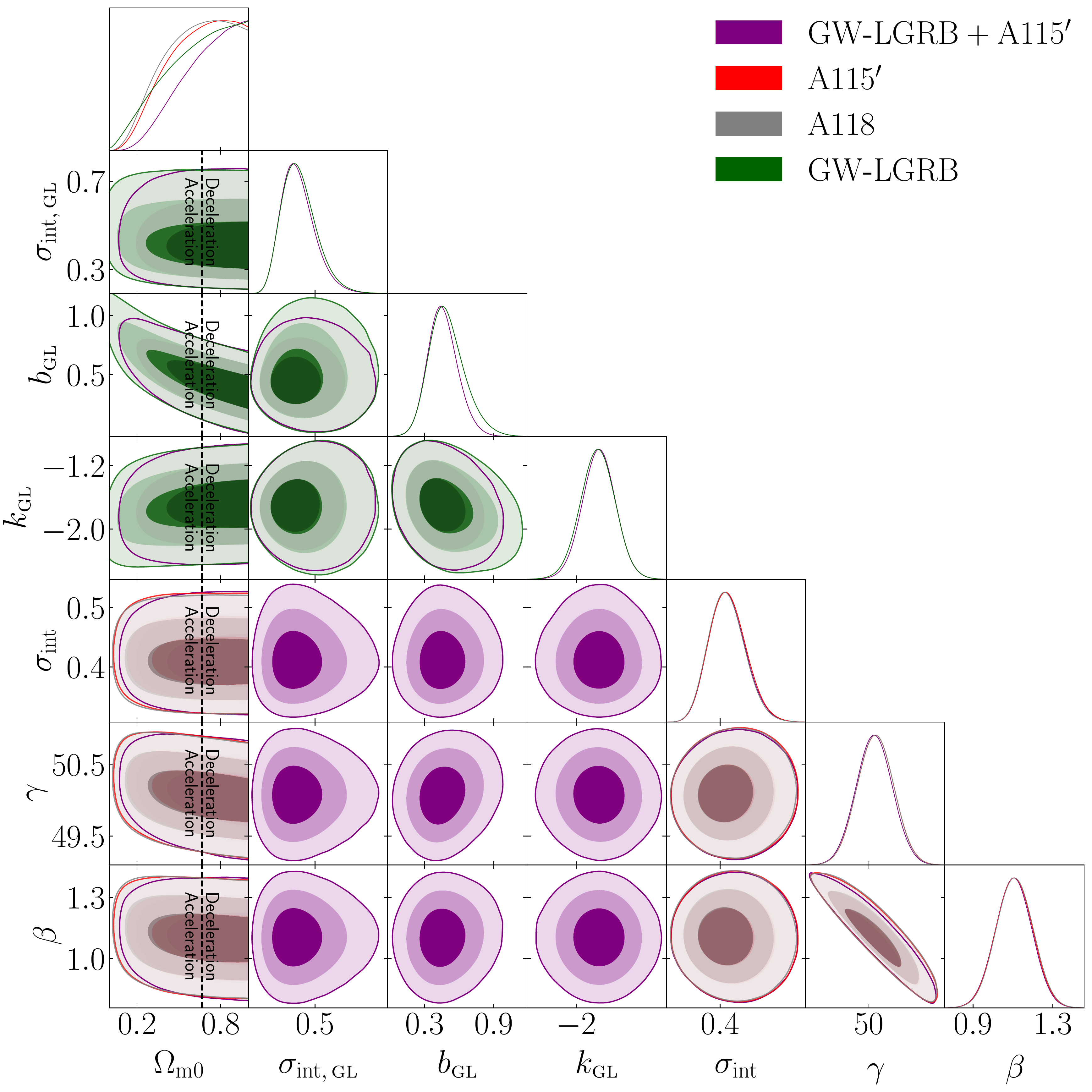}}
 \subfloat[Non-flat \lcdm]{%
    \includegraphics[width=3.45in,height=2.5in]{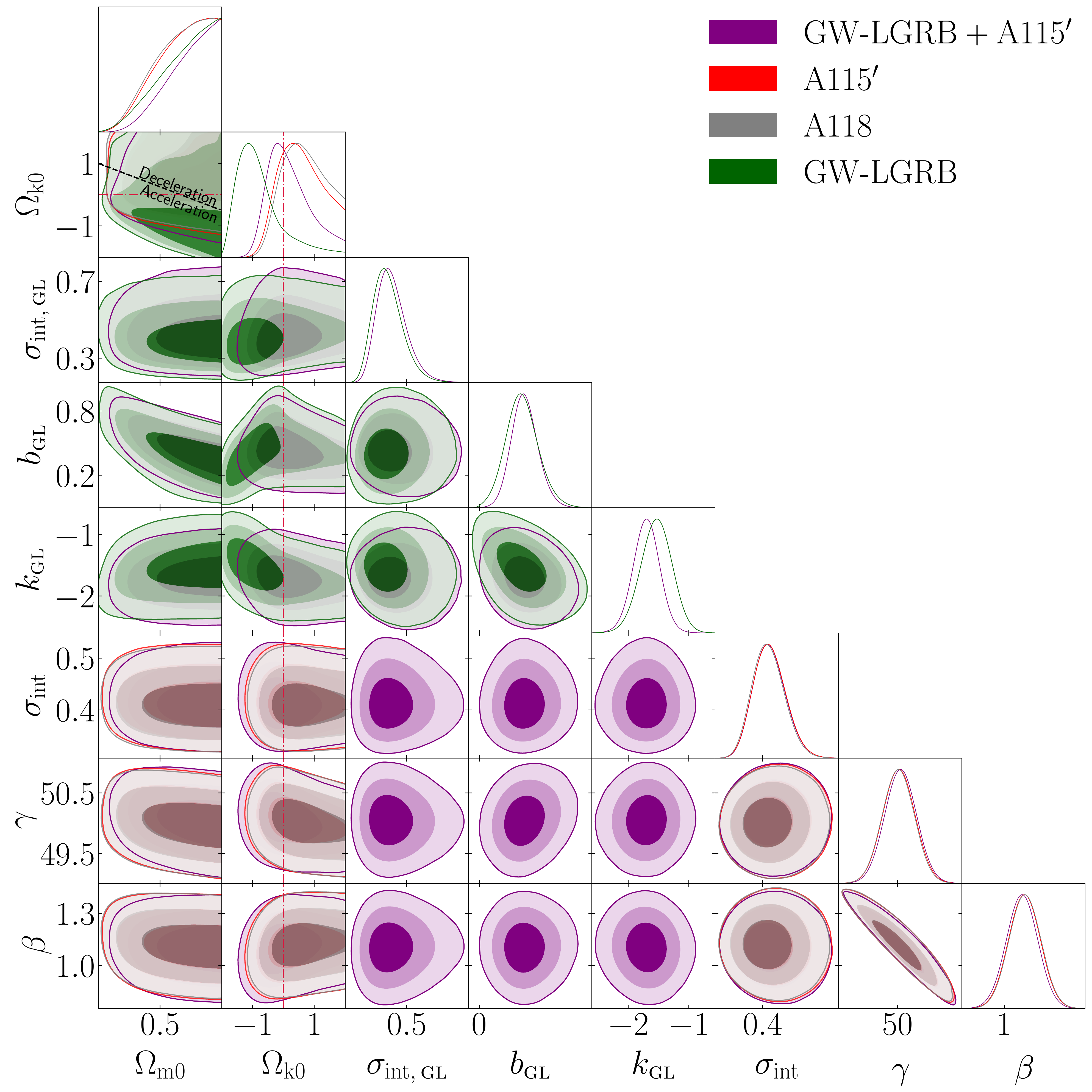}}\\
 \subfloat[Flat XCDM]{%
    \includegraphics[width=3.45in,height=2.5in]{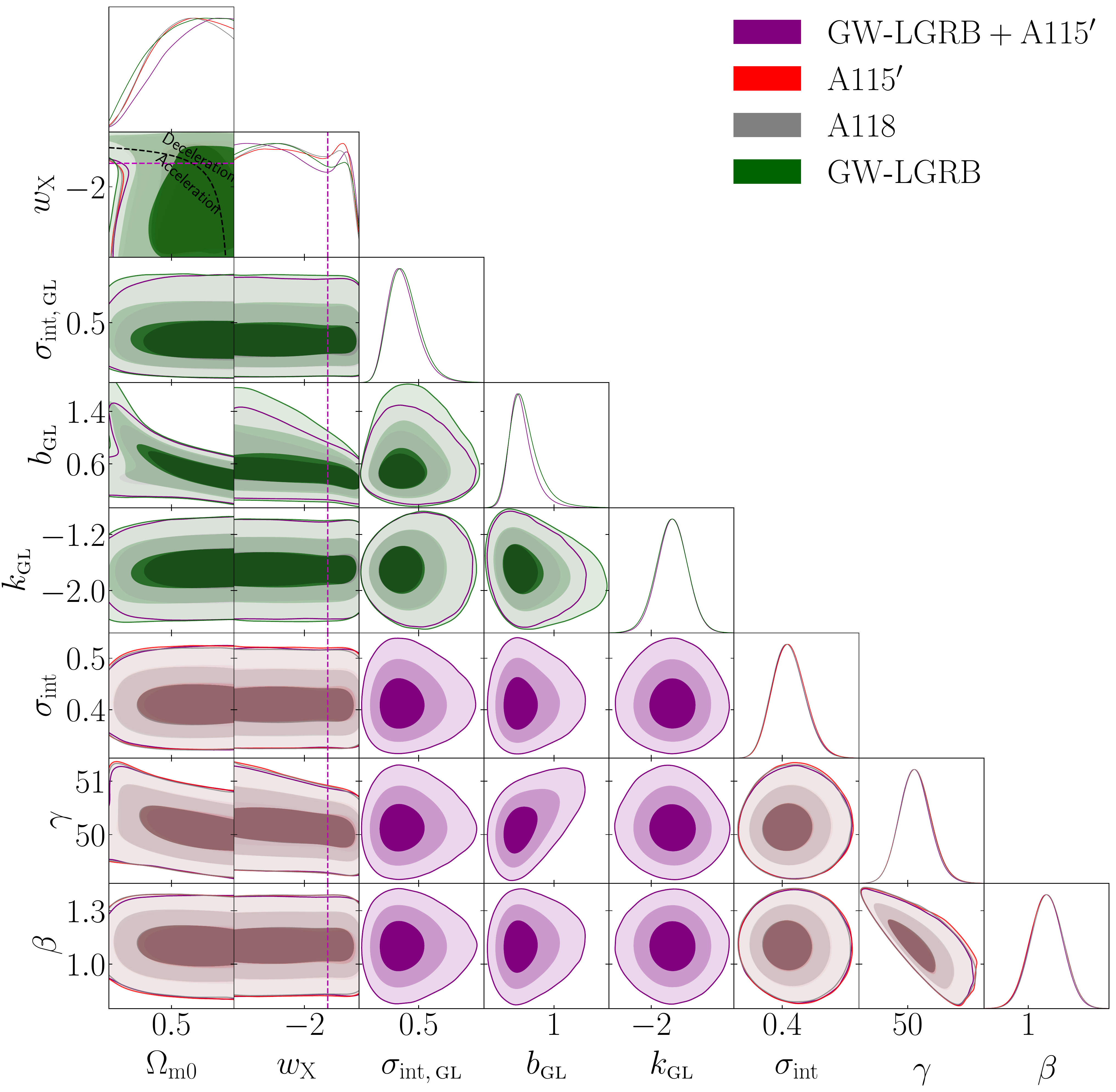}}
 \subfloat[Non-flat XCDM]{%
    \includegraphics[width=3.45in,height=2.5in]{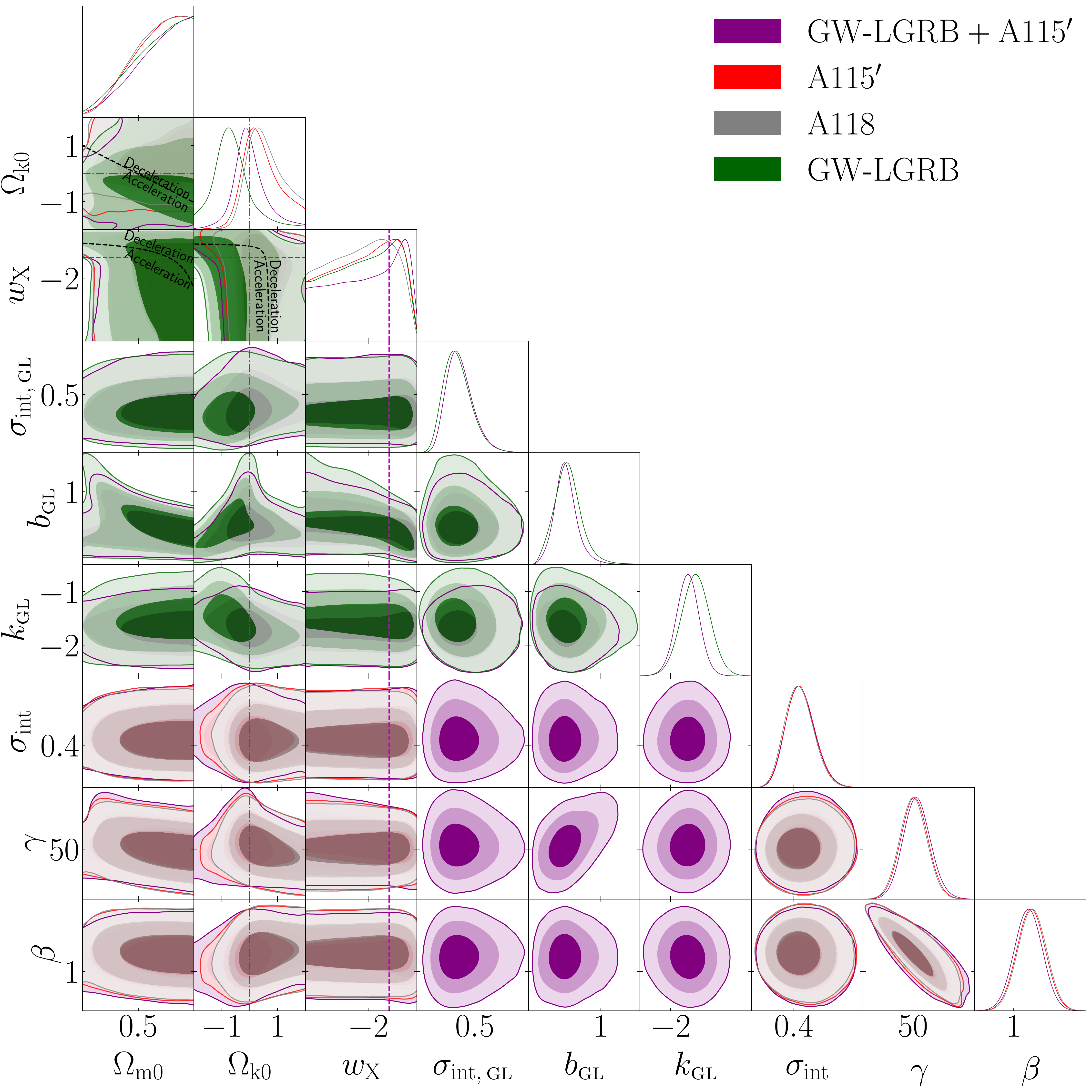}}\\
 \subfloat[Flat \pcdm]{%
    \includegraphics[width=3.45in,height=2.5in]{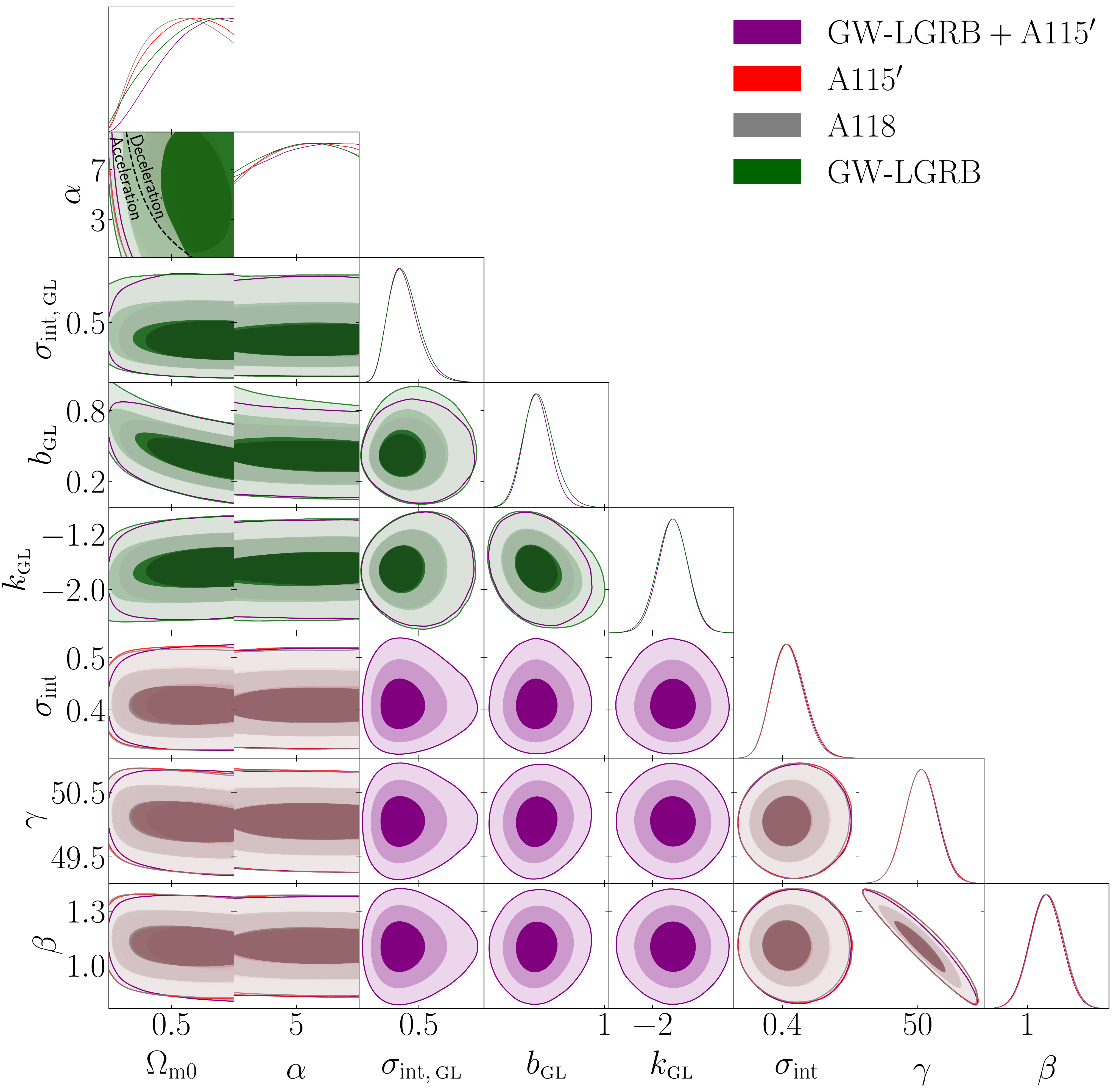}}
  \subfloat[Non-flat \pcdm]{%
     \includegraphics[width=3.45in,height=2.5in]{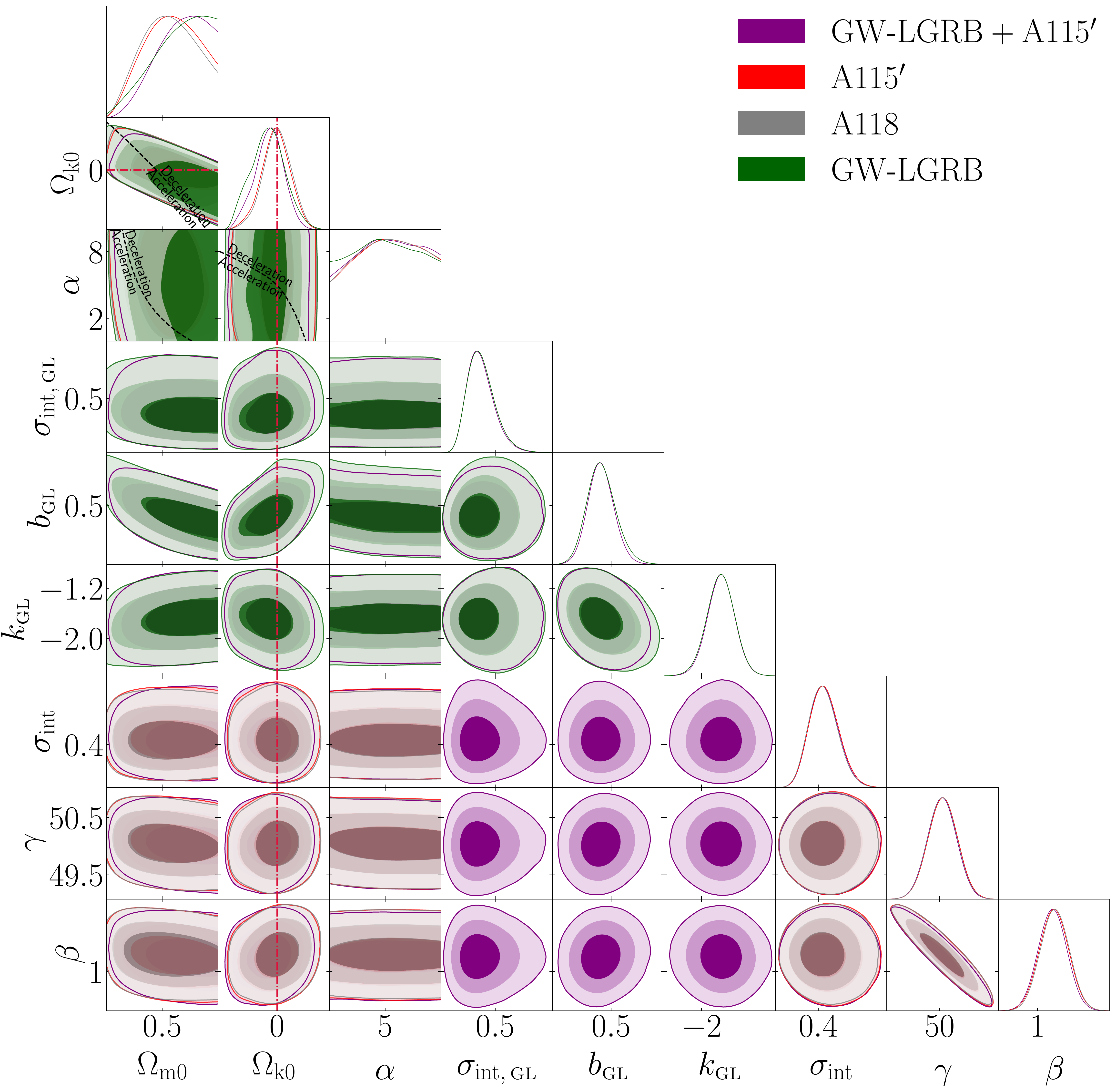}}\\
\caption{One-dimensional likelihoods and 1$\sigma$, 2$\sigma$, and 3$\sigma$ two-dimensional likelihood confidence contours from GW-LGRB (green), A118 (gray), A11$5'$ (red), and GW-LGRB + A11$5'$ (purple) data for all six models. The zero-acceleration lines are shown as black dashed lines, which divide the parameter space into regions associated with currently-accelerating and currently-decelerating cosmological expansion. In the non-flat XCDM and non-flat \pcdm\ cases, the zero-acceleration lines are computed for the third cosmological parameter set to the $H(z)$ + BAO data best-fitting values listed in Table \ref{tab:BFP2}. The crimson dash-dot lines represent flat hypersurfaces, with closed spatial hypersurfaces either below or to the left. The magenta lines represent $w_{\rm X}=-1$, i.e.\ flat or non-flat \lcdm\ models. The $\alpha = 0$ axes correspond to flat and non-flat \lcdm\ models in panels (e) and (f), respectively.}
\label{fig6}
\end{figure*}

The A118 data set was previously studied by \cite{Khadkaetal_2021b}. Here we analyze it along with the truncated A115 and A115$^{\prime}$ data sets, which are also used in joint analyses with the ML and GL data sets. The constraints from these data sets on the GRB correlation parameters and on the cosmological model parameters are presented in Tables \ref{tab:BFP2} and \ref{tab:1d_BFP2}. The corresponding posterior 1D probability distributions and 2D confidence regions of these parameters are shown in Figs. \ref{fig5} and \ref{fig6}, in gray (A118), red (A115 and A115$^{\prime}$), green (ML and GL), and purple (ML + A115 and GL + A115$^{\prime}$). Note that these analyses assume $H_0=70$ \hunit\ and $\Omega_{b}=0.05$.

The constraints from A115 data and from ML data, and from A115$^{\prime}$ data and from GL data, are not mutually inconsistent, so it is not unreasonable to examine joint ML + A115 and GL + A115$^{\prime}$ constraints. The ML data have the smallest intrinsic dispersion, $\sim 0.30-0.31$, with A115, A115$^{\prime}$, and GL having larger intrinsic dispersion, $\sim 0.40-0.43$ 

The constraints on the Amati parameters are quite cosmological-model-independent for these GRB data sets. In the A118 case, the slope $\beta$ ranges from a high of $1.121\pm0.091$ (non-flat XCDM) to a low of $1.106\pm0.089$ (flat XCDM), the intercept $\gamma$ ranges from a high of $50.15^{+0.26}_{-0.30}$ (flat XCDM) to a low of $50.01\pm0.26$ (non-flat \lcdm), and the intrinsic scatter $\sigma_{\rm int}$ ranges from a high of $0.412^{+0.028}_{-0.034}$ (flat XCDM) to a low of $0.411^{+0.027}_{-0.033}$ (flat \pcdm), with central values of each pair being $0.12\sigma$, $0.35\sigma$, and $0.02\sigma$ away from each other, respectively. 

In the A115 case, the slope $\beta$ ranges from a high of $1.124\pm0.092$ (non-flat \lcdm) to a low of $1.109\pm0.092$ (flat XCDM), the intercept $\gamma$ ranges from a high of $50.14^{+0.27}_{-0.31}$ (flat XCDM) to a low of $50.00\pm0.27$ (non-flat \lcdm), and the intrinsic scatter $\sigma_{\rm int}$ ranges from a high of $0.418^{+0.028}_{-0.034}$ (non-flat XCDM) to a low of $0.416^{+0.028}_{-0.034}$ (flat \pcdm), with central values of each pair being $0.12\sigma$, $0.34\sigma$, and $0.05\sigma$ away from each other, respectively.

In the A115$^{\prime}$ case, the slope $\beta$ ranges from a high of $1.117\pm0.090$ (non-flat \lcdm) to a low of $1.103\pm0.092$ (flat XCDM), the intercept $\gamma$ ranges from a high of $50.15^{+0.27}_{-0.30}$ (flat XCDM) to a low of $50.02\pm0.26$ (non-flat \lcdm), and the intrinsic scatter $\sigma_{\rm int}$ ranges from a high of $0.415^{+0.028}_{-0.034}$ (non-flat XCDM) to a low of $0.414^{+0.028}_{-0.034}$ (the others), with central values of each pair being $0.11\sigma$, $0.33\sigma$, and $0.02\sigma$ away from each other, respectively.

The lowest and highest values of $\beta$, $\gamma$, and $\sigma_{\rm int}$ from the A118, A115, and A115$^{\prime}$ cases differ from each other at $0.16\sigma$, $0.37\sigma$, and $0.16\sigma$, respectively. This implies that excluding three GRBs from A118 does not significantly affect the constraints on the Amati parameters. 

In the joint analysis of ML and A115 (ML + A115) data, $\beta$ ranges from a high of $1.113\pm0.091$ (non-flat \lcdm) to a low of $1.106\pm0.090$ (non-flat \pcdm), $\gamma$ ranges from a high of $50.13^{+0.27}_{-0.30}$ (flat XCDM) to a low of $50.03\pm0.25$ (flat and non-flat \pcdm), and $\sigma_{\rm int}$ ranges from a high of $0.417^{+0.028}_{-0.034}$ (flat XCDM) to a low of $0.416^{+0.027}_{-0.034}$ (flat \lcdm, and flat and non-flat \pcdm), with central values of each pair being $0.05\sigma$, $0.26\sigma$, and $0.02\sigma$ away from each other, respectively; also $k$ ranges from a high of $-1.012\pm0.088$ (flat XCDM) to a low of $-1.019\pm0.091$ (non-flat \lcdm), $b$ ranges from a high of $1.562^{+0.103}_{-0.220}$ (flat XCDM) to a low of $1.453^{+0.115}_{-0.139}$ (non-flat \pcdm), and $\sigma_{\mathrm{int,\,\textsc{ml}}}$ ranges from a high of $0.304^{+0.034}_{-0.052}$ (non-flat \lcdm) to a low of $0.300^{+0.033}_{-0.051}$ (flat XCDM), with central values of each pair being $0.06\sigma$, $0.44\sigma$, and $0.06\sigma$ away from each other, respectively. The lowest and highest values of $\beta$, $\gamma$, and $\sigma_{\rm int}$ from the A115 and ML + A115 cases differ from each other at $0.14\sigma$, $0.32\sigma$, and $0.05\sigma$, respectively; also those of $k$, $b$, and $\sigma_{\mathrm{int,\,\textsc{ml}}}$ differ from each other at $0.17\sigma$, $0.53\sigma$, and $0.09\sigma$, respectively.

In the joint analysis of GL and A115$^{\prime}$ (GL + A115$^{\prime}$) data, $\beta$ ranges from a high of $1.104\pm0.089$ (flat \lcdm) to a low of $1.096\pm0.089$ (non-flat \pcdm), $\gamma$ ranges from a high of $50.14^{+0.26}_{-0.30}$ (flat XCDM) to a low of $50.04\pm0.25$ (non-flat \pcdm), and $\sigma_{\rm int}$ ranges from a high of $0.415^{+0.027}_{-0.033}$ (non-flat XCDM) to a low of $0.413^{+0.027}_{-0.034}$ (flat \lcdm, flat XCDM, and non-flat \pcdm), with central values of each pair being $0.06\sigma$, $0.26\sigma$, and $0.05\sigma$ away from each other, respectively; also $k$ ranges from a high of $-1.682\pm0.208$ (non-flat XCDM) to a low of $-1.705\pm0.210$ (flat \lcdm), $b$ ranges from a high of $0.512^{+0.113}_{-0.210}$ (flat XCDM) to a low of $0.401^{+0.115}_{-0.128}$ (non-flat \pcdm), and $\sigma_{\mathrm{int,\,\textsc{gl}}}$ ranges from a high of $0.423^{+0.056}_{-0.090}$ (flat and non-flat \lcdm) to a low of $0.416^{+0.055}_{-0.088}$ (non-flat XCDM), with central values of each pair being $0.08\sigma$, $0.46\sigma$, and $0.07\sigma$ away from each other, respectively. The lowest and highest values of $\beta$, $\gamma$, and $\sigma_{\rm int}$ from the A115$^{\prime}$ and GL + A115$^{\prime}$ cases differ from each other at $0.17\sigma$, $0.30\sigma$, and $0.05\sigma$, respectively; also those of $k$, $b$, and $\sigma_{\mathrm{int,\,\textsc{gl}}}$ differ from each other at $0.52\sigma$, $0.47\sigma$, and $0.20\sigma$, respectively.

Judging from the constraints on the parameters of the Amati and Dainotti correlations, those from the joint analyses do not deviate much from the individual cases. We next focus on constraints on cosmological model parameters.

Similar to GRB data from Sec.\ \ref{subsec:MLSGL}, these data more favor currently accelerating cosmological expansion in the \lcdm\ and XCDM cases, but also more favor currently decelerating cosmological expansion in the \pcdm\ models, in the $\Om-\alpha$ and $\Om-\Ok$ parameter subspaces.

In the flat \lcdm\ model, the A118, A115, and A115$^{\prime}$ $2\sigma$ constraints on \om\ are mutually consistent (A115 with $\Om>0.241$), but they favor higher values of \om\ than do ML and GL while the joint ML + A115 and GL + A115$^{\prime}$ cases favor even higher values of $>0.298$ and $>0.339$, respectively. In fact, this is also true for the ML + GL case ($\Om>0.294$). 

In the non-flat \lcdm\ model, the A118, A115, and A115$^{\prime}$ $2\sigma$ constraints on \om\ are also mutually consistent, but they favor slightly higher values of \om\ than in the flat \lcdm\ case, with $\Om>0.299$ in the A115$^{\prime}$ non-flat \lcdm\ case. The joint ML + A115, GL + A115$^{\prime}$, and ML + GL cases favor higher \om\ $2\sigma$ limits of $>0.346$, $>0.381$, and $>0.338$, respectively. The A118, A115, ML + A115, A115$^{\prime}$, and GL + A115$^{\prime}$ data mildly favor open hypersurfaces in the non-flat \lcdm\ model, being less than $1\sigma$ away from flatness.

In the flat and non-flat XCDM parametrizations, the $2\sigma$ constraints on \om\ are mutually consistent in all cases, where in the flat XCDM parametrization, the $2\sigma$ limits are $\Om>0.181$ (A118), $>0.170$ (A115), and $>0.185$ (A115$^{\prime}$). The constraints on \wx\ are very loose, and thus affected by the \wx\ prior, and consistent with each other in all cases, and mildly favor phantom dark energy (but $\Lambda$ is less than $1\sigma$ away). In the non-flat XCDM parametrization, the A118, A115, ML + A115, A115$^{\prime}$, and GL + A115$^{\prime}$ data also mildly favor open hypersurfaces, with flatness being less than $1\sigma$ away.

In the flat \pcdm\ model, the A118, A115, and A115$^{\prime}$ constraints on \om\ are mutually consistent, with $2\sigma$ limits of $\Om>0.149$ (A118), $>0.145$ (A115), and $>0.159$ (A115$^{\prime}$), which are consistent with the other cases. These GRB data do not provide constraints on $\alpha$ in the flat \pcdm\ model, while in the non-flat \pcdm\ model, A118 and A115$^{\prime}$ provide constraints of $\alpha=5.203^{+3.808}_{-2.497}$ and $\alpha=5.215^{+3.853}_{-2.429}$, respectively. Note that $\alpha=0$ is still within $2\sigma$ for both cases. Similar trends hold for non-flat \pcdm\ \om\ constraints, but ML + A115 and GL + A115$^{\prime}$ data constraints posterior mean values are larger than for the individual data sets. The $2\sigma$ limits are $\Om>0.183$ (A118), $\Om=0.546^{+0.449}_{-0.384}$ (A115), $\Om>0.198$ (A115$^{\prime}$), $>0.251$ (ML + A115), and $>0.286$ (GL + A115$^{\prime}$). Except for the A115 data, non-flat \pcdm\ constraints favor closed hypersurfaces (unlike non-flat \lcdm\ and non-flat XCDM), but with flatness well within $1\sigma$ for all cases.

The $\Delta AIC$ and $\Delta BIC$ values with respect to the flat \lcdm\ model are listed in the last two columns of Table \ref{tab:BFP2}. In all cases (except for the GL case, which is discussed in Sec.\ \ref{subsec:MLSGL} above), the flat \lcdm\ model is the most favored model but the evidence against the other models are either weak or positive, except that, based on $BIC$, the evidence against non-flat XCDM and non-flat \pcdm\ are strong. For ML + A115 data the non-flat \pcdm\ model is very strongly disfavored with $\Delta BIC=10.03$.

In summary, while the joint analyses do slightly tighten the constraints, the improvements relative to those from A118 data alone are not significant.

\section{Conclusion}
\label{sec:conclusion}

We have used six different cosmological models in analyses of the three (ML, MS, and GL) Dainotti ($L_0-t_b$) correlation GRB data sets compiled by \cite{Wangetal_2021} and \cite{Huetal_2021}. We find for each data sets, as well as the MS + GL, ML + GL, and ML + MS combinations, that the GRB correlation parameters are independent of cosmological model. Our results thus indicate that these GRBs are standardizable through the Dainotti correlation and so can be used to constrain cosmological parameters, justifying the assumption made by  \cite{Wangetal_2021} and \cite{Huetal_2021}. These results also mean that the circularity problem does not affect cosmological parameter constraints derived from these GRB data.

In contrast to \cite{Wangetal_2021} and \cite{Huetal_2021} we do not use $H(z)$ data to calibrate these GRB data, instead we use these data to derive GRB only cosmological constraints. We find that ML, MS, GL, MS + GL, ML + GL, and ML + MS GRBs provide only weak restrictions on cosmological parameters. 

We have also used the more-restrictive  ML and GL Dainotti data sets in joint analyses with the largest available reliable compilation of Amati ($E_{\rm p}-E_{\rm iso}$) correlation A118 GRB data \citep{Khadkaetal_2021b}, but excluding three overlapping GRBs from the A118 data in the joint analyses. While the joint analyses do result in slightly tighter constraints, typically with larger lower limits on \om\ than those from the ML, GL, or A118 data alone, the improvements relative to the A118 data constraints are not significant.

Current GRB data provide quite weak constraints on cosmological parameters but do favor currently accelerated cosmological expansion in the \lcdm\ models and the XCDM parametrizations. We hope that in the near future there will be more and better-quality GRB measurements that will result in more restrictive GRB cosmological constraints. GRBs probe a very wide range of cosmological redshift space, a significant part of which is as yet unprobed, so it is worth putting effort into further developing GRB cosmological constraints. 

\section*{Acknowledgements}
We thank F. Y. Wang and J. P. Hu for valuable discussions. This research was supported in part by DOE grant DE-SC0011840. Part of the computation for this project was performed on the Beocat Research Cluster at Kansas State University. 

\section*{Data availability}
MD-LGRB data are available in \cite{Wangetal_2021} and MD-SGRB and GW-LGRB data are available in \cite{Huetal_2021}.




\bibliographystyle{mnras}
\bibliography{mybibfile} 

\begin{thebibliography}{}
\makeatletter
\relax
\def\mn@urlcharsother{\let\do\@makeother \do\$\do\&\do\#\do\^\do\_\do\%\do\~}
\def\mn@doi{\begingroup\mn@urlcharsother \@ifnextchar [ {\mn@doi@}
  {\mn@doi@[]}}
\def\mn@doi@[#1]#2{\def\@tempa{#1}\ifx\@tempa\@empty \href
  {http://dx.doi.org/#2} {doi:#2}\else \href {http://dx.doi.org/#2} {#1}\fi
  \endgroup}
\def\mn@eprint#1#2{\mn@eprint@#1:#2::\@nil}
\def\mn@eprint@arXiv#1{\href {http://arxiv.org/abs/#1} {{\tt arXiv:#1}}}
\def\mn@eprint@dblp#1{\href {http://dblp.uni-trier.de/rec/bibtex/#1.xml}
  {dblp:#1}}
\def\mn@eprint@#1:#2:#3:#4\@nil{\def\@tempa {#1}\def\@tempb {#2}\def\@tempc
  {#3}\ifx \@tempc \@empty \let \@tempc \@tempb \let \@tempb \@tempa \fi \ifx
  \@tempb \@empty \def\@tempb {arXiv}\fi \@ifundefined
  {mn@eprint@\@tempb}{\@tempb:\@tempc}{\expandafter \expandafter \csname
  mn@eprint@\@tempb\endcsname \expandafter{\@tempc}}}

\bibitem[\protect\citeauthoryear{{Amati}, {Guidorzi}, {Frontera}, {Della
  Valle}, {Finelli}, {Landi}  \& {Montanari}}{{Amati} et~al.}{2008}]{Amati2008}
{Amati} L.,  {Guidorzi} C.,  {Frontera} F.,  {Della Valle} M.,  {Finelli} F.,
  {Landi} R.,   {Montanari} E.,  2008, \mn@doi [\mnras]
  {10.1111/j.1365-2966.2008.13943.x}, \href
  {https://ui.adsabs.harvard.edu/abs/2008MNRAS.391..577A} {391, 577}

\bibitem[\protect\citeauthoryear{{Amati}, {Frontera}  \& {Guidorzi}}{{Amati}
  et~al.}{2009}]{Amati2009}
{Amati} L.,  {Frontera} F.,   {Guidorzi} C.,  2009, \mn@doi [\aap]
  {10.1051/0004-6361/200912788}, \href
  {https://ui.adsabs.harvard.edu/abs/2009A&A...508..173A} {508, 173}

\bibitem[\protect\citeauthoryear{{Amati}, {D'Agostino}, {Luongo}, {Muccino}  \&
  {Tantalo}}{{Amati} et~al.}{2019}]{Amati2019}
{Amati} L.,  {D'Agostino} R.,  {Luongo} O.,  {Muccino} M.,   {Tantalo} M.,
  2019, \mn@doi [\mnras] {10.1093/mnrasl/slz056}, \href
  {https://ui.adsabs.harvard.edu/abs/2019MNRAS.486L..46A} {486, L46}

\bibitem[\protect\citeauthoryear{{Arjona} \& {Nesseris}}{{Arjona} \&
  {Nesseris}}{2021}]{ArjonaNesseris2021}
{Arjona} R.,  {Nesseris} S.,  2021, \mn@doi [\prd]
  {10.1103/PhysRevD.103.103539}, \href
  {https://ui.adsabs.harvard.edu/abs/2021PhRvD.103j3539A} {103, 103539}

\bibitem[\protect\citeauthoryear{{Brinckmann} \& {Lesgourgues}}{{Brinckmann} \&
  {Lesgourgues}}{2019}]{Brinckmann2019}
{Brinckmann} T.,  {Lesgourgues} J.,  2019, \mn@doi [Physics of the Dark
  Universe] {10.1016/j.dark.2018.100260}, \href
  {https://ui.adsabs.harvard.edu/abs/2019PDU....24..260B} {24, 100260}

\bibitem[\protect\citeauthoryear{{Cao}, {Biesiada}, {Jackson}, {Zheng}, {Zhao}
  \& {Zhu}}{{Cao} et~al.}{2017}]{Cao_et_al2017a}
{Cao} S.,  {Biesiada} M.,  {Jackson} J.,  {Zheng} X.,  {Zhao} Y.,   {Zhu}
  Z.-H.,  2017, \mn@doi [\jcap] {10.1088/1475-7516/2017/02/012}, \href
  {http://ads.bao.ac.cn/abs/2017JCAP...02..012C} {2, 012}

\bibitem[\protect\citeauthoryear{{Cao}, {Ryan}  \& {Ratra}}{{Cao}
  et~al.}{2020}]{Caoetal_2020}
{Cao} S.,  {Ryan} J.,   {Ratra} B.,  2020, \mn@doi [\mnras]
  {10.1093/mnras/staa2190}, \href
  {https://ui.adsabs.harvard.edu/abs/2020MNRAS.tmp.2278C} {497, 3191}

\bibitem[\protect\citeauthoryear{{Cao}, {Ryan}  \& {Ratra}}{{Cao}
  et~al.}{2021a}]{Caoetal_2021c}
{Cao} S.,  {Ryan} J.,   {Ratra} B.,  2021a, \mn@doi [\mnras]
  {10.1093/mnras/stab3304}, \href
  {https://ui.adsabs.harvard.edu/abs/2021MNRAS.tmp.3017C} {}

\bibitem[\protect\citeauthoryear{{Cao}, {Ryan}, {Khadka}  \& {Ratra}}{{Cao}
  et~al.}{2021b}]{Caoetal_2021a}
{Cao} S.,  {Ryan} J.,  {Khadka} N.,   {Ratra} B.,  2021b, \mn@doi [\mnras]
  {10.1093/mnras/staa3748}, \href
  {https://ui.adsabs.harvard.edu/abs/2020MNRAS.tmp.3537C} {501, 1520}

\bibitem[\protect\citeauthoryear{{Cao}, {Ryan}  \& {Ratra}}{{Cao}
  et~al.}{2021c}]{Caoetal_2021b}
{Cao} S.,  {Ryan} J.,   {Ratra} B.,  2021c, \mn@doi [\mnras]
  {10.1093/mnras/stab942}, \href
  {https://ui.adsabs.harvard.edu/abs/2021MNRAS.504..300C} {504, 300}

\bibitem[\protect\citeauthoryear{{Cardone}, {Dainotti}, {Capozziello}  \&
  {Willingale}}{{Cardone} et~al.}{2010}]{Cardoneetal2010}
{Cardone} V.~F.,  {Dainotti} M.~G.,  {Capozziello} S.,   {Willingale} R.,
  2010, \mn@doi [\mnras] {10.1111/j.1365-2966.2010.17197.x}, \href
  {https://ui.adsabs.harvard.edu/abs/2010MNRAS.408.1181C} {408, 1181}

\bibitem[\protect\citeauthoryear{{Ch{\'a}vez}, {Terlevich}, {Terlevich},
  {Bresolin}, {Melnick}, {Plionis}  \& {Basilakos}}{{Ch{\'a}vez}
  et~al.}{2014}]{Chavez_2014}
{Ch{\'a}vez} R.,  {Terlevich} R.,  {Terlevich} E.,  {Bresolin} F.,  {Melnick}
  J.,  {Plionis} M.,   {Basilakos} S.,  2014, \mn@doi [\mnras]
  {10.1093/mnras/stu987}, \href
  {https://ui.adsabs.harvard.edu/abs/2014MNRAS.442.3565C} {442, 3565}

\bibitem[\protect\citeauthoryear{{Chen}, {Ratra}, {Biesiada}, {Li}  \&
  {Zhu}}{{Chen} et~al.}{2016}]{Chen_et_al_2016}
{Chen} Y.,  {Ratra} B.,  {Biesiada} M.,  {Li} S.,   {Zhu} Z.-H.,  2016, \mn@doi
  [\apj] {10.3847/0004-637X/829/2/61}, \href
  {http://adsabs.harvard.edu/abs/2016ApJ...829...61C} {829, 61}

\bibitem[\protect\citeauthoryear{{Chen}, {Kumar}  \& {Ratra}}{{Chen}
  et~al.}{2017}]{chen_etal_2017}
{Chen} Y.,  {Kumar} S.,   {Ratra} B.,  2017, \mn@doi [\apj]
  {10.3847/1538-4357/835/1/86}, \href
  {http://adsabs.harvard.edu/abs/2017ApJ...835...86C} {835, 86}

\bibitem[\protect\citeauthoryear{{D'Agostini}}{{D'Agostini}}{2005}]{D'Agostini_2005}
{D'Agostini} G.,  2005, preprint, \href
  {https://ui.adsabs.harvard.edu/abs/2005physics..11182D} {} (\mn@eprint
  {arXiv} {physics/0511182})

\bibitem[\protect\citeauthoryear{{DES Collaboration}}{{DES
  Collaboration}}{2019}]{DES_2019}
{DES Collaboration} 2019, \mn@doi [\prd] {10.1103/PhysRevD.99.123505}, \href
  {https://ui.adsabs.harvard.edu/abs/2019PhRvD..99l3505A} {99, 123505}

\bibitem[\protect\citeauthoryear{{Dainotti} \& {Del Vecchio}}{{Dainotti} \&
  {Del Vecchio}}{2017}]{DainottiaDelVecchio2017}
{Dainotti} M.~G.,  {Del Vecchio} R.,  2017, \mn@doi [\nar]
  {10.1016/j.newar.2017.04.001}, \href
  {https://ui.adsabs.harvard.edu/abs/2017NewAR..77...23D} {77, 23}

\bibitem[\protect\citeauthoryear{Dainotti, Cardone  \& Capozziello}{Dainotti
  et~al.}{2008}]{Dianotti_2008}
Dainotti M.~G.,  Cardone V.~F.,   Capozziello S.,  2008, \mn@doi [Monthly
  Notices of the Royal Astronomical Society: Letters]
  {10.1111/j.1745-3933.2008.00560.x}, 391, L79

\bibitem[\protect\citeauthoryear{Dainotti, Willingale, Capozziello, Cardone  \&
  Ostrowski}{Dainotti et~al.}{2010}]{Dainotti_2010}
Dainotti M.~G.,  Willingale R.,  Capozziello S.,  Cardone V.~F.,   Ostrowski
  M.,  2010, \mn@doi [The Astrophysical Journal]
  {10.1088/2041-8205/722/2/l215}, 722, L215

\bibitem[\protect\citeauthoryear{Dainotti, Cardone, Capozziello, Ostrowski  \&
  Willingale}{Dainotti et~al.}{2011}]{Dainotti_2011}
Dainotti M.~G.,  Cardone V.~F.,  Capozziello S.,  Ostrowski M.,   Willingale
  R.,  2011, \mn@doi [The Astrophysical Journal] {10.1088/0004-637x/730/2/135},
  730, 135

\bibitem[\protect\citeauthoryear{{Dainotti}, {Cardone}, {Piedipalumbo}  \&
  {Capozziello}}{{Dainotti} et~al.}{2013a}]{Dainottietal2013a}
{Dainotti} M.~G.,  {Cardone} V.~F.,  {Piedipalumbo} E.,   {Capozziello} S.,
  2013a, \mn@doi [\mnras] {10.1093/mnras/stt1516}, \href
  {https://ui.adsabs.harvard.edu/abs/2013MNRAS.436...82D} {436, 82}

\bibitem[\protect\citeauthoryear{{Dainotti}, {Petrosian}, {Singal}  \&
  {Ostrowski}}{{Dainotti} et~al.}{2013b}]{Dainottietal2013b}
{Dainotti} M.~G.,  {Petrosian} V.,  {Singal} J.,   {Ostrowski} M.,  2013b,
  \mn@doi [\apj] {10.1088/0004-637X/774/2/157}, \href
  {https://ui.adsabs.harvard.edu/abs/2013ApJ...774..157D} {774, 157}

\bibitem[\protect\citeauthoryear{{Dainotti}, {Nagataki}, {Maeda}, {Postnikov}
  \& {Pian}}{{Dainotti} et~al.}{2017}]{Dainottietal2017}
{Dainotti} M.~G.,  {Nagataki} S.,  {Maeda} K.,  {Postnikov} S.,   {Pian} E.,
  2017, \mn@doi [\aap] {10.1051/0004-6361/201628384}, \href
  {https://ui.adsabs.harvard.edu/abs/2017A&A...600A..98D} {600, A98}

\bibitem[\protect\citeauthoryear{{\MakeLowercase{D}e Cruz Perez}, {Sola
  Peracaula}, {Gomez-Valent}  \& {Moreno-Pulido}}{{\MakeLowercase{D}e Cruz
  Perez} et~al.}{2021}]{deCruzetal2021}
{\MakeLowercase{D}e Cruz Perez} J.,  {Sola Peracaula} J.,  {Gomez-Valent} A.,
  {Moreno-Pulido} C.,  2021, preprint, \href
  {https://ui.adsabs.harvard.edu/abs/2021arXiv211007569D} {} (\mn@eprint {}
  {2110.07569})

\bibitem[\protect\citeauthoryear{Demianski, Piedipalumbo, Sawant  \&
  Amati}{Demianski et~al.}{2021}]{Demianskietal_2021}
Demianski M.,  Piedipalumbo E.,  Sawant D.,   Amati L.,  2021, \mn@doi [\mnras]
  {10.1093/mnras/stab1669}, 506, 903

\bibitem[\protect\citeauthoryear{{Dhawan}, {Alsing}  \& {Vagnozzi}}{{Dhawan}
  et~al.}{2021}]{Dhawanetal2021}
{Dhawan} S.,  {Alsing} J.,   {Vagnozzi} S.,  2021, \mn@doi [\mnras]
  {10.1093/mnrasl/slab058}, \href
  {https://ui.adsabs.harvard.edu/abs/2021MNRAS.506L...1D} {506, L1}

\bibitem[\protect\citeauthoryear{{Di Valentino} et~al.,}{{Di Valentino}
  et~al.}{2021a}]{DiValentinoetal2021b}
{Di Valentino} E.,  et~al., 2021a, \mn@doi [Classical and Quantum Gravity]
  {10.1088/1361-6382/ac086d}, \href
  {https://ui.adsabs.harvard.edu/abs/2021CQGra..38o3001D} {38, 153001}

\bibitem[\protect\citeauthoryear{{Di Valentino}, {Melchiorri}  \& {Silk}}{{Di
  Valentino} et~al.}{2021b}]{DiValentinoetal2021a}
{Di Valentino} E.,  {Melchiorri} A.,   {Silk} J.,  2021b, \mn@doi [\apjl]
  {10.3847/2041-8213/abe1c4}, \href
  {https://ui.adsabs.harvard.edu/abs/2021ApJ...908L...9D} {908, L9}

\bibitem[\protect\citeauthoryear{{\MakeLowercase{E}BOSS
  Collaboration}}{{\MakeLowercase{E}BOSS Collaboration}}{2021}]{eBOSS_2020}
{\MakeLowercase{E}BOSS Collaboration} 2021, \mn@doi [\prd]
  {10.1103/PhysRevD.103.083533}, \href
  {https://ui.adsabs.harvard.edu/abs/2021PhRvD.103h3533A} {103, 083533}

\bibitem[\protect\citeauthoryear{{Efstathiou} \& {Gratton}}{{Efstathiou} \&
  {Gratton}}{2020}]{efstathiou_gratton_2020}
{Efstathiou} G.,  {Gratton} S.,  2020, \mn@doi [\mnras]
  {10.1093/mnrasl/slaa093}, \href
  {https://ui.adsabs.harvard.edu/abs/2020MNRAS.496L..91E} {496, L91}

\bibitem[\protect\citeauthoryear{{Fana Dirirsa} et~al.,}{{Fana Dirirsa}
  et~al.}{2019}]{Dirirsa2019}
{Fana Dirirsa} F.,  et~al., 2019, \mn@doi [\apj] {10.3847/1538-4357/ab4e11},
  \href {https://ui.adsabs.harvard.edu/abs/2019ApJ...887...13F} {887, 13}

\bibitem[\protect\citeauthoryear{{Farooq}, {Ranjeet Madiyar}, {Crandall}  \&
  {Ratra}}{{Farooq} et~al.}{2017}]{Farooq_Ranjeet_Crandall_Ratra_2017}
{Farooq} O.,  {Ranjeet Madiyar} F.,  {Crandall} S.,   {Ratra} B.,  2017,
  \mn@doi [\apj] {10.3847/1538-4357/835/1/26}, \href
  {http://adsabs.harvard.edu/abs/2017ApJ...835...26F} {835, 26}

\bibitem[\protect\citeauthoryear{{Foreman-Mackey}, {Hogg}, {Lang}  \&
  {Goodman}}{{Foreman-Mackey} et~al.}{2013}]{emcee}
{Foreman-Mackey} D.,  {Hogg} D.~W.,  {Lang} D.,   {Goodman} J.,  2013, \mn@doi
  [\pasp] {10.1086/670067}, \href
  {https://ui.adsabs.harvard.edu/abs/2013PASP..125..306F} {125, 306}

\bibitem[\protect\citeauthoryear{{Gonz{\'a}lez-Mor{\'a}n}
  et~al.,}{{Gonz{\'a}lez-Mor{\'a}n} et~al.}{2019}]{GonzalezMoran2019}
{Gonz{\'a}lez-Mor{\'a}n} A.~L.,  et~al., 2019, \mn@doi [\mnras]
  {10.1093/mnras/stz1577}, \href
  {https://ui.adsabs.harvard.edu/abs/2019MNRAS.487.4669G} {487, 4669}

\bibitem[\protect\citeauthoryear{{Gonz{\'a}lez-Mor{\'a}n}
  et~al.,}{{Gonz{\'a}lez-Mor{\'a}n} et~al.}{2021}]{GonzalezMoranetal2021}
{Gonz{\'a}lez-Mor{\'a}n} A.~L.,  et~al., 2021, \mn@doi [\mnras]
  {10.1093/mnras/stab1385}, \href
  {https://ui.adsabs.harvard.edu/abs/2021MNRAS.tmp.1358G} {}

\bibitem[\protect\citeauthoryear{{Handley}}{{Handley}}{2019}]{Handley2019}
{Handley} W.,  2019, \mn@doi [\prd] {10.1103/PhysRevD.100.123517}, \href
  {https://ui.adsabs.harvard.edu/abs/2019PhRvD.100l3517H} {100, 123517}

\bibitem[\protect\citeauthoryear{{Hu}, {Wang}  \& {Dai}}{{Hu}
  et~al.}{2021}]{Huetal_2021}
{Hu} J.~P.,  {Wang} F.~Y.,   {Dai} Z.~G.,  2021, \mn@doi [\mnras]
  {10.1093/mnras/stab2180}, \href
  {https://ui.adsabs.harvard.edu/abs/2021MNRAS.507..730H} {507, 730}

\bibitem[\protect\citeauthoryear{{Johnson}, {Sangwan}  \&
  {Shankaranarayanan}}{{Johnson} et~al.}{2021}]{Johnsonetal2021}
{Johnson} J.~P.,  {Sangwan} A.,   {Shankaranarayanan} S.,  2021, preprint,
  \href {https://ui.adsabs.harvard.edu/abs/2021arXiv210212367J} {} (\mn@eprint
  {} {2102.12367})

\bibitem[\protect\citeauthoryear{{Khadka} \& {Ratra}}{{Khadka} \&
  {Ratra}}{2020a}]{KhadkaRatra2020a}
{Khadka} N.,  {Ratra} B.,  2020a, \mn@doi [\mnras] {10.1093/mnras/staa101},
  \href {https://ui.adsabs.harvard.edu/abs/2020MNRAS.492.4456K} {492, 4456}

\bibitem[\protect\citeauthoryear{{Khadka} \& {Ratra}}{{Khadka} \&
  {Ratra}}{2020b}]{KhadkaRatra2020b}
{Khadka} N.,  {Ratra} B.,  2020b, \mn@doi [\mnras] {10.1093/mnras/staa1855},
  \href {https://ui.adsabs.harvard.edu/abs/2020MNRAS.497..263K} {497, 263}

\bibitem[\protect\citeauthoryear{{Khadka} \& {Ratra}}{{Khadka} \&
  {Ratra}}{2020c}]{KhadkaRatra2020c}
{Khadka} N.,  {Ratra} B.,  2020c, \mn@doi [\mnras] {10.1093/mnras/staa2779},
  \href {https://ui.adsabs.harvard.edu/abs/2020MNRAS.499..391K} {499, 391}

\bibitem[\protect\citeauthoryear{{Khadka} \& {Ratra}}{{Khadka} \&
  {Ratra}}{2021a}]{KhadkaRatra2021a}
{Khadka} N.,  {Ratra} B.,  2021a, preprint, \href
  {https://ui.adsabs.harvard.edu/abs/2021arXiv210707600K} {} (\mn@eprint {}
  {2107.07600})

\bibitem[\protect\citeauthoryear{{Khadka} \& {Ratra}}{{Khadka} \&
  {Ratra}}{2021b}]{KhadkaRatra2021b}
{Khadka} N.,  {Ratra} B.,  2021b, \mn@doi [\mnras] {10.1093/mnras/stab486},
  \href {https://ui.adsabs.harvard.edu/abs/2021MNRAS.502.6140K} {502, 6140}

\bibitem[\protect\citeauthoryear{{Khadka}, {Yu}, {Zaja{\v{c}}ek},
  {Martinez-Aldama}, {Czerny}  \& {Ratra}}{{Khadka}
  et~al.}{2021a}]{Khadkaetal_2021a}
{Khadka} N.,  {Yu} Z.,  {Zaja{\v{c}}ek} M.,  {Martinez-Aldama} M.~L.,  {Czerny}
  B.,   {Ratra} B.,  2021a, \mn@doi [\mnras] {10.1093/mnras/stab2807}, \href
  {https://ui.adsabs.harvard.edu/abs/2021MNRAS.508.4722K} {508, 4722}

\bibitem[\protect\citeauthoryear{{Khadka}, {Luongo}, {Muccino}  \&
  {Ratra}}{{Khadka} et~al.}{2021b}]{Khadkaetal_2021b}
{Khadka} N.,  {Luongo} O.,  {Muccino} M.,   {Ratra} B.,  2021b, \mn@doi [\jcap]
  {10.1088/1475-7516/2021/09/042}, \href
  {https://ui.adsabs.harvard.edu/abs/2021JCAP...09..042K} {2021, 042}

\bibitem[\protect\citeauthoryear{{KiDS Collaboration}}{{KiDS
  Collaboration}}{2021}]{KiDSCollaboration2021}
{KiDS Collaboration} 2021, \mn@doi [\aap] {10.1051/0004-6361/202039805}, \href
  {https://ui.adsabs.harvard.edu/abs/2021A&A...649A..88T} {649, A88}

\bibitem[\protect\citeauthoryear{{Li}, {Du}  \& {Xu}}{{Li}
  et~al.}{2020}]{li_etal_2020}
{Li} E.-K.,  {Du} M.,   {Xu} L.,  2020, \mn@doi [\mnras]
  {10.1093/mnras/stz3308}, \href
  {https://ui.adsabs.harvard.edu/abs/2020MNRAS.491.4960L} {491, 4960}

\bibitem[\protect\citeauthoryear{{Li}, {Keeley}, {Shafieloo}, {Zheng}, {Cao},
  {Biesiada}  \& {Zhu}}{{Li} et~al.}{2021}]{Lietal2021}
{Li} X.,  {Keeley} R.~E.,  {Shafieloo} A.,  {Zheng} X.,  {Cao} S.,  {Biesiada}
  M.,   {Zhu} Z.-H.,  2021, \mn@doi [\mnras] {10.1093/mnras/stab2154}, \href
  {https://ui.adsabs.harvard.edu/abs/2021MNRAS.507..919L} {507, 919}

\bibitem[\protect\citeauthoryear{{Lian}, {Cao}, {Biesiada}, {Chen}, {Zhang}  \&
  {Guo}}{{Lian} et~al.}{2021}]{Lian_etal_2021}
{Lian} Y.,  {Cao} S.,  {Biesiada} M.,  {Chen} Y.,  {Zhang} Y.,   {Guo} W.,
  2021, \mn@doi [\mnras] {10.1093/mnras/stab1373}, \href
  {https://ui.adsabs.harvard.edu/abs/2021MNRAS.505.2111L} {505, 2111}

\bibitem[\protect\citeauthoryear{Luongo \& Muccino}{Luongo \&
  Muccino}{2021}]{galaxies9040077}
Luongo O.,  Muccino M.,  2021, \mn@doi [Galaxies] {10.3390/galaxies9040077}, 9

\bibitem[\protect\citeauthoryear{{Luongo}, {Muccino}, {Colg{\'a}in},
  {Sheikh-Jabbari}  \& {Yin}}{{Luongo} et~al.}{2021}]{Luongoetal2021}
{Luongo} O.,  {Muccino} M.,  {Colg{\'a}in} E.~{\'O}.,  {Sheikh-Jabbari} M.~M.,
   {Yin} L.,  2021, preprint, \href
  {https://ui.adsabs.harvard.edu/abs/2021arXiv210813228L} {} (\mn@eprint {}
  {2108.13228})

\bibitem[\protect\citeauthoryear{{Lusso} et~al.,}{{Lusso}
  et~al.}{2020}]{Lussoetal2020}
{Lusso} E.,  et~al., 2020, \mn@doi [\aap] {10.1051/0004-6361/202038899}, \href
  {https://ui.adsabs.harvard.edu/abs/2020A&A...642A.150L} {642, A150}

\bibitem[\protect\citeauthoryear{{Mania} \& {Ratra}}{{Mania} \&
  {Ratra}}{2012}]{Mania_2012}
{Mania} D.,  {Ratra} B.,  2012, \mn@doi [Physics Letters B]
  {10.1016/j.physletb.2012.07.011}, \href
  {https://ui.adsabs.harvard.edu/abs/2012PhLB..715....9M} {715, 9}

\bibitem[\protect\citeauthoryear{{Ooba}, {Ratra}  \& {Sugiyama}}{{Ooba}
  et~al.}{2018a}]{ooba_etal_2018a}
{Ooba} J.,  {Ratra} B.,   {Sugiyama} N.,  2018a, \mn@doi [\apj]
  {10.3847/1538-4357/aad633}, \href
  {http://adsabs.harvard.edu/abs/2018ApJ...864...80O} {864, 80}

\bibitem[\protect\citeauthoryear{{Ooba}, {Ratra}  \& {Sugiyama}}{{Ooba}
  et~al.}{2018b}]{ooba_etal_2018b}
{Ooba} J.,  {Ratra} B.,   {Sugiyama} N.,  2018b, \mn@doi [\apj]
  {10.3847/1538-4357/aadcf3}, \href
  {http://adsabs.harvard.edu/abs/2018ApJ...866...68O} {866, 68}

\bibitem[\protect\citeauthoryear{{Ooba}, {Ratra}  \& {Sugiyama}}{{Ooba}
  et~al.}{2018c}]{ooba_etal_2018c}
{Ooba} J.,  {Ratra} B.,   {Sugiyama} N.,  2018c, \mn@doi [\apj]
  {10.3847/1538-4357/aaec6f}, \href
  {http://adsabs.harvard.edu/abs/2018ApJ...869...34O} {869, 34}

\bibitem[\protect\citeauthoryear{{Ooba}, {Ratra}  \& {Sugiyama}}{{Ooba}
  et~al.}{2019}]{ooba_etal_2019}
{Ooba} J.,  {Ratra} B.,   {Sugiyama} N.,  2019, \mn@doi [\apss]
  {10.1007/s10509-019-3663-4}, \href
  {https://ui.adsabs.harvard.edu/abs/2019Ap&SS.364..176O} {364, 176}

\bibitem[\protect\citeauthoryear{{Park} \& {Ratra}}{{Park} \&
  {Ratra}}{2018}]{park_ratra_2018}
{Park} C.-G.,  {Ratra} B.,  2018, \mn@doi [\apj] {10.3847/1538-4357/aae82d},
  \href {http://adsabs.harvard.edu/abs/2018ApJ...868...83P} {868, 83}

\bibitem[\protect\citeauthoryear{{Park} \& {Ratra}}{{Park} \&
  {Ratra}}{2019a}]{park_ratra_2019a}
{Park} C.-G.,  {Ratra} B.,  2019a, \mn@doi [\apss] {10.1007/s10509-019-3567-3},
  \href {https://ui.adsabs.harvard.edu/abs/2019Ap&SS.364...82P} {364, 82}

\bibitem[\protect\citeauthoryear{{Park} \& {Ratra}}{{Park} \&
  {Ratra}}{2019b}]{park_ratra_2019b}
{Park} C.-G.,  {Ratra} B.,  2019b, \mn@doi [\apss] {10.1007/s10509-019-3627-8},
  \href {https://ui.adsabs.harvard.edu/abs/2019Ap&SS.364..134P} {364, 134}

\bibitem[\protect\citeauthoryear{{Park} \& {Ratra}}{{Park} \&
  {Ratra}}{2019c}]{park_ratra_2019c}
{Park} C.-G.,  {Ratra} B.,  2019c, \mn@doi [\apj] {10.3847/1538-4357/ab3641},
  \href {https://ui.adsabs.harvard.edu/abs/2019ApJ...882..158P} {882, 158}

\bibitem[\protect\citeauthoryear{{Park} \& {Ratra}}{{Park} \&
  {Ratra}}{2020}]{park_ratra_2020}
{Park} C.-G.,  {Ratra} B.,  2020, \mn@doi [\prd] {10.1103/PhysRevD.101.083508},
  \href {https://ui.adsabs.harvard.edu/abs/2020PhRvD.101h3508P} {101, 083508}

\bibitem[\protect\citeauthoryear{{Pavlov}, {Westmoreland}, {Saaidi}  \&
  {Ratra}}{{Pavlov} et~al.}{2013}]{pavlov13}
{Pavlov} A.,  {Westmoreland} S.,  {Saaidi} K.,   {Ratra} B.,  2013, \mn@doi
  [\prd] {10.1103/PhysRevD.88.123513}, \href
  {http://adsabs.harvard.edu/abs/2013PhRvD..88l3513P} {88, 123513}

\bibitem[\protect\citeauthoryear{{Peebles}}{{Peebles}}{1984}]{peeb84}
{Peebles} P.~J.~E.,  1984, \mn@doi [\apj] {10.1086/162425}, \href
  {http://adsabs.harvard.edu/abs/1984ApJ...284..439P} {284, 439}

\bibitem[\protect\citeauthoryear{{Peebles} \& {Ratra}}{{Peebles} \&
  {Ratra}}{1988}]{peebrat88}
{Peebles} P.~J.~E.,  {Ratra} B.,  1988, \mn@doi [\apjl] {10.1086/185100}, \href
  {http://adsabs.harvard.edu/abs/1988ApJ...325L..17P} {325, L17}

\bibitem[\protect\citeauthoryear{{Perivolaropoulos} \&
  {Skara}}{{Perivolaropoulos} \& {Skara}}{2021}]{PerivolaropoulosSkara2021}
{Perivolaropoulos} L.,  {Skara} F.,  2021, preprint, \href
  {https://ui.adsabs.harvard.edu/abs/2021arXiv210505208P} {} (\mn@eprint {}
  {2105.05208})

\bibitem[\protect\citeauthoryear{{Planck Collaboration}}{{Planck
  Collaboration}}{2020}]{planck2018b}
{Planck Collaboration} 2020, \mn@doi [\aap] {10.1051/0004-6361/201833910},
  \href {https://ui.adsabs.harvard.edu/abs/2020A&A...641A...6P} {641, A6}

\bibitem[\protect\citeauthoryear{{Rana}, {Jain}, {Mahajan}  \&
  {Mukherjee}}{{Rana} et~al.}{2017}]{rana_jain_mahajan_mukherjee_2017}
{Rana} A.,  {Jain} D.,  {Mahajan} S.,   {Mukherjee} A.,  2017, \mn@doi [\jcap]
  {10.1088/1475-7516/2017/03/028}, \href
  {http://adsabs.harvard.edu/abs/2017JCAP...03..028R} {3, 028}

\bibitem[\protect\citeauthoryear{{Ratra} \& {Peebles}}{{Ratra} \&
  {Peebles}}{1988}]{ratpeeb88}
{Ratra} B.,  {Peebles} P.~J.~E.,  1988, \mn@doi [\prd]
  {10.1103/PhysRevD.37.3406}, \href
  {http://adsabs.harvard.edu/abs/1988PhRvD..37.3406R} {37, 3406}

\bibitem[\protect\citeauthoryear{{Rezaei}, {Peracaula}  \&
  {Malekjani}}{{Rezaei} et~al.}{2021}]{Rezaeietal2021}
{Rezaei} M.,  {Peracaula} J.~S.,   {Malekjani} M.,  2021, \mn@doi [\mnras]
  {10.1093/mnras/stab3117}, \href
  {https://ui.adsabs.harvard.edu/abs/2021MNRAS.tmp.2893R} {}

\bibitem[\protect\citeauthoryear{{Risaliti} \& {Lusso}}{{Risaliti} \&
  {Lusso}}{2015}]{RisalitiLusso2015}
{Risaliti} G.,  {Lusso} E.,  2015, \mn@doi [\apj] {10.1088/0004-637X/815/1/33},
  \href {https://ui.adsabs.harvard.edu/abs/2015ApJ...815...33R} {815, 33}

\bibitem[\protect\citeauthoryear{{Risaliti} \& {Lusso}}{{Risaliti} \&
  {Lusso}}{2019}]{RisalitiLusso2019}
{Risaliti} G.,  {Lusso} E.,  2019, \mn@doi [Nature Astronomy]
  {10.1038/s41550-018-0657-z}, \href
  {https://ui.adsabs.harvard.edu/abs/2019NatAs...3..272R} {3, 272}

\bibitem[\protect\citeauthoryear{{Ryan}, {Doshi}  \& {Ratra}}{{Ryan}
  et~al.}{2018}]{Ryan_1}
{Ryan} J.,  {Doshi} S.,   {Ratra} B.,  2018, \mn@doi [\mnras]
  {10.1093/mnras/sty1922}, \href
  {https://ui.adsabs.harvard.edu/abs/2018MNRAS.480..759R} {480, 759}

\bibitem[\protect\citeauthoryear{{Ryan}, {Chen}  \& {Ratra}}{{Ryan}
  et~al.}{2019}]{Ryanetal2019}
{Ryan} J.,  {Chen} Y.,   {Ratra} B.,  2019, \mn@doi [\mnras]
  {10.1093/mnras/stz1966}, \href
  {https://ui.adsabs.harvard.edu/abs/2019MNRAS.488.3844R} {488, 3844}

\bibitem[\protect\citeauthoryear{{Salvaterra} et~al.,}{{Salvaterra}
  et~al.}{2009}]{Salvaterraetal2009}
{Salvaterra} R.,  et~al., 2009, \mn@doi [\nat] {10.1038/nature08445}, \href
  {https://ui.adsabs.harvard.edu/abs/2009Natur.461.1258S} {461, 1258}

\bibitem[\protect\citeauthoryear{{Samushia} \& {Ratra}}{{Samushia} \&
  {Ratra}}{2010}]{samushia_ratra_2010}
{Samushia} L.,  {Ratra} B.,  2010, \mn@doi [\apj]
  {10.1088/0004-637X/714/2/1347}, \href
  {http://adsabs.harvard.edu/abs/2010ApJ...714.1347S} {714, 1347}

\bibitem[\protect\citeauthoryear{{Sangwan}, {Tripathi}  \& {Jassal}}{{Sangwan}
  et~al.}{2018}]{Sangwanetal2018}
{Sangwan} A.,  {Tripathi} A.,   {Jassal} H.~K.,  2018, preprint, \href
  {https://ui.adsabs.harvard.edu/abs/2018arXiv180409350S} {} (\mn@eprint {}
  {1804.09350})

\bibitem[\protect\citeauthoryear{{Scolnic} et~al.,}{{Scolnic}
  et~al.}{2018}]{scolnic_et_al_2018}
{Scolnic} D.~M.,  et~al., 2018, \mn@doi [\apj] {10.3847/1538-4357/aab9bb},
  \href {http://adsabs.harvard.edu/abs/2018ApJ...859..101S} {859, 101}

\bibitem[\protect\citeauthoryear{{Singh}, {Sangwan}  \& {Jassal}}{{Singh}
  et~al.}{2019}]{singh_etal_2019}
{Singh} A.,  {Sangwan} A.,   {Jassal} H.~K.,  2019, \mn@doi [\jcap]
  {10.1088/1475-7516/2019/04/047}, \href
  {https://ui.adsabs.harvard.edu/abs/2019JCAP...04..047S} {2019, 047}

\bibitem[\protect\citeauthoryear{{Sinha} \& {Banerjee}}{{Sinha} \&
  {Banerjee}}{2021}]{SinhaBanerjee2021}
{Sinha} S.,  {Banerjee} N.,  2021, \mn@doi [\jcap]
  {10.1088/1475-7516/2021/04/060}, \href
  {https://ui.adsabs.harvard.edu/abs/2021JCAP...04..060S} {2021, 060}

\bibitem[\protect\citeauthoryear{{Sol{\`a} Peracaula}, {de Cruz P{\'e}rez}  \&
  {G{\'o}mez-Valent}}{{Sol{\`a} Peracaula}
  et~al.}{2018}]{SolaPeracaulaetal2018}
{Sol{\`a} Peracaula} J.,  {de Cruz P{\'e}rez} J.,   {G{\'o}mez-Valent} A.,
  2018, \mn@doi [\mnras] {10.1093/mnras/sty1253}, \href
  {https://ui.adsabs.harvard.edu/abs/2018MNRAS.478.4357S} {478, 4357}

\bibitem[\protect\citeauthoryear{{Sol{\`a} Peracaula}, {G{\'o}mez-Valent}  \&
  {de Cruz P{\'e}rez}}{{Sol{\`a} Peracaula}
  et~al.}{2019}]{SolaPercaulaetal2019}
{Sol{\`a} Peracaula} J.,  {G{\'o}mez-Valent} A.,   {de Cruz P{\'e}rez} J.,
  2019, \mn@doi [Physics of the Dark Universe] {10.1016/j.dark.2019.100311},
  \href {https://ui.adsabs.harvard.edu/abs/2019PDU....25..311S} {25, 100311}

\bibitem[\protect\citeauthoryear{{Tanvir} et~al.,}{{Tanvir}
  et~al.}{2009}]{Tanviretal2009}
{Tanvir} N.~R.,  et~al., 2009, \mn@doi [\nat] {10.1038/nature08459}, \href
  {https://ui.adsabs.harvard.edu/abs/2009Natur.461.1254T} {461, 1254}

\bibitem[\protect\citeauthoryear{{Ure{\~n}a-L{\'o}pez} \&
  {Roy}}{{Ure{\~n}a-L{\'o}pez} \& {Roy}}{2020}]{UrenaLopezRoy2020}
{Ure{\~n}a-L{\'o}pez} L.~A.,  {Roy} N.,  2020, \mn@doi [\prd]
  {10.1103/PhysRevD.102.063510}, \href
  {https://ui.adsabs.harvard.edu/abs/2020PhRvD.102f3510U} {102, 063510}

\bibitem[\protect\citeauthoryear{{Vagnozzi}, {Di Valentino}, {Gariazzo},
  {Melchiorri}, {Mena}  \& {Silk}}{{Vagnozzi} et~al.}{2021a}]{Vagnozzi_2020a}
{Vagnozzi} S.,  {Di Valentino} E.,  {Gariazzo} S.,  {Melchiorri} A.,  {Mena}
  O.,   {Silk} J.,  2021a, \mn@doi [Physics of the Dark Universe]
  {10.1016/j.dark.2021.100851}, \href
  {https://ui.adsabs.harvard.edu/abs/2021PDU....3300851V} {33, 100851}

\bibitem[\protect\citeauthoryear{{Vagnozzi}, {Loeb}  \& {Moresco}}{{Vagnozzi}
  et~al.}{2021b}]{Vagnozzi_2020b}
{Vagnozzi} S.,  {Loeb} A.,   {Moresco} M.,  2021b, \mn@doi [\apj]
  {10.3847/1538-4357/abd4df}, \href
  {https://ui.adsabs.harvard.edu/abs/2021ApJ...908...84V} {908, 84}

\bibitem[\protect\citeauthoryear{{Velasquez-Toribio} \&
  {Fabris}}{{Velasquez-Toribio} \& {Fabris}}{2020}]{Velasquez-Toribio_2020}
{Velasquez-Toribio} A.~M.,  {Fabris} J.~C.,  2020, \mn@doi [European Physical
  Journal C] {10.1140/epjc/s10052-020-08785-z}, \href
  {https://ui.adsabs.harvard.edu/abs/2020EPJC...80.1210V} {80, 1210}

\bibitem[\protect\citeauthoryear{{Wang}, {Dai}  \& {Liang}}{{Wang}
  et~al.}{2015}]{Wangetal2015}
{Wang} F.~Y.,  {Dai} Z.~G.,   {Liang} E.~W.,  2015, \mn@doi [\nar]
  {10.1016/j.newar.2015.03.001}, \href
  {https://ui.adsabs.harvard.edu/abs/2015NewAR..67....1W} {67, 1}

\bibitem[\protect\citeauthoryear{{Wang}, {Wang}, {Cheng}  \& {Dai}}{{Wang}
  et~al.}{2016}]{Wang_2016}
{Wang} J.~S.,  {Wang} F.~Y.,  {Cheng} K.~S.,   {Dai} Z.~G.,  2016, \mn@doi
  [\aap] {10.1051/0004-6361/201526485}, \href
  {https://ui.adsabs.harvard.edu/abs/2016A&A...585A..68W} {585, A68}

\bibitem[\protect\citeauthoryear{{Wang}, {Hu}, {Zhang}  \& {Dai}}{{Wang}
  et~al.}{2021}]{Wangetal_2021}
{Wang} F.~Y.,  {Hu} J.~P.,  {Zhang} G.~Q.,   {Dai} Z.~G.,  2021, preprint,
  \href {https://ui.adsabs.harvard.edu/abs/2021arXiv210614155W} {} (\mn@eprint
  {} {2106.14155})

\bibitem[\protect\citeauthoryear{{Wei}}{{Wei}}{2018}]{wei_2018}
{Wei} J.-J.,  2018, \mn@doi [\apj] {10.3847/1538-4357/aae696}, \href
  {http://adsabs.harvard.edu/abs/2018ApJ...868...29W} {868, 29}

\bibitem[\protect\citeauthoryear{{Xu}, {Chen}, {Xu}  \& {Cao}}{{Xu}
  et~al.}{2021}]{Xuetal2021}
{Xu} T.,  {Chen} Y.,  {Xu} L.,   {Cao} S.,  2021, preprint, \href
  {https://ui.adsabs.harvard.edu/abs/2021arXiv210902453X} {} (\mn@eprint {}
  {2109.02453})

\bibitem[\protect\citeauthoryear{{Yang}, {Banerjee}  \& {{\'O}
  Colg{\'a}in}}{{Yang} et~al.}{2020}]{Yangetal2020}
{Yang} T.,  {Banerjee} A.,   {{\'O} Colg{\'a}in} E.,  2020, \mn@doi [\prd]
  {10.1103/PhysRevD.102.123532}, \href
  {https://ui.adsabs.harvard.edu/abs/2020PhRvD.102l3532Y} {102, 123532}

\bibitem[\protect\citeauthoryear{{Yu}, {Ratra}  \& {Wang}}{{Yu}
  et~al.}{2018}]{yu_etal_2018}
{Yu} H.,  {Ratra} B.,   {Wang} F.-Y.,  2018, \mn@doi [\apj]
  {10.3847/1538-4357/aab0a2}, \href
  {http://adsabs.harvard.edu/abs/2018ApJ...856....3Y} {856, 3}

\bibitem[\protect\citeauthoryear{{Zhai}, {Blanton}, {Slosar}  \&
  {Tinker}}{{Zhai} et~al.}{2017}]{Zhaietal2017}
{Zhai} Z.,  {Blanton} M.,  {Slosar} A.,   {Tinker} J.,  2017, \mn@doi [\apj]
  {10.3847/1538-4357/aa9888}, \href
  {https://ui.adsabs.harvard.edu/abs/2017ApJ...850..183Z} {850, 183}

\bibitem[\protect\citeauthoryear{{Zhao} \& {Xia}}{{Zhao} \&
  {Xia}}{2021}]{ZhaoXia2021}
{Zhao} D.,  {Xia} J.-Q.,  2021, \mn@doi [European Physical Journal C]
  {10.1140/epjc/s10052-021-09491-0}, \href
  {https://ui.adsabs.harvard.edu/abs/2021EPJC...81..694Z} {81, 694}

\bibitem[\protect\citeauthoryear{{Zheng}, {Cao}, {Biesiada}, {Li}, {Liu}  \&
  {Liu}}{{Zheng} et~al.}{2021}]{Zhengetal2021}
{Zheng} X.,  {Cao} S.,  {Biesiada} M.,  {Li} X.,  {Liu} T.,   {Liu} Y.,  2021,
  \mn@doi [Science China Physics, Mechanics, and Astronomy]
  {10.1007/s11433-020-1664-9}, \href
  {https://ui.adsabs.harvard.edu/abs/2021SCPMA..6459511Z} {64, 259511}

\makeatother
\end{thebibliography}







\bsp	
\label{lastpage}
\end{document}